%% file: project.tex
\documentclass[sigconf, nonacm]{acmart}

\newcommand\paperdoi{XX.XX/XXX.XX}

\newcommand\paperavailabilityurl{https://github.com/VCityTeam/ConVer-G}
\newcommand\paperdatasetdoi{https://zenodo.org/records/17198832}

\newcommand\paperpagestyle{plain} 

\usepackage{listings}
\usepackage{xcolor}
\usepackage{multirow}
\usepackage{stfloats}
\usepackage{graphicx}
\usepackage{pifont}
\usepackage[table]{xcolor}

\definecolor{valuecolor}{HTML}{fff6cb}
\definecolor{idcolor}{HTML}{ffe8cb}
\definecolor{condensedcolor}{HTML}{ffcaca}
\definecolor{unboundcolor}{HTML}{bfafc1}
\definecolor{RowGray}{HTML}{EBEBEB}

\definecolor{blazegraphcolor}{HTML}{B3D8F8}
\definecolor{jenacolor}{HTML}{E6E6FA}
\definecolor{quaqueflatcolor}{HTML}{FFD580}
\definecolor{quaquecondensedcolor}{HTML}{90ee90}

\begin{document}
\title{Condensed Representation for Snapshot-Based RDF Graphs}

\include{content/style-listings}

\author{Jey Puget Gil}
\email{jey.puget-gil@liris.cnrs.fr}
\orcid{0009-0006-6198-7488}
\affiliation{%
  \streetaddress{20, Avenue Albert Einstein}
  \institution{Universite Claude Bernard Lyon 1, CNRS, INSA Lyon, LIRIS, UMR 5205}
  \city{Villeurbanne}
  \country{FRANCE}
  \postcode{69621}
}

\author{Emmanuel Coquery}
\email{emmanuel.coquery@univ-lyon1.fr}
\orcid{}
\affiliation{%
  \streetaddress{20, Avenue Albert Einstein}
  \institution{Universite Claude Bernard Lyon 1, CNRS, INSA Lyon, LIRIS, UMR 5205}
  \city{Villeurbanne}
  \country{FRANCE}
  \postcode{69621}
}

\author{John Samuel}
\email{john.samuel@cpe.fr}
\orcid{0000-0001-8721-7007}
\affiliation{%
  \streetaddress{20, Avenue Albert Einstein}
  \institution{CPE Lyon, CNRS, INSA Lyon, Universite Claude Bernard Lyon 1, Université Lumière Lyon 2, Ecole Centrale de Lyon, LIRIS, UMR 5205}
  \city{Villeurbanne}
  \country{FRANCE}
  \postcode{69621}
}

\author{Gilles Gesquière}
\email{gilles.gesquiere@univ-lyon2.fr}
\orcid{0000-0001-7088-1067}
\affiliation{%
  \institution{Université Lumière Lyon 2, CNRS, Université Claude Bernard Lyon 1, INSA Lyon, Ecole Centrale de Lyon, LIRIS, UMR 5205}
  \streetaddress{20, Avenue Albert Einstein}
  \city{Villeurbanne}
  \country{FRANCE}
  \postcode{69621}
}

\renewcommand{\shortauthors}{Puget Gil et al.}

\include{content/macros}

\include{content/abstract}

\maketitle

\pagestyle{\paperpagestyle}

\ifdefempty{\paperavailabilityurl}{}{
\begingroup\small\noindent\raggedright\textbf{Paper Artifact Availability:}\\
The source code has been made available at \url{\paperavailabilityurl}.
\ifdefempty{\paperdatasetdoi}{}{The detailed results are available at \url{\paperdatasetdoi}.}
\endgroup
}



\begin{teaserfigure}
  \includegraphics[width=\textwidth]{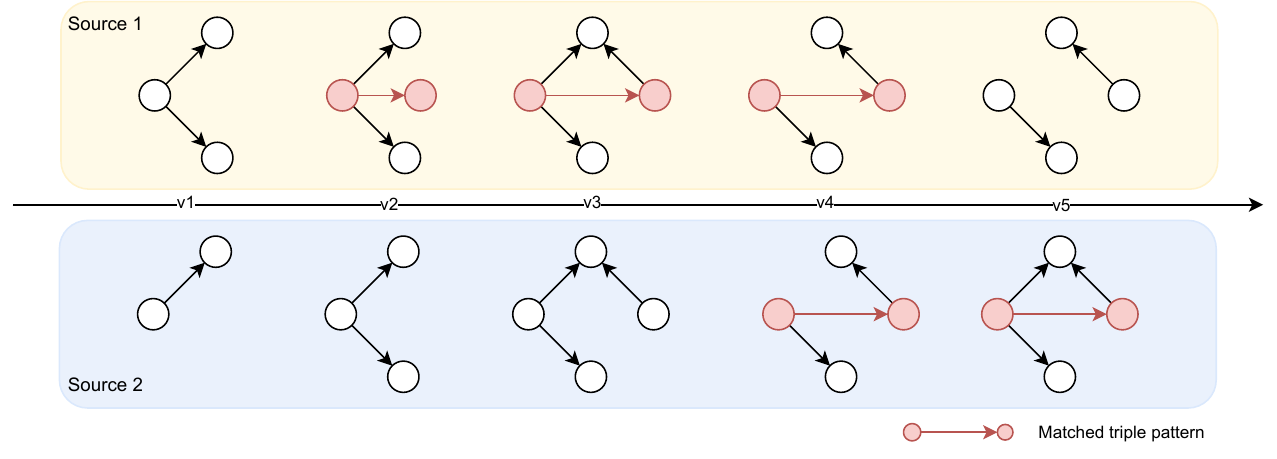}
  \caption{Matched triple pattern on versioned RDF graphs}
  \Description{Matched triple pattern on versioned RDF graphs}
  \label{fig:teaser}
\end{teaserfigure}

\input{content/chapters/introduction}
\input{content/chapters/state-of-the-art}
\input{content/chapters/notations}
\input{content/chapters/problem}
\input{content/chapters/results}
\input{content/chapters/discussion}
\input{content/acknowledments}


\bibliographystyle{ACM-Reference-Format}
\bibliography{bibliography}

\input{content/appendix}

\end{document}

%% file: content/style-listings.tex
\lstdefinelanguage{SPARQL}{
  basicstyle=\small\ttfamily,
  columns=fullflexible,
  breaklines=true,
  sensitive=true,
  tabsize = 2,
  showstringspaces=false,
  morecomment=[l][\color{green}]{\#},       
  morecomment=[n][\color{red}]{<}{>}, 
  morestring=[b][\color{orange}]{\"},  
  keywordsprefix=?,
  classoffset=0,
  keywordstyle=\color{darkgray},
  morekeywords={},
  classoffset=1,
  keywordstyle=\color{purple},
  morekeywords={rdf,vers},
  classoffset=2,
  keywordstyle=\color{blue},
  morekeywords={
    SELECT,CONSTRUCT,DESCRIBE,ASK,WHERE,FROM,NAMED,PREFIX,BASE,OPTIONAL,
    FILTER,GRAPH,LIMIT,OFFSET,SERVICE,UNION,EXISTS,NOT,BINDINGS,MINUS,a,GROUP,BY,ORDER,ASC,DESC
  }
}

%% file: content/macros.tex
\def\ojoin{\setbox0=\hbox{$\bowtie$}%
  \rule[0.175ex]{.25em}{.5pt}\llap{\rule[0.85ex]{.25em}{.5pt}}}
\def\leftouterjoin{\mathbin{\ojoin\mkern-6.25mu\bowtie}}

\newcommand{\literal}{
    l
}
\newcommand{\literals}{
    \MakeUppercase{\literal}
}
\newcommand{\literalset}{
    \mathcal{\literals}
}
\newcommand{\iri}{
    i
}
\newcommand{\iris}{
    \MakeUppercase{\iri}
}
\newcommand{\irisets}{
    \mathcal{\iris}
}
\newcommand{\term}{
    t
}
\newcommand{\termfs}{
    \MakeUppercase{\term}
}
\newcommand{\termfset}{
    \mathcal{\termfs}
}
\newcommand{\termf}[1]{
    \term_{#1}
}
\newcommand{\subject}{
    s
}
\newcommand{\subjectterm}{
    \termf{\subject}
}
\newcommand{\predicate}{
    p
}
\newcommand{\predicateterm}{
    \termf{\predicate}
}
\newcommand{\ngraph}{
    g
}
\newcommand{\object}{
    o
}
\newcommand{\objectterm}{
    \termf{\object}
}
\newcommand{\namedgraphterm}{
    \termf{\ngraph}
}
\newcommand{\termset}{
    \mathcal{\MakeUppercase{\term}}
}
\newcommand{\namedgraphs}{
    \termset_{\ngraph}
}
\newcommand{\interpretT}[1]{
    \llbracket #1 \rrbracket
}
\newcommand{\applyEBV}[2]{
    #1 \sim_{EBV} #2
}
\newcommand{\nnamedgraph}[1]{
    \termf{\ngraph#1}
}
\newcommand{\triple}{
    tr
}
\newcommand{\triples}{
    \MakeUppercase{\triple}
}
\newcommand{\tripleset}{
    \mathcal{\triples}
}
\newcommand{\quadf}{
    q
}
\newcommand{\quadfs}{
    \MakeUppercase{\quadf}
}
\newcommand{\quadset}{
    \mathcal{\quadfs}
}
\newcommand{\rdftriple}{
    (\subjectterm, \predicateterm, \objectterm)
}
\newcommand{\rdfquad}{
    (\subjectterm, \predicateterm, \objectterm, \namedgraphterm)
}
\newcommand{\mping}{
    m
}
\newcommand{\mapping}[1][]{
    \mping_{#1}
}
\newcommand{\mappings}{
    \MakeUppercase{\mping}
}
\newcommand{\mappingset}{
    \mathcal{\mappings}
}
\newcommand{\solutionsequence}{
    seq
}

\newcommand{\query}{
    qu
}
\newcommand{\queries}{
    \MakeUppercase{\query}
}
\newcommand{\queryset}{
    \mathcal{\queries}
}

\newcommand{\vers}[1]{
    \versions(#1)
}
\newcommand{\version}{
    v
}
\newcommand{\versions}{
    \MakeUppercase{\version}
}
\newcommand{\versionset}{
    \mathcal{\versions}
}
\newcommand{\fsversions}{
    \versions^\condense
}
\newcommand{\vi}{
    {\version}{\iri}
}
\newcommand{\vis}{
    \MakeUppercase{\vi}
}
\newcommand{\viset}{
    \mathcal{\vis}
}
\newcommand{\vistg}[1]{
    \vis^{-1}_{\namedgraphterm}(#1)
}
\newcommand{\visv}[1]{
    \vis^{-1}_{\version}(#1)
}

\newcommand{\vifunction}{
    \vis: \versionset \times \irisets \to \viset
}
\newcommand{\versionedquad}{
    {\version}{\quadf}
}
\newcommand{\versionedquads}{
    \MakeUppercase{\versionedquad}
}
\newcommand{\versionedquadset}{
    \mathcal{\versionedquads}
}

\newcommand{\versionedrdfquad}{
    (\subjectterm, \predicateterm, \objectterm, \namedgraphterm, \version)
}
\newcommand{\domain}[1]{
    \mathcal{D}(#1)
}
\newcommand{\rdfdataset}{
    d
}
\newcommand{\flatten}{
    F
}
\newcommand{\condense}{
    C
}
\newcommand{\flatmodel}{
    \rdfdataset_{\flatten}
}
\newcommand{\condensedmodel}{
    \rdfdataset_{\condense}
}

\newcommand{\algebra}{
    A
}
\newcommand{\algebraf}{
    \algebra_{\flatten}
}
\newcommand{\algebrac}{
    \algebra_{\condense}
}
\newcommand{\algebras}{
    \mathcal{\algebra}
}
\newcommand{\algebracset}{
    \mathcal{\algebra}_{\condense}
}
\newcommand{\algebrafset}{
    \mathcal{\algebra}_{\flatten}
}
\newcommand{\cpt}{
    \sim
}
\newcommand{\cptf}{
    \cpt_{\flatten}
}
\newcommand{\cptc}{
    \cpt_{\condense}
}
\newcommand{\cptffunc}[2]{
    #1 \cptf #2
}
\newcommand{\cptcfunc}[2]{
    #1 \cptc #2
}
\newcommand{\eval}{
    eval
}
\newcommand{\evalE}[1]{
    \eval_{\expression}(#1)
}
\newcommand{\evalLE}[1]{
    \eval_{\expressionset}(#1)
}
\newcommand{\evalf}[1]{
    \eval_\flatten(#1)
}
\newcommand{\evalc}[1]{
    \eval_\condense(#1)
}
\newcommand{\join}{
    \bowtie
}
\newcommand{\joinf}[1]{
    \join_\flatten(#1)
}
\newcommand{\leftouterjoinf}[1]{
    \leftouterjoin_\flatten(#1)
}
\newcommand{\joinc}[1]{
    \join_\condense(#1)
}
\newcommand{\leftouterjoinc}[1]{
    \leftouterjoin_\condense(#1)
}
\newcommand{\union}{
    \cup
}
\newcommand{\unionf}[1]{
    \union_\flatten(#1)
}
\newcommand{\unionc}[1]{
    \union_\condense(#1)
}
\newcommand{\diff}{
    \setminus
}
\newcommand{\diffc}[1]{
    \diff_\condense(#1)
}
\newcommand{\difff}[1]{
    \diff_\flatten(#1)
}
\newcommand{\merge}{
    \curlyvee
}
\newcommand{\mergef}[1]{
    \merge_\flatten(#1)
}
\newcommand{\mergec}[1]{
    \merge_\condense(#1)
}
\newcommand{\vals}[1][]{
    \valuer_{\mapping[#1]}
}
\newcommand{\valsfunc}[2]{
    \vals[#1](#2)
}
\newcommand{\expand}{
    \Xi
}
\newcommand{\expandfunc}[1]{
    \expand(#1)
}
\newcommand{\permut}{
    \pi
}
\newcommand{\permutfunc}[1]{
    \permut(#1)
}
\newcommand{\triplepattern}{
    TP
}
\newcommand{\triplepatternf}[1]{
    \triplepattern_{\flatten}(#1)
}
\newcommand{\quadpattern}{
    QP
}
\newcommand{\quadpatternc}[1]{
    \quadpattern_{\condense}(#1)
}
\newcommand{\quadpatternf}[1]{
    \quadpattern_{\flatten}(#1)
}
\newcommand{\filter}{
    \sigma
}
\newcommand{\filterf}[1]{
    \filter_\flatten(#1)
}
\newcommand{\filterc}[1]{
    \filter_\condense(#1)
}
\newcommand{\projection}{
    \Pi
}
\newcommand{\expression}{
    e
}
\newcommand{\expressions}{
    \MakeUppercase{\expression}
}
\newcommand{\expressionset}{
    \mathcal{\expressions}
}
\newcommand{\variables}[1]{
    vars(#1)
}
\newcommand{\group}{
    Gr
}
\newcommand{\groupF}{
    \group_\flatten
}
\newcommand{\groupFFunc}[1][\expressionset_n, \Omega]{
    \groupF(#1)
}
\newcommand{\groupC}{
    \group_\condense
}
\newcommand{\groupCFunc}[1][\expressionset, \Omega]{
    \groupC(#1)
}
\newcommand{\scalarvals}{
    \mathcal{S}
}
\newcommand{\aggFunc}{
    AggFun
}
\newcommand{\aggFuncF}{
    \aggFunc_{\flatten}
}
\newcommand{\aggFuncC}{
    \aggFunc_{\condense}
}
\newcommand{\aggregation}{
    Agg
}
\newcommand{\aggregationF}[1][]{
    \aggregation_{\flatten#1}
}
\newcommand{\aggregationC}[1][]{
    \aggregation_{\condense#1}
}
\newcommand{\aggregationFFunc}[1][\expressionset_{n}, \aggFuncF, \scalarvals, \groupF]{
    \aggregationF(#1)
}
\newcommand{\aggregationCFunc}[1][\expressionset_{n}, \aggFuncC, \scalarvals, \groupC]{
    \aggregationC(#1)
}
\newcommand{\expressionsagg}{
    \mathcal{AE}
}
\newcommand{\aggregateJoin}{
    AggJ
}
\newcommand{\aggregateJoinF}{
    \aggregateJoin_{\flatten}
}
\newcommand{\aggregateJoinC}{
    \aggregateJoin_{\condense}
}
\newcommand{\aggregateJoinFFunc}[1][\expressionsagg_{n}]{
    \aggregateJoinF(#1)
}
\newcommand{\aggregateJoinCFunc}[1][\expressionsagg_{n}]{
    \aggregateJoinC(#1)
}
\newcommand{\card}{
    card
}
\newcommand{\var}{
    var
}
\newcommand{\vars}{
    \MakeUppercase{\var}
}
\newcommand{\varset}{
    \mathcal{\vars}
}

\newcommand{\environement}{
    \Gamma
}
\newcommand{\environements}{
    \Delta
}
\newcommand{\representation}{
    r
}
\newcommand{\downgrade}[3]{
    \downarrow^{#1}_{#2}(#3)
}
\newcommand{\translateresult}[2]{
    \downarrow^{#1}_{#2}
}
\newcommand{\translatevalue}[2]{
    \translate_{#1}(#2)
}
\newcommand{\representations}{
    \MakeUppercase{\representation}
}
\newcommand{\representationset}{
    \mathcal{\representations}
}
\newcommand{\uprepresentationFunc}{
    \downarrow_{\representation}
}

\newcommand{\lowerenv}{
    \sqcap_{\environements}
}

\newcommand{\infrp}{
    <_{\representation}
}
\newcommand{\leqrp}{
    \leq_{\representation}
}

\newcommand{\geqrp}{
    \geq_{\representation}
}

\newcommand{\optrepr}{
    \representations^{*}
}

\newcommand{\idtoval}{
{\idr}to{\valuer}
}

\newcommand{\idtovalfunc}[1]{
    \idtoval(#1)
}

\newcommand{\optreprfunc}{
    \optrepr: \expressionset \to (\varset \to \representationset)
}

\newcommand{\valuer}{\nu}
\newcommand{\idr}{\iota}
\newcommand{\condensedr}{\zeta}
\newcommand{\unboundgraphr}{\psi}
\newcommand{\equivalent}{
    \cong
}
\newcommand{\mvaluer}[2]{
    \mapping[#1]^{#2\valuer}
}
\newcommand{\midr}[2]{
    \mapping[#1]^{#2\idr}
}
\newcommand{\mcondensedr}[2]{
    \mapping[#1]^{#2\condensedr}
}
\newcommand{\munboundgraphr}[2]{
    \mapping[#1]^{#2\unboundgraphr}
}
\newcommand{\dvaluer}{
    value_{\representations}
}
\newcommand{\translate}{
    \tau
}

\newcommand{\concatenate}[2]{
    #1 \concatenatefunc #2
}
\newcommand{\concatenatefunc}{
    \mathbin\Vert
}
\newcommand{\bigconcatenatefunc}[1]{
    {\big|\big|}_{#1}
}

%% file: content/abstract.tex
\begin{abstract}
    Evolving phenomena, often complex, can be represented using knowledge graphs, which have the capability to model heterogeneous data from multiple sources.
    Nowadays, a considerable amount of sources delivering periodic updates to knowledge graphs in various domains is openly available.
    The evolution of data is of interest to knowledge graph management systems, and therefore it is crucial to organize these constantly evolving data to make them easily accessible and exploitable for analysis.
    In this article, we will present and formalize the condensed representation of these evolving graphs and propose a new solution called QuaQue that allows querying across multiple versions of graphs and we also present the results of our benchmark comparing our solution against existing approaches.
\end{abstract}

%% file: content/chapters/introduction.tex
\section{Introduction and motivations}
\subsection{Data evolution and versioning}

Time is a fundamental aspect of data, and the need for temporal data management\cite{jensen2009temporal} has been recognized for a long time. Temporal snapshots of data help to understand the evolution of data over time and are essential for various applications, such as historical data analysis, trend analysis, and data quality assessment.

While our requirement emerged from urban data management, where temporal data management is highly significant. The challenge of managing evolving data is present in many domains, including but not limited to urban studies, biomedical research, and software engineering. In these fields, data is frequently updated, extended, or corrected, and it is crucial to track and analyze these changes effectively. For example, in biomedical research, tracking changes in ontologies or patient records over time is important; in software engineering, managing the evolution of code or documentation as RDF graphs is increasingly common. The growing availability of open and periodically released data enables researchers to study complex phenomena with incremental changes\cite{chaturvedi_managing_2017,kutzner_citygml_2020,samuel_representation_2020,eriksson_comparison_2021}. A comprehensive understanding of dynamic elements—such as population growth, infrastructure development, or evolving scientific knowledge—is crucial for effective management and analysis in these domains.

To address these challenges, we propose the use of versioning to refer to data at different points or periods of time. The versioning of scientific data enables the exploitation and reproducibility of experiments and analyses\cite{klump2021versioning}. Versioning captures the state of the data at a given point or period in time, facilitating reproducibility, integrity, and the ability to restore previous versions in case of errors. It also supports experimentation with different analytical approaches by allowing easy switching between data versions.

Knowledge graphs, and in particular RDF graphs, are increasingly used to represent evolving data in a variety of domains\cite{vinasco2021towards}. By incorporating versioning into these knowledge graphs, it becomes possible to capture concurrent viewpoints and the evolution of entities over time\cite{samuel_representation_2020}. This enables the representation of different states of the data, providing a comprehensive view of its evolution. Our approach leverages knowledge graphs from heterogeneous sources (concurrent viewpoints) at different periods of time (hereafter, referred to as versions).

\subsection{Improving the comprehension of data evolution}
In many domains, data is continuously evolving as new information becomes available, corrections are made, or entities change over time. This constant evolution creates challenges for data management systems that need to track and analyze these changes effectively. Additionally, the growing volume of data collected from multiple sources and at different temporal granularities adds complexity to the task of managing and understanding data evolution.

To achieve effective management of evolving data and ensure reproducibility of experiments and analyses, we identified the following requirements for our system:

\begin{itemize}
    \item \textbf{R1: Managing and querying multiple graphs and their different versions}: Our system should allow users to query multiple knowledge graphs from one or more sources and access their different versions simultaneously. This capability enables comprehensive analysis and comparison of data evolution over time.
    \item \textbf{R2: Linking versioned data with metadata}: It is crucial to establish a connection between versioned data and their associated metadata. This linking provides valuable context and additional information about the data, enhancing its usability and interpretability. For example, when tracking changes to a biomedical record or a software artifact, metadata could include the source, method, and date of observation, allowing users to better understand and validate the data's provenance and reliability.
    \item \textbf{R3: Isolating and querying individual graph versions}: Our system must provide a mechanism to isolate and query specific versions of a knowledge graph independently. This capability allows users to focus their analysis on particular versions without interference from other versions, enabling precise examination of the data at specific snapshots.
\end{itemize}

By addressing these requirements, our system aims to empower researchers and practitioners in diverse fields to gain deeper insights into data dynamics, facilitate informed decision-making, and ensure the reproducibility and reliability of scientific experiments and analyses. Although our focus in this article is on the formalization of versioned RDF graphs, the principles and mechanisms we present are generic and can be applied to any domain where RDF data evolves over time.

In this article, to address the above requirements, we present the formalization of flat and condensed representations and querying of evolving knowledge graphs. We also present our proposed solution that uses and compares both representations using benchmark data. The state of the art in entity versioning is detailed in Section~\ref{section:stateoftheart}. Section~\ref{section:notations} presents the necessary preliminaries and notations. Section~\ref{section:contributions} presents the contributions of our work: datasets using flat and condensed models, algebra for handling query operators and the resulting data in flat and condensed representations. In Section~\ref{section:benchmarks}, we briefly present our implementation—\textbf{QuaQue} (part of the ConVer-G system)\cite{gil2024convergconcurrentversioningknowledge}—and compare the results obtained for flat and condensed representations. Section~\ref{section:discussions} concludes the article with a discussion of the limitations and possible enhancements to our proposed work.

%% file: content/chapters/state-of-the-art.tex
\section{State of the art - Versioning}
\label{section:stateoftheart}
Resources evolve over time and hence the statements associated to describe them.
Version control systems \cite{ZOLKIFLI2018408}, for example, handle this evolution of data (or code) by associating a tag and a timestamp to a snapshot of the data at a given instant of time.
This supplementary information associated to a snapshot, referred to as a version can be used to not only represent the state of the data at any given point of time, but also to understand the evolution of data over a period of time.
As mentioned by \citet{bayoudhi2020survey}, versioning is a critical aspect of management in a project, allowing developers to track changes to the elements, manage different versions, and collaborate with other team members.
For the sake of familiarity, we will use the word \emph{versioning} to this process of adding information like timestamp, tag to snapshots of one or more resources and their associated statements at any given moment of time.
In this section, we will take a look at different approaches of versioning. Table~\ref{table:code-data-versioning} summarizes these results.

\begin{table}
    \caption{Comparison of code and data versioning}
    \label{table:code-data-versioning}
    \begin{tabular}{ |l|p{2.5cm}|p{2.6cm}| }
        \hline
        \textbf{Characteristics} & \textbf{Code versioning}          & \textbf{Data versioning}                      \\
        \hline
        \textbf{Object}          & Files                             & Datasets (structured, unstructured, etc.)     \\        \hline
        \textbf{Granularity}     & Lines of code, files              & Data, entire datasets                         \\        \hline
        \textbf{Primary goals}   & History, collaboration, branching & Reproducibility, provenance, experimentation, \\        \hline
        \textbf{Tools}           & Git, Mercurial, SVN               & DVC, Pachyderm, Machine Learning platforms    \\        \hline
    \end{tabular}
\end{table}

\subsection{Code versioning}
Version control systems \cite{ZOLKIFLI2018408} such as Git, Mercurial, Subversion, CVS, and Bazaar are commonly used for managing the evolution of source code.
However, these solutions are not specifically designed for data versioning, especially for structured data \cite{sveen_geomdiff_2020} and have certain limitations for efficient querying \cite{Bulteau2023,Yilmaz2025,samuel2018urbanco2fab}.
Apart from some exceptions, like Git LFS, they don't provide adequate support for large files and versioned data queries.
Therefore, they may not be the ideal choice for managing data versioning in a project.

\subsection{Dataset versioning}
\subsubsection{Independent copies (IC)}
This approach allows querying data as it existed in a specific version.
\emph{Example: Which buildings were present in the city at a particular version?}

A common approach to dataset versioning is to maintain independent copies (snapshots) of the data for each version.
Tools such as Qri \cite{QRI}, SemVersion \cite{volkel2005semversion}, GeoGig \cite{GeoGig}, and UrbanCo2Fab \cite{samuel2018urbanco2fab} follow this strategy, often leveraging Git-based mechanisms.
While this method simplifies version management and enables straightforward retrieval of any version by checking out its corresponding snapshot, it has notable limitations.
In particular, querying across multiple versions simultaneously is inefficient, as each query requires materializing the full state of the relevant versions.
This restricts the ability to perform cross-version analysis or answer queries that span several versions.
Similarly, machine learning data management tools such as \href{https://dvc.org/}{DVC}, \href{https://dagshub.com/}{DagsHub}, and \href{https://wandb.ai/}{Weights and Biases} provide robust tracking and collaboration features for structured data, but do not support efficient querying across multiple snapshots, making them unsuitable for use cases requiring multi-version queries.

\subsubsection{Timestamp-based versioning (TB)}
This representation allows querying data as it existed at a specific point in time.
\emph{Example: Which buildings were present in the city at a particular version?}

A common approach to data versioning is to associate each record with a timestamp or a validity interval, as seen in temporal or bitemporal tables \cite{10.1145/2380776.2380786}.
Tools such as ConVer-G \cite{gil2024convergconcurrentversioningknowledge}, Drydra \cite{anderson2016transaction}, RDF-TX \cite{gao2016rdf}, v-RDFCSA \cite{7786197}, and x-RDF-3X \cite{10.14778/1920841.1920877} adopt this strategy.
This method enables querying data as it existed at a specific point in time or within a given interval.
While effective for linear histories, timestamp-based versioning is limited when handling concurrent changes, such as branches or forks, which are typical in collaborative environments and version control systems.
As a result, it may not fully support scenarios requiring representation of multiple, diverging timelines.

\subsubsection{Change-based versioning (CB)}
This representation enables retrieval of the differences between two versions.
\emph{Example: Which buildings were added or removed between two versions?}

Change-based versioning systems, such as Delta Lake and Apache Hudi store changes made to data as changes, allowing for efficient retrieval of any version or the differences between versions.
In the context of RDF, tools such as R43ples \cite{graube2014r43ples}, R\&WBase \cite{vander2013r}, and \href{https://www.stardog.com/}{Stardog} follow this paradigm.
A prominent system for versioning RDF data is OSTRICH \cite{taelman2019triple}.

However, OSTRICH has notable limitations for our needs.
It only supports linear version histories, lacking support for branching or concurrent versions, which are essential for modeling multiple viewpoints in urban evolution.
Additionally, its aggregated changesets grow with each new version, resulting in slow ingestion for archives with long histories.

\subsubsection{Fragment-based versioning (FB)}
This approach enables random access to different versions by managing indexes of file fragments for each version.
This archiving policy stores snapshots of each changed fragment of an archive.
These fragments can be defined at any level of granularity, such as individual graphs within a dataset.
An index is maintained to reference which specific fragments belong to any given version of the dataset.
This approach avoids the need to store full, independent copies for each version, and in contrast to change-based systems, it does not require the reconstruction of versions by applying a series of change operations.
The Quit Store implements this policy by treating modified files within its Git repository as the fragments to be versioned.

\subsection{Querying versioned data}
This subsection reviews how different systems implement versioning, the types of queries they support, and the consequences for managing knowledge graphs.
The following analysis compares existing solutions and highlights the need for advanced versioning models that address complex, multi-version scenarios.
We adopt the query types defined by Ruben Taelman in the OSTRICH system~\cite{taelman2019triple}, which offer a comprehensive framework for querying versioned data.

\subsubsection{Version materialization (VM)}
We consider version materialization as a means to create and manage distinct versions of knowledge graphs.
This approach allows for the reconstruction of any specific version. This is particularly linked with the concept of (IC).

\subsubsection{Delta materialization (DM)}
Delta materialization is a technique used in databases and data management systems to track and retrieve changes in query results between two versions of data.

Instead of recomputing the entire result of a query for both versions, DM focuses on computing the delta—the set of changes (additions and deletions) in the query result that occurred between two versions.
This is especially useful when data changes incrementally and you want to know only what has changed, not the full result.

\subsubsection{Version query (VQ)}
A version query annotates each result of query $Q$ with the specific versions of the RDF archive $A$ in which that result appears.
This allows users to see not only the query results, but also the history of when each result was valid across different versions of the data.

\subsubsection{Cross-version query (CV)}
A cross-version query typically refers to a database or information retrieval operation that compares or joins the results of two queries executed on different versions of a dataset.
This allows you to analyze changes, differences, or relationships between data across versions.

\subsubsection{Cross-delta (CD)}
A cross-delta query combines the results of multiple changesets of a dataset.
It is similar to a delta materialization (DM) query but supports more complex comparisons than just identifying additions and deletions.
Instead of only showing what has changed, a CD query can analyze how specific values have evolved between versions.
For instance, it can be used to compare the value of a particular attribute for a given entity across two versions, enabling a more detailed analysis of data evolution.

OSTRICH supports all query types, but its native capabilities are limited to single triple pattern queries.
To execute full SPARQL queries, an additional query engine such as GLENDA \cite{pelgrin2023glenda} is required.
Furthermore, OSTRICH relies on a linear sequence of integer revision numbers for versioning, whereas our use case demands a more flexible model capable of representing arbitrary relationships between versions, such as those found in a version graph.

This work builds on concepts from version control systems \cite{Swierstra2014,QRI,arndt-2020-dissertation,ZOLKIFLI2018408,taelman2019triple} and extends them to knowledge graphs.
We introduce a "condensed representation" for querying with SPARQL, where the graph is modeled as a set of quads annotated with the versions in which they are valid.
Our main contributions include the formalization of this condensed representation for evolving knowledge graphs and the development of an algebra for query operators and the resulting data in both flat and condensed forms.

Table~\ref{table:versioning-tools-queries} in Appendix is an extension of Olivier Paul Pelgrin's survey of versioning tools for RDF data \cite{pelgrinefficient} which summarizes the capabilities of different versioning tools with respect to the types of queries they support.

%% file: content/chapters/notations.tex
\section{Preliminaries and notations}
\label{section:notations}

In this section, we will present the key concepts and notations related to representing knowledge graphs in the form of RDF and querying them using the SPARQL query language.

\subsection{RDF context representation}
For the sake of simplicity, we considered the following RDF context representation based on the W3C definitions \cite{W3CConcepts}:

\subsubsection{Literal}
A literal denotes resources of the real world.
\begin{itemize}
    \item $\literal, \literal_1, \literal_2\dots$ a literal
    \item $\literals, \literals_1, \literals_2\dots$ a finite set of literals
    \item $\literalset$ the infinite set of all literals
\end{itemize}

\subsubsection{IRI}
Internationalized Resource Identifier (IRI) is a Unicode string, analogous to Uniform Resource Identifiers (URI), and used to identify resources (e.g., \texttt{http://example.org/resource/Paris} is an IRI identifying the resource "Paris").
\begin{itemize}
    \item $\iri, \iri_1, \iri_2\dots$ an IRI
    \item $\iris, \iris_1, \iris_2\dots$ a finite set of IRI
    \item $\irisets$ the infinite set of all IRIs
\end{itemize}

\subsubsection{Blank Node}\label{blank-node}
Not all resources and the relationships can be explicitly named and are represented as blank nodes indicating the existence of something. To handle blank nodes, we use Skolemization \cite{hogan_everything_2014}, a technique that replaces blank nodes with generated IRIs. This allows us to treat blank nodes as regular resources with globally unique identifiers, enabling better integration and querying of RDF data.

\subsubsection{Terms}
Terms are used to refer to IRIs or literals.
\begin{itemize}
    \item $\term, \term_1, \term_2\dots$ a term
    \item $\termfs, \termfs_1, \termfs_2\dots$ a finite set of terms
    \item $\termfset$ the infinite set of all terms
    \item $\termfset = \{ \literalset, \irisets \}$ where:
          \begin{itemize}
              \item $\literalset \varsubsetneq \termfset$
              \item $\irisets \varsubsetneq \termfset$
              \item $\literalset \cap \irisets = \emptyset$
          \end{itemize}
\end{itemize}

\subsubsection{Triples}
An RDF triple is a statement about a resource that consists of three parts: a subject, an object, and a predicate. The subject is an IRI that represents the resource being described. The object can be an IRI or a literal that represents the value of the property. The predicate is an IRI that represents the relationship between the subject and the object.

\begin{figure}[h]
    \includegraphics[width=0.45\textwidth]{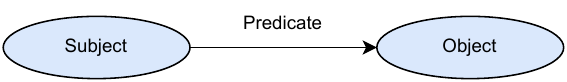}
    \caption{Example of a RDF triple}
    \Description{Example of a RDF triple}
    \label{fig:triple-exemple}
\end{figure}

\begin{itemize}
    \item $\triple = \rdftriple$ an RDF triple where:
          \begin{itemize}
              \item $\subjectterm \in \irisets$ is a subject
              \item $\predicateterm \in \irisets$ is a predicate
              \item $\objectterm \in \termfset$ is an object
          \end{itemize}
    \item $\triples, \triples_1, \triples_2\dots$ a finite set of triples
    \item $\tripleset$ the infinite set of all triples
\end{itemize}

\subsubsection{Graphs}
An RDF graph is a set of RDF triples. It represents a collection of resources and the relationships between them. An RDF graph can be named or unnamed by a resource. An unnamed graph is called the default graph.

\subsubsection{Quads}
A quad is a statement that consists of four parts: all triple elements (the subject, the predicate, the object), and the graph name. The graph name is an IRI that represents the context in which the triple is defined.

\begin{figure}[h]
    \includegraphics[width=0.45\textwidth]{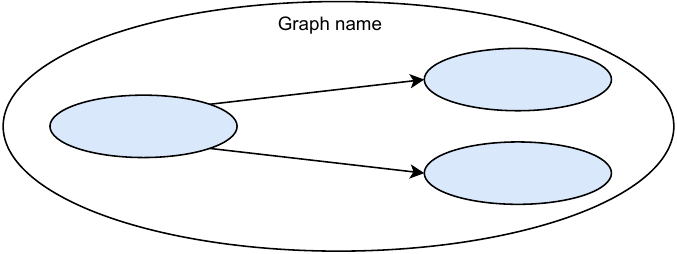}
    \caption{Example of a named graph}
    \Description{Example of a named graph}
    \label{fig:quad-exemple}
\end{figure}

\begin{itemize}
    \item $\quadf = \rdfquad$ an RDF quad
    \item $\quadfs, \quadfs_1, \quadfs_2\dots$ a finite set of quads
    \item $\quadset$ the infinite set of all quads
    \item $\namedgraphterm \in \namedgraphs$ where $\namedgraphs \subseteq \irisets$, $\namedgraphterm$ is the graph name and $\namedgraphs$ is the set of all graph names
\end{itemize}

\subsubsection{Datasets}
A dataset $\rdfdataset$ is a collection of graphs. A dataset can contain zero or more named graphs, each of which contains a set of RDF triples. A dataset can also contain a default graph, which is a set of triples that are not associated with any specific graph name.

\subsubsection{Versioning}
A version is a snapshot of the RDF data at a given point in time. We define the following versioning representation:

\begin{itemize}
    \item $\version, \version_1, \version_2\dots$ a version
    \item $\versions, \versions_1, \versions_2\dots$ a finite set of versions
    \item $\versionset$ the infinite set of all versions
\end{itemize}

Because the RDF data model is atemporal as explained in the "RDF and Change over Time" paragraph of the W3C RDF 1.1 Concepts and Abstract Syntax \cite{W3CConcepts}, we consider the versioning of RDF data as a set of versions.

\subsection{SPARQL queries}
SPARQL \cite{W3CSPARQL} is a declarative programming language used to query data stores containing a set of triples or quads.

\subsubsection{Variables}
A SPARQL variable is a placeholder for a value that is not known beforehand. It allows querying for information without having to specify the exact values the user is looking for. Variables are denoted by a question mark "?" or a dollar sign "\$", immediately followed by a string without any white space characters.

\begin{itemize}
    \item $\var, \var_1, \var_2\dots$ a SPARQL variable
    \item $\vars, \vars_1, \vars_2\dots$ a finite set of SPARQL variables
    \item $\varset$ the infinite set of all SPARQL variables
\end{itemize}

\subsubsection{Query}
A SPARQL query is a formal language expression used to retrieve and manipulate data stored in RDF format, according to the W3C definition \cite{W3CSPARQL}.
\begin{itemize}
    \item $\query, \query_1, \query_2\dots$ a SPARQL query
    \item $\queries, \queries_1, \queries_2\dots$ a finite set of SPARQL queries
    \item $\queryset$ the infinite set of all SPARQL queries
\end{itemize}

\paragraph{SPARQL expression}
A SPARQL expression refers to a part of a SPARQL query that computes or filters values within the query. We note:
\begin{itemize}
    \item $\expression, \expression_1, \expression_2\dots$ a SPARQL expression
    \item $\expressions, \expressions_1, \expressions_2\dots$ a finite set of SPARQL expressions
    \item $\expressionset$ the infinite set of all SPARQL expressions
\end{itemize}

SPARQL expression can be:
\begin{itemize}
    \item A term $\term$ (e.g., $10$) is an expression if $\term \in \termfset$
    \item A variable $\var$ (e.g., $?x$) is an expression if $\var \in \vars$
    \item A function $f(\expression_1, \dots, \expression_n)$ is an expression, if $f$ is a function and $\forall i \in \{1, \dots, n\}, \expression_i$ is a SPARQL expression (e.g., $AVG(\expression)$ where $\expression$ is a SPARQL expression)
    \item $\expression_1 \neq \expression_2$ is an expression if $\expression_1$ and $\expression_2$ are expressions
\end{itemize}

SPARQL expressions are used to:

\begin{itemize}
    \item Combine values from different RDF triples to form more complex patterns (e.g., triple patterns, basic graph patterns, or subqueries)
    \item Filter results based on conditions (e.g., using the \texttt{FILTER} keyword)
    \item Transform data using functions like mathematical operations, string manipulation, or logical operators (e.g., \texttt{CONCAT}, \texttt{STRLEN}...)
\end{itemize}

\paragraph{Solution mapping}
A solution mapping is a mapping between variables and RDF terms. It is used to represent the results of a SPARQL query. We note $\mapping: \varset \to \termfset$ a solution mapping that maps a set of variables to a set of RDF terms and $\mappingset$ an infinite set of solution mappings.

\paragraph{Solution sequence}
A solution sequence is a sequence of solution mappings. We denote a solution sequence as $\solutionsequence = [\mapping[1], \dots, \mapping[n]]$ where each $\mapping[i]$ is a solution mapping.

%% file: content/chapters/problem.tex
\section{Contributions}
\label{section:contributions}

In this section, we present our contributions (Figure~\ref{fig:models-translation-exemple}). Firstly, we present the key extensions to the RDF model discussed in Section~\ref{section:notations} and the resulting dataset models. We discuss in detail the flat and condensed models of representation with examples. We then show how these datasets can be queried with the existing operators of the SPARQL query. Finally, we present the necessary formalisms and theorems to show how such queries can be evaluated. 

\input{content/chapters/problem/definitions}

\input{content/chapters/problem/models}

\input{content/chapters/problem/algebra}

\input{content/chapters/problem/evaluation}

%% file: content/chapters/problem/definitions.tex
\subsection{RDF versioning definitions}

\subsubsection{Versioned IRI}
In order to represent the versioning of IRI, we introduce the concept of a versioned IRI. A versioned IRI is an IRI that is associated with a version. We define the following versioned IRI representation:

\begin{itemize}
    \item Let $\vifunction$ be a bijective function.
    \item We note $\vi, \vi_1, \vi_2\dots$ the elements of $\viset$, referred to as versioned IRI
    \item We note $\vis, \vis_1, \vis_2\dots$ a finite set of versioned IRI
\end{itemize}

Let $\version \in \versionset$, $\namedgraphterm \in \irisets$ and $vi \in \viset$. We define the reverse functions of $\vis$:
\begin{equation}
    \begin{split}
        \visv{\vi}     & = \version \Leftrightarrow \exists \namedgraphterm \vis(\version, \namedgraphterm) = \vi  \\
        \vistg{\vi} & = \namedgraphterm \Leftrightarrow \exists \version \vis(\version, \namedgraphterm) = \vi
    \end{split}
\end{equation}

\subsubsection{Versioned quad}
A versioned quad $\versionedquad = \versionedrdfquad$ is a quintuplet where the quad $\rdfquad$ is valid at the version $\version$. We denote $\versionedquads, \versionedquads_1, \versionedquads_2\dots$ a finite set of versioned quads and $\versionedquadset$ an infinite set of versioned quads.

\subsubsection{Versioned dataset}
A versioned dataset is a dataset that contains a set of versioned graphs. A versioned graph is a graph that contains a set of RDF triples associated to a versioned named graph built with a version and a graph. By convention, default graphs are assigned a special name, such as \texttt{default}, to distinguish them from named graphs.

\subsubsection{Dataset representation}
Consider a RDF context containing data and its associated metadata. We define the dataset function such that holds all the versioned quads stored in the database: $\rdfdataset = 2^{\versionedquadset}$. We note $\vers{\rdfdataset}$ the set of versions in the dataset.

\begin{figure}[h]
    \includegraphics[width=0.47\textwidth]{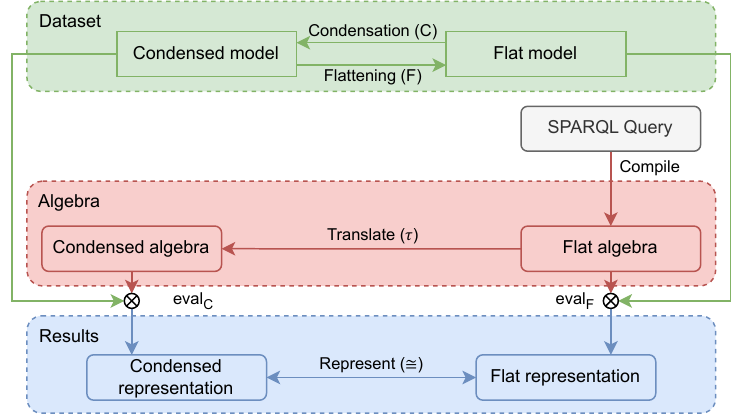}
    \caption{Dataset, algebra and results representation}
    \Description{Dataset, algebra and results representation}
    \label{fig:models-translation-exemple}
\end{figure}

The figure~\ref{fig:models-translation-exemple} provides an overview of the different levels of representation and processing within our framework. It highlights the relationships between the dataset, query, algebra, evaluation, and results, as well as the possible transformations between flat and condensed forms at each stage. The main aspects illustrated in the figure are as follows:
\begin{itemize}
    \item \textbf{Dataset level:} the figure illustrates that the dataset can be represented in either a flat or condensed form, and that it is possible to convert between these two representations using the \texttt{condensation} and \texttt{flattening} functions.
    \item \textbf{Query:} A SPARQL query is represented in a flat algebra, which is the original query written by the user. The \texttt{compile} function returns the flat algebra of the query.
    \item \textbf{Algebra level:} it shows the representation of the algebra in both flat and condensed forms, as well as the translation process from the flat algebra (derived from the original SPARQL query) to the condensed algebra using the \texttt{translate} function.
    \item \textbf{Evaluation level:} it illustrates the evaluation of the algebra. Taking the associated dataset and algebra, it shows the evaluation of the algebra in both flat and condensed forms. The evaluation process is represented by the \texttt{eval} relationship, which indicates that the evaluation can be performed on either representation.
    \item \textbf{Results level:} it depicts the representation of results in both flat and condensed forms, and the \texttt{represent} relationship that enables conversion between these representations, again using the \texttt{condense} and \texttt{flatten} functions.
\end{itemize}

%% file: content/chapters/problem/models.tex
\subsection{Dataset models}

Let us explore the different dataset models that we will use in our work. The following sections define a flat model and a condensed model for dataset versioning. To illustrate these concepts, we begin with an example that will be referenced throughout the paper for demonstration purposes.

\begin{example}
    Let's assume that we have a dataset representing concurrent points of view about the height of some buildings. In Table \ref{dataset-2v-2ng}, we present two versions (v:1 and v:2) with quads. The column \emph{Named graph} consists of two named graphs corresponding to the data coming from two sources named, Gr-Lyon and IGN.

    \begin{table}[h]
        \caption{Flat model of the versioned dataset}
        \label{dataset-2v-2ng}
        \begin{tabular}{ |l|l|l|l|l| }
            \hline
            \textbf{Subject} & \textbf{Predicate} & \textbf{Object} & \textbf{Named graph}        & \textbf{Version}     \\
            \hline
            ex:bldg\#1       & height             & 10.5            & \multirow{2}{*}{ng:Gr-Lyon} & \multirow{3}{*}{v:1} \\
            \cline{1-3}
            ex:bldg\#2       & height             & 9.1             &                             &                      \\
            \cline{1-4}
            ex:bldg\#1       & height             & 11              & ng:IGN                      &                      \\
            \hline
            ex:bldg\#1       & height             & 10.5            & ng:IGN                      & \multirow{3}{*}{v:2} \\
            \cline{1-4}
            ex:bldg\#1       & height             & 10.5            & \multirow{2}{*}{ng:Gr-Lyon} &                      \\
            \cline{1-3}
            ex:bldg\#3       & height             & 15              &                             &                      \\
            \hline
        \end{tabular}
    \end{table}

    Table~\ref{dataset-2v-2ng-vi} represents all the versioned IRI of the dataset.
    In this example, each IRI starting with \texttt{vi:} are images (results) of the pair (version, named graph) produced by the function $\vis$ (Versioned IRI).

    \begin{table}[h]
        \caption{Versioned IRI of the dataset}
        \label{dataset-2v-2ng-vi}
        \begin{tabular}{ |l|l|l| }
            \hline
            \textbf{Version} & \textbf{Named graph} & \textbf{Versioned IRI} \\
            \hline
            \multirow{2}{2em}{v:1}
                             & ng:Gr-Lyon           & vi:1                   \\
            \cline{2-3}
                             & ng:IGN               & vi:2                   \\
            \hline
            \multirow{2}{2em}{v:2}
                             & ng:Gr-Lyon           & vi:3                   \\
            \cline{2-3}
                             & ng:IGN               & vi:4                   \\
            \hline
        \end{tabular}
    \end{table}
\end{example}

\subsubsection{Flat model}
We define the \textit{flat model} as a representation of a dataset as a set of quintuplets. In this model, each triple is augmented with two additional elements: the name of the graph corresponding to the data source, and the version information.

\begin{example}
    Table \ref{dataset-2v-2ng} represents the flat model of the dataset. $\flatmodel: \irisets \times \irisets \times \termfset \times \irisets \times \versionset$ designates the relation such as the quad $\rdfquad$ is valid in the version $\version$.

    The supplementary information related to the named graphs and versions is stored separately in a metadata table. Table \ref{dataset-flat-metadata} shows this metadata for the flat model in Table \ref{dataset-2v-2ng}. For example, in Table~\ref{dataset-2v-2ng-vi} the first row indicate that vi:1 is a versioned IRI of the named graph Gr-Lyon and belongs to version 1. It could also be used to store supplementary information such as provenance, authorship, or other metadata relevant to the versioned dataset.

    \begin{table}[h]
        \caption{Metadata of the flat model}
        \label{dataset-flat-metadata}
        \begin{tabular}{ |l|l|l|l|l| }
            \hline
            \textbf{Subject}      & \textbf{Predicate} & \textbf{Object} & \textbf{N. graph}         & \textbf{Version}     \\
            \hline
            \multirow{2}{*}{vi:1} & version-of         & ng:Gr-Lyon      & \multirow{8}{*}{ng:Metadata} & \multirow{8}{*}{v:0} \\
            \cline{2-3}
                                  & in-version         & 1               &                              &                      \\
            \cline{1-3}
            \multirow{2}{*}{vi:2} & version-of         & ng:IGN          &                              &                      \\
            \cline{2-3}
                                  & in-version         & 1               &                              &                      \\
            \cline{1-3}
            \multirow{2}{*}{vi:3} & version-of         & ng:Gr-Lyon      &                              &                      \\
            \cline{2-3}
                                  & in-version         & 2               &                              &                      \\
            \cline{1-3}
            \multirow{2}{*}{vi:4} & version-of         & ng:IGN          &                              &                      \\
            \cline{2-3}
                                  & in-version         & 2               &                              &                      \\
            \hline
        \end{tabular}
    \end{table}
\end{example}

\subsubsection{Condensed model}

$\condensedmodel: \mathbb{N} \times \mathbb{N} \times \mathbb{N} \times \mathbb{N} \to 2^\versionset$ is a function that takes four identifiers that are bound to an RDF quad and returns a set of finite elements of $\versionset$ such that $2^\versionset$ represents a set of finite versions where the quad is valid.

\begin{example}
    The table \ref{dataset-condensed-dataset} represents the condensed model of the versioned dataset (simplified by replacing the identifiers by their values). The first row, for example states that the triple present in the named graph Gr-Lyon is valid in versions v:1 and v:2. Each entry of this table can be replaced by the identifiers listed in Appendix~\ref{ids-to-values}.

    \begin{table}[h]
        \caption{Condensed model of the versioned dataset}
        \label{dataset-condensed-dataset}
        \begin{tabular}{ |l|l|l|l|l| }
            \hline
            \textbf{Subject} & \textbf{Predicate} & \textbf{Object} & \textbf{N. Graph} & \textbf{Versions} \\
            \hline
            ex:bldg\#1       & height             & 10.5            & ng:Gr-Lyon           & \{v:1, v:2\}      \\
            \hline
            ex:bldg\#2       & height             & 9.1             & ng:Gr-Lyon           & \{v:1\}           \\
            \hline
            ex:bldg\#1       & height             & 11              & ng:IGN               & \{v:1\}           \\
            \hline
            ex:bldg\#3       & height             & 15              & ng:Gr-Lyon           & \{v:2\}           \\
            \hline
            ex:bldg\#1       & height             & 10.5            & ng:IGN               & \{v:2\}           \\
            \hline
        \end{tabular}
    \end{table}

    Metadata plays a role in our approach, particularly for managing concurrent snapshots or concurrent viewpoints. By associating a constant version \texttt{v:0} to all metadata, we ensure that version information and relationships between versions remain consistent and accessible across the entire dataset. This is especially important when dealing with branching version histories or multiple concurrent viewpoints, as the metadata helps track and organize these complex relationships. While this means metadata is not condensed and can only be represented in the flat model, this design choice provides a stable foundation for managing versioned data across different temporal and conceptual dimensions.
\end{example}

\subsubsection{Model relationships}
\begin{definition}[Condensation]
    Consider a flat dataset $\flatmodel = \irisets \times \irisets \times \termfset \times \irisets \times \versionset$. We can condense the dataset $\flatmodel$ into the dataset $\condensedmodel = \irisets \times \irisets \times \termfset \times \irisets \to 2^\versionset$. The condensation function $\condense$ takes a quad and returns the set of versions in which that quad exists. This function is defined as:
    $$\condense = \rdfquad \mapsto \{ \version \mid \versionedrdfquad \in \flatmodel \}$$

    The domain of $\condense$ is the set of quads in the dataset $\flatmodel$ where the quad occurs at a set of versions such that: $\domain{\condense} = \{ \rdfquad \mid \exists{\version} \versionedrdfquad \in \flatmodel \}$.
\end{definition}

\begin{definition}[Flattening]
    Let $\condensedmodel = \irisets \times \irisets \times \termfset \times \irisets \to 2^\versionset$ be a dataset. We can flatten the dataset $\condensedmodel$ into the dataset: $\flatmodel = \irisets \times \irisets \times \termfset \times \irisets \times \versionset$. The flattening function $\flatten$ is defined as:
    $$\flatten = \{ \versionedrdfquad \mid \version \in \condensedmodel\rdfquad \}$$

    The domain of $\flatten$ is the set of versioned quads in the dataset $\condensedmodel$ where the quad exists in each version according to the condensed model: $\domain{\flatten} = \{ \versionedrdfquad \mid \exists{\version} \in \condensedmodel\rdfquad \}$.
\end{definition}

\begin{theorem}[$\flatten \circ \condense (\flatmodel) = \flatmodel$]
    Given a flat model $\flatmodel$, its image by the composition of the $\flatten$ and the $\condense$ functions is equal to the flat model.
\end{theorem}

\begin{proof}
    We want to prove the following equivalence: $\flatten \circ \condense (\flatmodel) = \flatmodel$.
    This means we need to prove:
    \begin{itemize}
        \item On the one hand, that $\flatten \circ \condense (\flatmodel) \subseteq \flatmodel$;
        \item On the other hand, that $\flatmodel \subseteq \flatten \circ \condense (\flatmodel)$.
    \end{itemize}
    In other words, we need to show that every element of $\flatten \circ \condense (\flatmodel)$ belongs to $\flatmodel$, and conversely, every element of $\flatmodel$ belongs to $\flatten \circ \condense (\flatmodel)$.

    Let the versioned quad $\versionedrdfquad \in \flatmodel$ such that \\
    $\version \in \condense(\flatmodel)\rdfquad$, then $\versionedrdfquad \in \flatten(\condense(\flatmodel))$.

    Let $\versionedrdfquad \in \flatten(\condense(\flatmodel))$, then $\version \in \condense(\flatmodel)\rdfquad = \{ \version' \mid (\subjectterm, \predicateterm, \objectterm, \namedgraphterm, \version') \in \flatmodel \}$. This proves that $\flatten \circ \condense (\flatmodel) = \flatmodel$.
\end{proof}

\begin{theorem}[$\condense \circ \flatten (\condensedmodel) = \condensedmodel$]
    Given a condensed model $\condensedmodel$, its image by the composition of the $\condense$ and the $\flatten$ functions is equal to the condensed model.
\end{theorem}

\begin{proof}
    We want to prove the following equivalence: $\condense \circ \flatten (\condensedmodel) = \condensedmodel$.
    This means we must prove:
    \begin{itemize}
        \item For every element of $\condense \circ \flatten (\condensedmodel)$, there exists a corresponding element in $\condensedmodel$;
        \item For every element of $\condensedmodel$, there exists a corresponding element in $\condense \circ \flatten (\condensedmodel)$.
    \end{itemize}
    In other words, we show equality by proving both inclusions.

    Let $\fsversions = \condense(\flatten(\condensedmodel))\rdfquad$, so $\fsversions = \{ \version \mid \versionedrdfquad \in \flatten(\condensedmodel) \} = \{ \version \mid \version \in \condensedmodel\rdfquad \} = \condensedmodel\rdfquad$.
\end{proof}

%% file: content/chapters/problem/algebra.tex
\subsection{SPARQL algebra}

A SPARQL query algebra provides a formal framework for representing SPARQL queries \cite{cyganiak2005relational}. It also defines the execution semantics by translating the high-level query syntax into a series of operations that an RDF query engine can process. We use $\algebras$ to denote the SPARQL algebras set. This paper will present two types of algebra: a condensed algebra and a flat algebra that will be respectively noted as $\algebracset$ and $\algebrafset$. Based on the SPARQL algebra proposed by \cite[Cyganiak, Richard, A relational algebra for SPARQL]{cyganiak2005relational}, we modified the SPARQL Algebra to take into account graph versioning to allow us to query them. An initial proposition was made by \cite[Cuevas, Hogan]{cuevas2020versioned} to represent queries over RDF Archives. We have taken this proposition as a basis to define what we call \textit{the flat SPARQL algebra}.

\subsubsection{Flat SPARQL algebra}
The flat SPARQL algebra contains operators that define its expressive power.
For example, the quad pattern $\quadpatternf{?\subject, ?\predicate, ?\object, ?\ngraph}$ retrieves triples matching the triple pattern $\triplepatternf{?\subject, ?\predicate, ?\object}$ within a versioned graph $?\ngraph$, and is considered syntactic sugar for the graph operator, as shown by \cite[Cuevas, Hogan]{cuevas2020versioned}.
The \texttt{JOIN} operator $\joinf{\expression_1, \expression_2}$ combines results from two expressions based on common variable bindings, while the \texttt{FILTER} operator $\filterf{\expressionset, \expression}$ restricts results according to filter functions over variable bindings.
The \texttt{UNION} operator $\unionf{\expression_1, \expression_2}$ merges results from two expressions into a single set.
The \texttt{DIFF} operator $\difff{\expression_1, \expression_2, \expression_3}$ computes the difference between two sets of solutions, conditioned by an additional expression, and is essential for defining the \texttt{OPTIONAL} operator.
The \texttt{OPTIONAL} operator $\leftouterjoinf{\expression_1, \expression_2, \expression_3}$ extends the join by including all results from the first expression and, where possible, matching results from the second, returning nulls otherwise.
Finally, the aggregation operator groups results based on variables and supports operations such as count, sum, average, min, and max, typically used with the \texttt{GROUP BY} clause in SPARQL queries.

\subsubsection{Condensed algebra}
The condensed algebra provides operators tailored for querying over all versions of a graph in a compact representation. The quad pattern $\quadpatternc{?\subject, ?\predicate, ?\object, ?\ngraph}$ retrieves triples matching the triple pattern $\triplepatternf{?\subject, ?\predicate, ?\object}$ across all versioned graphs $?\ngraph$. The condensed algebra includes optimized aggregation operators that can process data across versions. For example, aggregation operators can leverage optimizations such as count-views (as defined by Sarah Cohen \cite{cohen2006rewriting}) to efficiently compute aggregate values over versioned data without having to materialize each version separately.

\subsubsection{Variable representation}
SPARQL queries are composed of operators and contain variables whose representations need to be defined. The variable representation is used to determine its behavior in the query and how it is processed in the algebra.

\begin{figure}[h!]
    \centering
    \includegraphics[width=0.45\textwidth]{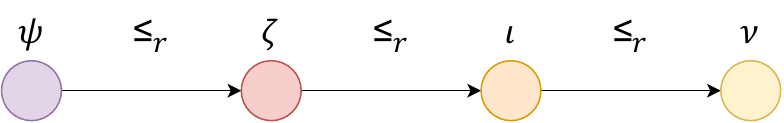}
    \caption{Hierarchy of variable representations in SPARQL algebra}
    \Description{A diagram showing the hierarchy and relationships between the variable representations: value, id, condensed, and unbound graph.}
    \label{fig:variable-representations}
\end{figure}

It is important to note that in our current formalization, we do not handle cases where there are variable clashes. Variable clashes occur when two variables with the same name but different scopes or contexts are used within the same query, leading to ambiguity and potential errors in query evaluation. By not addressing these clashes, we assume that all variables are uniquely named within their respective scopes. This simplification allows us to focus on the core aspects of our model and its algebraic semantics without the added complexity of variable disambiguation. However, in a more robust formalization, mechanisms to detect and resolve such clashes would be necessary to ensure the correctness and reliability of query results.
In our formalization, each variable in a SPARQL query is assigned a \emph{variable representation} that determines how its value is interpreted and manipulated during query evaluation. Figure~\ref{fig:variable-representations} illustrates the hierarchy and relationships between the different representations.

A variable representation is a type that is assigned to a variable in a SPARQL query.
\begin{itemize}
    \item We note:
          \begin{itemize}
              \item $\representation, \representation_1, \representation_2,\dots$ a variable representation
              \item $\representations, \representations_1, \representations_2,\dots$ a set of variable representations
              \item $\representationset$ an infinite set of variable representations
          \end{itemize}
    \item We define four variable representations:
          \begin{itemize}
              \item \textbf{value $\valuer$}, where variable's bound to a term
              \item \textbf{id $\idr$}, where variable's bound to a term identifier
              \item \textbf{condensed $\condensedr$}, where variable's bound to a list of versioned named graph identifiers
              \item \textbf{unbound graph $\unboundgraphr$}, where variable's bound to the list of all versioned named graph identifiers
          \end{itemize}
\end{itemize}

\begin{definition}[Variable representation]
    Let a bounded total order of variable representations defined as follows: $(\{ \valuer, \idr, \condensedr, \unboundgraphr \}, \leqrp)$ where:
    \begin{itemize}
        \item $\{ \valuer, \idr, \condensedr, \unboundgraphr \}$ is the set of elements
        \item $\leqrp$ is the total order relation with:
              \begin{itemize}
                  \item $\unboundgraphr \leqrp \condensedr \leqrp \idr \leqrp \valuer$
                  \item $\valuer$ is the greatest element
                  \item $\unboundgraphr$ is the least element
              \end{itemize}
    \end{itemize}
    This total order serves as a guide to determine the variable representation of the output of a SPARQL operator.
    For any two representations $r_1, r_2$, we write $r_1 \infrp r_2$ to denote their minimum according to $\leqrp$.
\end{definition}

\begin{definition}[Environment]
    Let the environment $\environement: \varset \to \representationset$ be a partial function that maps a variable to a variable representation. It is represented as a set of pairs $(\var, \representation)$ where $\var$ is a variable and $\representation$ is a variable representation. We note $\domain{\environement}$ the domain of the function $\environement$.
\end{definition}

\begin{definition}[Domain of a set of environments]
    Let $\environements$ be a set of environments. We note:
    \begin{itemize}
        \item $\environements, \environements_1, \environements_2,\dots$ a finite set of environments
        \item $\domain{\environements} = \{ \domain{\environement} \mid \environement \in \environements \}$ the set of domains of all environments in $\environements$
    \end{itemize}
\end{definition}

\begin{definition}[Lower environment of a variable]
    $\lowerenv: 2^{\environements} \to \varset \to \environements$ is a function that takes a set of environments and a variable and returns the environment where the variable is reduced to the lowest representation. It is defined as follows:
    \begin{equation}
        \begin{split}
            \domain{\lowerenv(\environements)} = & \cap \{ \domain{\environement} \mid \environement \in \environements \} \\
            \lowerenv(\environements)(\var) = & \representation \mid \forall \environement \in \environements, \\
            & \representation \leqrp \environement(\var) \land \forall \representation', (\forall \environement \in \environements \representation' \leqrp \environement(\var)) \Rightarrow \representation' \leqrp \representation
        \end{split}
    \end{equation}
\end{definition}

\begin{definition}[Logical provability of an environment]
    Logical provability, denoted by $\vdash$, is a relation between an environment $\environement$ and an expression $\expression$ such that $\environement \vdash \expression$ holds if and only if, under the assumptions and variable representations specified by $\environement$, the expression $\expression$ is well-typed and can be evaluated according to the typing rules of the algebra.
\end{definition}

\begin{lemma}[Greatest Environment Typing]
    Let $\environements$ a set of environments and $\expression$ is an expression. If $\forall \environement \in \environements, \environement \vdash \expression$, then $\geqrp(\environements) \vdash \expression$.
\end{lemma}

In our approach, SPARQL operators impose constraints on variable representations. Let's take the \texttt{FILTER} operator as an example. When using comparison operators (like <, >, =, !=) in a FILTER clause, we can only compare actual values - we cannot compare condensed graph identifiers ($\condensedr$) or internal identifiers ($\idr$) directly. Therefore, when using FILTER, we must promote any variables of type $\condensedr$ or $\idr$ to the $\valuer$ representation by resolving them to their actual values. This promotion ensures the variables can be properly compared in filter conditions.
In the case of the \texttt{JOIN} operator, it is necessary for the joined variables to be of the same type in order to apply the appropriate join operation. If the variables are of different types, they need to be promoted to a common type.

\begin{definition}[Optimal representation]
    We note that $\optrepr(\expression)$ is the optimal environment for the expression $\expression$. It is the environment where all the variables are promoted to the lowest representation. We define the function $\optreprfunc$ such that:
    \begin{equation}
        \optrepr(\expression) = \lowerenv(\{ \environement \mid \environement \vdash \expression \})
    \end{equation}
\end{definition}

\begin{example}
    Let's consider the SPARQL query \ref{ontology-based-query-specific-ng} below.

    \begin{lstlisting}[language=SPARQL, caption={Get all the building heights for a specific named graph}, label={ontology-based-query-specific-ng}]
    SELECT ?height WHERE {
            GRAPH ?vng {
                ?s building:height ?height
            }
            ?vng v:is-version-of ng:Gr-Lyon .
    }
    \end{lstlisting}

    The optimal environment at the \texttt{quad pattern} operator is $\{ (?s, \idr), (?height, \idr), (?vng, \condensedr) \}$.
\end{example}

\begin{definition}[Value translation]
    \label{def:value-translation}
    Let $\translatevalue{\representation \leadsto \representation'}{\dvaluer(\representation)}$ be a function that returns the set of values in the representation $\representation'$ that are equivalent to the value in the representation $\representation$. The function $\translatevalue{\representation \leadsto \representation'}{\dvaluer(\representation)}$ is defined as:
    \begin{equation}
        \begin{split}
            \translatevalue{\unboundgraphr \leadsto \condensedr}{\bullet} = & \{ (n, \versions) \mid \idtovalfunc{n} \in \namedgraphs \land \versions \subseteq \vers{\rdfdataset} \} \\
            \translatevalue{\condensedr \leadsto \idr}{(n, \versions)} = & \{ n' \mid \exists \version \in \versions : \vis(\version, \idtovalfunc{n}) = \idtovalfunc{n'} \} \\
            \translatevalue{\idr \leadsto \valuer}{n} = & \{ \idtovalfunc{n} \}
        \end{split}
    \end{equation}
\end{definition}

\begin{definition}[Type downgrade of a variable's mapping]
    Let $\downgrade{\var}{\representation \to \representation'}{\mapping}$ be a function that returns the condensed mappings where the variable $\var$ is reduced to the representation $\representation'$. The function $\downgrade{\var}{\representation \to \representation'}{\mapping}$ is defined as:
    \begin{equation}
        \begin{split}
            \downgrade{\var}{\environement \to \environement'}{\mapping} & = [\mapping' \mid  \\
            & \begin{cases}
                \mapping'(\var') = \mapping(\var)                                                                      & \text{if } \var \neq \var' \\
                \mapping'(\var) \in \translatevalue{\environement(\var) \leadsto \environement'(\var)}{\mapping(\var)} & \text{otherwise}
            \end{cases} \\
            ]&
        \end{split}
    \end{equation}
\end{definition}

\begin{definition}[Type reduction of a solution sequence]
    Let $\downgrade{\var}{\representation \to \representation'}{\solutionsequence}$ be a function that returns the condensed solution sequence where the variable $\var$ is reduced to the representation $\representation'$. The function $\downgrade{\var}{\representation \to \representation'}{\solutionsequence}$ is defined as:
    \begin{equation}
        \downgrade{\var}{\environement \to \environement'}{\solutionsequence} = \bigconcatenatefunc{\mapping \in \solutionsequence}(\downgrade{\var}{\representation \to \representation'}{\mapping})
    \end{equation}
\end{definition}

\begin{definition}[Condensed solution sequence's variable translation]
    The translation function $\translateresult{\var}{\environement \to \environement'}$ is used to translate the variable representation of a solution sequence from an environment $\environement$ to another environment $\environement'$. It is defined as follows:
    \begin{equation}
        \begin{split}
            \translateresult{\var}{\environement \to \environement'}(\solutionsequence) = & \translateresult{\var}{\environement \to \environement'}([\mapping[1], \dots, \mapping[n]]) \\
            = & \concatenate{\downgrade{\var}{\environement(\var) \to \environement'(\var)}{\mapping[1]}}{ \concatenate{\dots}{\downgrade{\var}{\environement(\var) \to \environement'(\var)}{\mapping[n]}}} \\
            = & \bigconcatenatefunc{\mapping \in \solutionsequence}(\downgrade{\var}{\environement(\var) \to \environement'(\var)}{\mapping})
        \end{split}
    \end{equation}
    Where $\concatenate{}{}$ is the concatenation operator that concatenates two solution sequences, and $\bigconcatenatefunc{}$ is the function that concatenates all the solution sequences in the list.
\end{definition}

\begin{definition}[Environment transformation]
    The environment transformation function $\uprepresentationFunc: ((\varset \to \representationset) \times (\varset \to \representationset)) \to (\algebracset \to \algebracset)$ is used to transform the variable representation to a target representation. This function takes two environments (i.e., two ways of associating variables to representations) and constructs a transformation that converts all variables from the old environment to the new environment. For each variable whose representation changes, a specific translation operation is applied to that variable. All these operations are then composed to obtain the overall transformation. It is defined as follows:
    \begin{equation}
        \begin{split}
            \uprepresentationFunc(\environement, \environement') = & \quad \bigcirc_{\{\var \mid \environement(\var) \neq \environement'(\var)\}} \translateresult{\var}{\environement \to \environement'} \text{ if } \environement \infrp \environement'
        \end{split}
    \end{equation}
\end{definition}

\subsubsection{Metadata operators}
The metadata operators are used to manipulate and query the metadata associated with versioned named graphs. These operators work together to provide fine-grained control over which versions and graphs are included in query results. They are particularly useful when working with versioned RDF datasets where we need to access data from specific versions or named graphs. A detailed ontology of the versioning operators is shown in Figure \ref{vers:ontology} in the appendix.

\paragraph{Versioned IRI restriction operator}
Given a SPARQL \texttt{GRAPH} operator, we can restrict the versioned graph to match a specific versioned IRI. The versioned IRI restriction operator \text{vi-restriction} is used in conjunction with the \texttt{GRAPH} keyword in SPARQL queries.

$vi-restr$ is a function that takes a versioned IRI and returns the singleton that contains the couple $(n, \versions)$ where $n$ is the identifier of the named graph, $\versions$ the singleton containing the variable associated to the versioned named graph, and $\idtoval$ the function that converts an RDF identifier to its value.

The expression of this operator is defined as follows:
\begin{equation}
    \begin{split}
        \text{vi-restr} & : \vis \to 2^{(\mathbb{N} \times 2^{\versionset})} \\
        \text{vi-restr} & : \vi \mapsto \{ (n, \{ \version \}) \mid \vis(\version, \namedgraphterm) = \vi \land \idtovalfunc{n} = \namedgraphterm \}
    \end{split}
\end{equation}

\begin{example}
    We can restrict the versioned graph using a versioned IRI shown in the SPARQL query \ref{ontology-based-query-specific-vng} defined below. This query retrieves all the building heights for a specific versioned named graph \texttt{<vng:Gr-Lyon-v1>}.

    \begin{lstlisting}[language=SPARQL, caption={Get all the building heights for a specific versioned graph}, label={ontology-based-query-specific-vng}]
    SELECT ?height WHERE {
        GRAPH <vng:Gr-Lyon-v1> {
            ?s building:height ?height
        }
    }
    \end{lstlisting}
\end{example}

\paragraph{Graph restriction operator}
Given a SPARQL \texttt{GRAPH} operator, we can restrict the graph by setting the named graph. The graph restriction operator $\text{vngs-of}$ is used in conjunction with the triple pattern inside the metadata graph in SPARQL queries.

\begin{example}
    We can restrict the graph by setting the named graph linked to the versioned graph shown in the SPARQL query \ref{ontology-based-query-specific-ng} previously defined. This query retrieves all the building heights for a specific named graph \texttt{ng:Gr-Lyon}.

    Given the flat dataset in Table \ref{dataset-2v-2ng}, the result of this query is: $[\{height \mapsto 10.5\}, \{height \mapsto 9.1\}, \{height \mapsto 10.5\}, \{height \mapsto 15\}]$.
    \begin{equation}
        \begin{split}
            \text{vngs-of} & : \namedgraphs \to \mathbb{N} \times 2^{\versionset} \\
            \text{vngs-of} & : \namedgraphterm \mapsto \{(n, \versions_{\namedgraphterm}) \mid \idtovalfunc{n} = \namedgraphterm \land \versions_{\namedgraphterm} = \{ \version \in \vers{\rdfdataset} \mid \namedgraphterm \text{ present in } \version \}
            \}
        \end{split}
    \end{equation}
\end{example}

\paragraph{Version restriction operator}
Given a SPARQL \texttt{GRAPH} operator, we can restrict the version by setting the version. The version restriction operator $\text{vngs-in}$ is used in conjunction with the triple pattern inside the metadata graph in SPARQL queries.
\begin{equation}
    \begin{split}
        \text{vngs-in} & : \versionset \to 2^{(\mathbb{N} \times 2^{\versionset})} \\
        \text{vngs-in} & : \version \mapsto \{(n, \{\version\}) \mid  \idtovalfunc{n} = \namedgraphterm \land \namedgraphterm \in \namedgraphs
        \}
    \end{split}
\end{equation}

\begin{example}
    We can restrict the version by setting the version linked to the versioned graph described in the SPARQL query \ref{ontology-based-query-specific-v} below. This query retrieves all the building heights for a specific version \texttt{v:1}.

    \begin{lstlisting}[language=SPARQL, caption={Get all building heights for a specific version}, label={ontology-based-query-specific-v}]
    SELECT ?height WHERE {
         GRAPH ?vng {
              ?s building:height ?height
         }
         ?vng v:in-version v:1 .
    }
    \end{lstlisting}

    The result of this query is: $[\{height \mapsto 10.5\}, \{height \mapsto 9.1\}]$.
\end{example}

\subsubsection{Typing rules}
\paragraph{Expression typing}
$\environement \vdash \expression:\representation$, if assuming that the variables are representable as defined in $\environement$, $\expression$ can evaluate to a value that can be represented by $\representation$.

The expression typing rules determine the variable representation of a SPARQL expression. We define the SPARQL expressions typing rules. A premise (judgment) is a statement or an assertion on a given abstract syntax tree. An inference rule denoted by $(Rule)$ is given by $\frac{P_1, \dots, P_n}{Conc}$ where $P_1, \dots, P_n$ are the premises and $Conc$ is the conclusion (composed of one judgment).
\begin{equation}
    \begin{split}
        & \text{Terms}: \frac{}{\environement \vdash \expression: \valuer} \text{ where } \expression \in \termfset                                                                                           \\
        & \text{Variables}: \frac{}{\environement \vdash x: \representation} \text{ where } x \in \vars \text{ and } \environement(x) = \representation                                                                         \\
        & \text{Functions}: \frac{\Gamma_1 \vdash \expression_1: \valuer, \dots, \Gamma_n \vdash \expression_n: \valuer}{\environement \vdash f(\expression_1, \dots, \expression_n): \valuer} \text{ where } f \text{ is a function} \\
        & \text{Equality}: \frac{\environement \vdash \expression_1: \representation, \environement \vdash \expression_2: \representation}{\environement \vdash \expression_1 = \expression_2: \valuer} \text{ where } \idr \leqrp \representation
    \end{split}
\end{equation}

The equality rule is used to determine the variable representation of the equality and inequality operators. Those operators take two expressions as input and returns a boolean value (considered as a $\valuer$). The two expressions needs to be of the same type and at least an $\idr$ to be able to compare them.

\paragraph{Operators typing}
The typing rules are used to determine the variable representation of the output of a SPARQL operator. In this paper we define typing rules for selected SPARQL operators.
\begin{equation}
    \begin{split}
        & (\quadpattern \text{ with IRI}): \frac{}{\environement \vdash \quadpattern(s, p, o, x)} \text{ and } x \in \termfset                                                                                                   \\
        & (\quadpattern \text{ with variable}): \frac{}{\environement \vdash \quadpattern(s, p, o, x)} \text{ if } x \in \domain{\environement}
    \end{split}
\end{equation}

\begin{example}
    \begin{lstlisting}[language=SPARQL, caption={Concatenation of the building names by height where the height is greater than 10}, label={sparql:filter}]
    SELECT ?height (GROUP_CONCAT(CONCAT("B.", ?b); separator=",") AS ?concatbuilding)  WHERE {
        GRAPH ?vng {
            ?b height ?height .
        }
        FILTER(?height > 10)
    } GROUP BY ?height
    \end{lstlisting}

    For certain aggregations, we need to convert the aggregated variables into $\valuer$ to be able to perform the aggregation operations. For example, for a sum operation, we need to convert the $\idr$ variables into $\valuer$ to be able to perform the operation. Let's consider the SPARQL query in Listing \ref{sparql:filter}. The grouping attributes are \texttt{?height} and the aggregate is \texttt{GROUP\_CONCAT(CONCAT("B.", ?b); separator=",")}. The variable \texttt{?height} is of type $\idr$ and the variable \texttt{?b} is of type $\valuer$.
\end{example}

The equation below presents several typing rules for algebraic operations in formal logic. Each rule describes how an expression can be typed based on its environment and the sub-expressions or values it contains. For each rule, the premises are the conditions that must be met for the rule to apply, and the conclusion is the type of the expression.
\begin{equation}
    \begin{split}
        & Join (\join) :  \frac{\environement \vdash \expression_1, \environement \vdash \expression_2}{\environement \vdash \join(\expression_1, \expression_2)}                                                                                                              \\
        & Filter (\filter):  \frac{\environement \vdash \expression, \environement \vdash c:\valuer}{\environement \vdash \filter(\expression, c)}                                                                \\
        & Union (\union): \frac{\environement \vdash \expression_1, \environement \vdash \expression_2}{\environement \vdash \union(\expression_1, \expression_2)}                                                          \\
        & Diff (\diff): \frac{\environement \vdash \expression_1, \environement \vdash \expression_2, \environement \vdash \expression_3}{\environement \vdash \diff(\expression_1, \expression_2, \expression_3)}                                                \\
        & Optional (\leftouterjoin): \frac{\environement \vdash \expression_1, \environement \vdash \expression_2, \environement \vdash \expression_3}{\environement \vdash \leftouterjoin(\expression_1, \expression_2, \expression_3)}                                                    \\
        & Projection (\projection):  \frac{\environement \vdash \expression}{\environement \vdash \projection(\expression, \{\var_1, \dots, \var_n\})}  \\
        & Group By (\group): \frac{\forall 1 \leq i \leq m, \environement \vdash \expression_i:\valuer, \environement \vdash \expression}{\environement \vdash \group(\{ \expression_1', \dots, \expression_m' \}, \expression)} \\
        & Aggregation (\aggregation): \frac{\forall 1 \leq i \leq n, \environement \vdash \expression_i:\valuer, \environement \vdash \group}{\environement \vdash \aggregation(\{\expression_1, \dots, \expression_n\}, \aggFunc, \scalarvals, \group)} \\
        & Aggregate Join (\aggregateJoin): \frac{\forall 1 \leq i \leq k, \environement \vdash \aggregation_i}{\environement \vdash \aggregateJoin(\aggregation_1, \dots, \aggregation_k)}
    \end{split}
\end{equation}

\begin{example}
    Let's consider the SPARQL query above \ref{sparql:filter}. The term \texttt{10} in the FILTER expression is of type $\valuer$.
\end{example}

\begin{example}
    Let's consider the SPARQL query in Listing \ref{sparql:filter}. The function \texttt{FILTER} is of type $\valuer$ because the expression \texttt{?height > 10} is of type $\valuer$. The variable \texttt{?height} is of type $\valuer$ because it needs to compare it with the term \texttt{10}. In this case, the function ">" is a function that takes two terms as input and returns a boolean value (considered as a $\valuer$).
\end{example}

\subsubsection{Operators translation}
As shown in Figure~\ref{fig:models-translation-exemple}, the dataset is stored in a condensed model. In order to evaluate a SPARQL query on this condensed dataset, we need to translate the input SPARQL query to another query language in order to make it compatible with the condensed model. For this purpose, we will now define the translation functions for the SPARQL operators of the flat algebra to the condensed algebra. The translation of these operators are done with the function $\translate$ and is defined as:

The translation of the flat quad pattern to the condensed quad pattern is straightforward. It doesn't require any variable type transformation.
\begin{equation}
    \begin{split}
        \translate(\quadpatternf{\subject, \predicate, \object, \ngraph}) =& \quadpatternc{\subject, \predicate, \object, \ngraph}
    \end{split}
\end{equation}

The translation of the flat join operator is done by promoting the variables to the lowest common representation between the two expressions.
\begin{equation}
    \begin{split}
        \translate(\joinf{\expression_1, \expression_2}) =& \joinc{
            \uprepresentationFunc(\optrepr(\expression_1), \environement) (\translate(\expression_1)), \uprepresentationFunc(\optrepr(\expression_2), \environement) (\translate(\expression_2))} \\
        & \text{where } \environement = \optrepr(\joinf{\expression_1, \expression_2})
    \end{split}
\end{equation}

The translation of the flat filter operator is straightforward and doesn't require any variable type transformation because they are already in the lowest common representation in the flat algebra.
\begin{equation}
    \translate(\filterf{\expressionset, \expression}) = \filterc{\expressionset, \translate(\expression)}
\end{equation}

The translation of the flat optional operator is done by promoting the variables to the lowest common representation between the two joined expressions and the filter expression.
\begin{equation}
    \begin{split}
        \translate(\leftouterjoinf{\expression_1, \expression_2, \expression_3}) =& \leftouterjoinc{ \\
            & \uprepresentationFunc(\optrepr(\expression_1), \environement) (\translate(\expression_1)), \\
            & \uprepresentationFunc(\optrepr(\expression_2), \environement) (\translate(\expression_2)), \\
            & \uprepresentationFunc(\optrepr(\expression_3), \environement) (\translate(\expression_3))} \\
        & \text{where } \environement = \optrepr(\leftouterjoinf{\expression_1, \expression_2, \expression_3})
    \end{split}
\end{equation}

The translation of the flat diff operator preserves the three-argument structure without variable promotion, as the conditions are already evaluated on $\valuer$-typed variables.
\begin{equation}
    \translate(\difff{\expression_1, \expression_2, \expression_3}) = \diffc{\translate(\expression_1), \translate(\expression_2), \translate(\expression_3)}
\end{equation}

The translation of the flat union operator is done by promoting the variables to the lowest common representation such that the variables are of the same type.
\begin{equation}
    \begin{split}
        \translate(\unionf{\expression_1, \expression_2}) =& \unionc{
            \uprepresentationFunc(\optrepr(\expression_1), \environement) (\translate(\expression_1)), \uprepresentationFunc(\optrepr(\expression_2), \environement) (\translate(\expression_2))} \\
        & \text{where } \environement = \optrepr(\unionf{\expression_1, \expression_2})
    \end{split}
\end{equation}

The translation of the flat group operator requires promoting the grouping variables to their lowest common representation, ensuring compatibility across all group expressions. The translation is defined as follows:
\begin{equation}
    \begin{split}
        \translate(\groupFFunc[\expressionset_n, \expression]) = & \groupC( \\
        & \{ \uprepresentationFunc(\optrepr(\expression_1), \environement) (\translate(\expression_1)), \dots, \uprepresentationFunc(\optrepr(\expression_n), \environement) (\translate(\expression_n)) \}, \\
        & \translate(\expression))
    \end{split}
\end{equation}

The translation of the flat aggregation operator involves promoting each aggregated expression to the appropriate representation in the target environment. The translation is defined as follows:
\begin{equation}
    \begin{split}
        \translate(\aggregationFFunc[\expressionset_n, \aggFuncF, \scalarvals, \groupF]) = & \aggregationC([\uprepresentationFunc(\optrepr(\expression_1), \environement) (\translate(\expression_1)), \\
            & \dots, \uprepresentationFunc(\optrepr(\expression_n), \environement) (\translate(\expression_n))], \\
        & \translate(\aggFuncF), \scalarvals, \translate(\groupF))
    \end{split}
\end{equation}

The translation of the flat aggregate join operator is performed by translating each aggregated expression individually and then combining them using the condensed aggregate join operator. The translation is defined as follows:
\begin{equation}
    \begin{split}
        \translate(\aggregateJoinF(\aggregationF[1], \dots, \aggregationF[n])) = \aggregateJoinC(\translate(\aggregationF[1]), \dots, \translate(\aggregationF[n]))
    \end{split}
\end{equation}

The translation of the flat projection operator preserves the projected variable set unchanged; the variable representation of each variable is determined by the sub-expression before projection.
\begin{equation}
    \translate(\projection_\flatten(\expression, \{\var_1, \dots, \var_n\})) = \projection_\condense(\translate(\expression), \{\var_1, \dots, \var_n\})
\end{equation}

%% file: content/chapters/problem/evaluation.tex
\subsection{SPARQL query evaluation}

\subsubsection{Algebra semantics}
\paragraph{Solution mapping}
We define four partial functions:
\begin{itemize}
    \item $\mvaluer{}{}: \varset \to \dvaluer(\valuer)$ that maps a set of variables to a set of RDF terms as defined in the "Solution mappings" of SPARQL Query language for RDF \cite{W3CQuery}.
    \item $\midr{}{}: \varset \to \dvaluer(\idr)$ that maps a set of variables to a set of term identifiers.
    \item $\mcondensedr{}{}: \varset \to \dvaluer(\condensedr)$ that maps a set of variables to a set of pairs composed of a term identifier and finite sets of versions. This solution mapping is called versioned solution mapping.
    \item $\munboundgraphr{}{}: \varset \to \dvaluer(\unboundgraphr)$ that maps a set of variables to the singleton that contains all the versioned named graphs denoted $\{ \bullet \}$.
\end{itemize}

We note $\mapping$ the function that returns the instances of a variable in a solution mapping in the representation $\representation$. The function $\mapping$ is defined as: $\mapping = \dvaluer(\representation)$ where $\dvaluer$ is the domain of the solution mapping such that $\dvaluer(\valuer) = \termfset$, $\dvaluer(\idr) = \mathbb{N}$, $\dvaluer(\condensedr) = \mathbb{N} \times 2^{\versionset}$ and $\dvaluer(\unboundgraphr) = \{ \bullet \}$. We consider a mapping valid when all variables appear only in one representation. The domain of each representation is disjoint such that $\dvaluer(\valuer) \cap \dvaluer(\idr) = \emptyset$, $\dvaluer(\valuer) \cap \dvaluer(\condensedr) = \emptyset$, $\dvaluer(\valuer) \cap \dvaluer(\unboundgraphr) = \emptyset$, $\dvaluer(\idr) \cap \dvaluer(\condensedr) = \emptyset$, $\dvaluer(\idr) \cap \dvaluer(\unboundgraphr) = \emptyset$ and $\dvaluer(\condensedr) \cap \dvaluer(\unboundgraphr) = \emptyset$.

\begin{definition}[Expression]
    An expression is composed of terms, variables and filter functions. We inductively define $\expressionset$ as:
    \begin{equation}
        \begin{split}
            \begin{cases}
                \term \in \expressionset                                       & \text{ if } \term \in \termfset                            \\
                \var \in \expressionset                                        & \text{ if } \var \in \vars                                 \\
                g(\expression_1, \dots, \expression_n)^{\bigcup_{i}{\vars'_i}} & \text{ if } \expression_i \in \expressionset               \\
                                                                               & \text{ and } g \text{ is a filter function with arity } n. \\
                                                                               & \text{ OR/2, AND/2 and NOT/1}                              \\
                                                                               & \text{ are filter functions}
            \end{cases}
        \end{split}
    \end{equation}
\end{definition}

\paragraph{Flat algebra semantics}
The function $\evalf{\flatmodel, \expression}$ evaluates a SPARQL expression $\expression$ over a flat dataset $\flatmodel$ and returns a solution sequence $[\mvaluer{1}{}, \dots, \mvaluer{n}{}]$.



\begin{definition}[Evaluation of an expression]
    We build $values$ according to the interpretation of the "Effective Boolean Value" \cite{W3CSPARQL} that is used to calculate the result of a filter expression such that $\term \in values$ if $\applyEBV{\term}{true} \lor \applyEBV{\term}{false}$. We consider also that $null \in values$.

    The function $\interpretT{\term}: value$ is a function that returns the value of a term $\term$ and the function $\interpretT{g/n}: value^n \to value$ is a function that returns the value of a filter function $g/n$ applied to $n$ values. They are defined as follows:
    \begin{equation}
        \begin{split}
            \evalE{\term,\mvaluer{}{}} =                                   & \interpretT{\term}                                                                                                                \\
            \evalE{\var,\mvaluer{}{}} =                                    & \begin{cases}
                                                                                 \mvaluer{}{}(\var) & \text{if } \var \in \domain{\mvaluer{}{}} \\
                                                                                 null               & \text{otherwise}
                                                                             \end{cases}                                                          \\
            \evalE{g(\expression_1, \dots, \expression_n), \mvaluer{}{}} = & \interpretT{g/n}(\evalE{\expression_1, \mvaluer{}{}}, \dots, \evalE{\expression_n, \mvaluer{}{}})                                 \\
            \evalE{\expression, \mapping} =                                & \evalE{\expression, \mvaluer{}{}}                                                                                                 \\
            \evalE{\card, \mvaluer{}{}} =                                  & 1                                                                                                                                 \\
            \evalE{\card, \mapping} =                                      & \prod_{\var \in \mcondensedr{}{}}|\mcondensedr{}{}(\var)|  \times \prod_{\var \in \munboundgraphr{}{}}|\munboundgraphr{}{}(\var)|
        \end{split}
    \end{equation}

    \begin{example}
        Let's consider the SPARQL query \ref{sparql:filter}. The interpretation of the filter expression is $\interpretT{?height > 10} = \interpretT{?height} > 10$. Then, the evaluation of the filter is $\evalE{?height > 10, \mvaluer{}{}} = \interpretT{?height} > 10$. The result of the filter is $values = \{ true, false \}$.
    \end{example}
\end{definition}

\begin{lemma}[Invariance under mapping equivalence]
    \label{lemma:evalE}
    Let $\mapping[1], \mapping[2]$ be two mappings such that:
    \begin{equation}
        \begin{split}
            \forall{var} \in VAR, (\mapping[1](var) = \mapping[2](var) \lor          \\
            (var \notin \domain{\mapping[1]} \land var \notin \domain{\mapping[2]})) \\
            \forall{\expression_i} \in \expressionset, \evalE{\mapping[1]} = \evalE{\mapping[2]}
        \end{split}
    \end{equation}
\end{lemma}

\begin{proof}[Invariance under mapping equivalence]
    We proceed by induction on the structure of the expression $\expressionset$. First, consider the case where the expression is a variable $\var$ and $\var \notin \domain{\mapping[1]}$:
    \begin{equation}
        \begin{split}
            \var \notin dom(\mapping[2]) \land \evalE{\var, \mapping[1]} = null \land \evalE{\var, \mapping[2]} = null
        \end{split}
    \end{equation}

    Next, consider the case where the expression is a variable $\var$ and $\var \in \domain{\mapping[1]}$:
    \begin{equation}
        \begin{split}
            \evalE{\var, \mapping[1]} = \mapping[1](\var) \land \evalE{\var, \mapping[2]} = \mapping[2](\var) \land \\
            \mapping[1](\var) = \mapping[2](\var)
        \end{split}
    \end{equation}
\end{proof}

\begin{definition}[Evaluation of a list of expressions]
    We define the function $\evalLE{\expressionset_n, \mapping}$ that evaluates a list of expressions $\expressionset_n$ as:
    \begin{equation}
        \begin{split}
            \evalLE{\expressionset_n, \mapping} & = \evalLE{\{ \expression_1, \dots, \expression_n \}, \mapping}                \\
                                                & = [ \evalE{\expression_1, \mapping}, \dots, \evalE{\expression_n, \mapping} ]
        \end{split}
    \end{equation}
\end{definition}

\begin{definition}[Flat grouping function]
    The function $\groupFFunc$ is defined as:
    \begin{equation}
        \begin{split}
            \groupFFunc = & \{ \evalLE{\expressionset_n, \mvaluer{}{}} \to [ \mvaluer{}{\prime} \mid \mvaluer{}{\prime} \in \Omega,                   \\
                          & \evalLE{\expressionset_n, \mvaluer{}{\prime}} = \evalLE{\expressionset_n, \mvaluer{}{}} ] \mid \mvaluer{}{} \in \Omega \}
        \end{split}
    \end{equation}
\end{definition}

\begin{example}
    After compiling query~\ref{sparql:filter} and applying it to the dataset~\ref{dataset-2v-2ng}, the function $\groupFFunc = \groupFFunc[( ?height ), \Omega]$, where $\Omega$ is shown in Table~\ref{tab:groupFFunc-evaluation}.

    \begin{table}[h!]
        \centering
        \caption{Value of $\Omega$ before evaluating the flat group operation for query~\ref{sparql:filter} on dataset~\ref{dataset-2v-2ng}.}
        \label{tab:groupFFunc-evaluation}
        \begin{tabular}{|c|c|c|c|c|c|}
            \hline
            \cellcolor{unboundcolor}$\unboundgraphr$ & \cellcolor{condensedcolor}$\condensedr$ & \cellcolor{idcolor}$\idr$ & \multicolumn{3}{c|}{\cellcolor{valuecolor}$\valuer$}               \\
            \hline
            \rowcolor{RowGray}
                                                     &                                         &                           &
            \textbf{?b}                              &
            \textbf{?height}                         &
            \textbf{?vng}                                                                                                                                                                       \\
            \hline
                                                     &                                         &                           & ex:bldg\#1                                           & 10.5 & vi:1 \\
                                                     &                                         &                           & ex:bldg\#1                                           & 10.5 & vi:4 \\
                                                     &                                         &                           & ex:bldg\#1                                           & 10.5 & vi:3 \\
                                                     &                                         &                           & ex:bldg\#1                                           & 11   & vi:2 \\
                                                     &                                         &                           & ex:bldg\#3                                           & 15   & vi:3 \\
            \hline
        \end{tabular}
    \end{table}
\end{example}

\begin{definition}[Flat aggregation function]
    We define the function $\aggregationFFunc$ that creates the results list of an aggregation function. It is defined as:
    \begin{equation}
        \begin{split}
             & \aggregationFFunc =                                                                              \\
             & \aggFuncF([\evalLE{\expressionset_n, \mapping} \mid \mapping \in \domain{\groupF}], \scalarvals)
        \end{split}
    \end{equation}
    Where $\expressionset$ is the list of expressions, $\aggFuncF$ is the aggregation function and $\scalarvals$ a partial function passed to $\aggFuncF$ representing the parameters of the aggregation function.
\end{definition}

\begin{definition}[Flat aggregate join function]
    \label{def:aggregate-join-func}
    The result of the aggregation is a mapping of the variables to the result of the aggregation function.
    We note $\expressionsagg::= \aggregation(\expressionset_n, \aggFunc, \scalarvals, \group)$ the expressions of aggregation.
    $\aggregateJoinFFunc = [ agg_i \to val_i \mid k \in \{ k' \mid \exists 1 \leq j \leq n, k' \in \domain{\aggregationF[j]} \}, \forall 1 \leq i \leq n, (k \to val_i) \in \aggregationF[i] ]$.
\end{definition}

\begin{example}
    The result of the aggregate join ($\aggregateJoinFFunc$) can be represented as follows:

    \begin{table}[h!]
        \centering
        \caption{Result of $\aggregateJoinFFunc$ for the example.}
        \begin{tabular}{|c|c|c|c|c|}
            \hline
            \cellcolor{unboundcolor}$\unboundgraphr$ & \cellcolor{condensedcolor}$\condensedr$ & \cellcolor{idcolor}$\idr$ & \multicolumn{2}{c|}{\cellcolor{valuecolor}$\valuer$}                                          \\
            \hline
                                                     &                                         &                           & \textbf{?1}                                          & \textbf{?2}                            \\
            \hline
                                                     &                                         &                           & 10.5                                                 & B.ex:bldg\#1,B.ex:bldg\#1,B.ex:bldg\#1 \\
                                                     &                                         &                           & 11                                                   & B.ex:bldg\#1                           \\
                                                     &                                         &                           & 15                                                   & B.ex:bldg\#3                           \\
            \hline
        \end{tabular}
    \end{table}
\end{example}

\begin{definition}[Flat compatibility function]
    $\cptffunc{\mvaluer{1}{}}{\mvaluer{2}{}}$ is a function that checks if two solution mappings are compatible. It is defined as:
    \begin{equation}
        \begin{split}
            \cptffunc{\mvaluer{1}{}}{\mvaluer{2}{}} & = \forall{\var} \in \domain{\mvaluer{1}{}} \cap \domain{\mvaluer{2}{}}, \mvaluer{1}{}(\var) = \mvaluer{2}{}(\var)
        \end{split}
    \end{equation}
\end{definition}

In order to evaluate the join of two quads patterns, we need to define the merge function $\merge_\flatten$ of two mappings. The combination function is defined as:
\begin{equation}
    \begin{split}
        \mergef{\mvaluer{1}{}, \mvaluer{2}{}} & = \mvaluer{}{} \mid \cptffunc{\mvaluer{1}{}}{\mvaluer{2}{}} \land \forall{x} \in \domain{\mvaluer{1}{}} \cap \domain{\mvaluer{2}{}}, \\
        \mvaluer{}{}(x) =                     & \begin{cases}
                                                    \mvaluer{1}{}(x), & \text{if } x \in \domain{\mvaluer{1}{}} \\
                                                    \mvaluer{2}{}(x), & \text{otherwise}
                                                \end{cases}
    \end{split}
\end{equation}

The evaluation functions of those graph patterns are defined in the appendix \ref{appendix:graph-patterns}.

\paragraph{Condensed algebra semantics}
The function $\evalc{\condensedmodel, \expressions}$ evaluates a SPARQL expression $\expressions$ over a condensed dataset $\condensedmodel$ and produces a sequence of condensed solution mappings $[\mapping[1], \dots, \mapping[n]]$.

\begin{definition}[Condensed grouping function]
    The grouping function $\groupCFunc$ is defined as:
    \begin{equation}
        \begin{split}
            \groupCFunc = & \{ \evalLE{\expressionset_n, \mapping} \to [ \mapping' \mid                                                                    \\
                          & \mapping' \in \Omega, \evalLE{\expressionset_n, \mapping'} = \evalLE{\expressionset_n, \mapping} ] \mid \mapping \in \Omega \}
        \end{split}
    \end{equation}

    \begin{example}
        After compiling query~\ref{sparql:filter} and applying it to the dataset~\ref{dataset-condensed-dataset}, the function $\groupCFunc = \groupCFunc[( ?height ), \Omega]$, where $\Omega$ is shown in Table~\ref{tab:groupCFunc-evaluation}.
        \begin{table}[h!]
            \centering
            \caption{Value of $\Omega$ before evaluating the condensed group operation for query~\ref{sparql:filter} on dataset~\ref{dataset-condensed-dataset}.}

            \label{tab:groupCFunc-evaluation}
            \begin{tabular}{|c|c|c|c|c|c|}
                \hline
                \multicolumn{1}{|c|}{\cellcolor{unboundcolor}$\unboundgraphr$} & \multicolumn{1}{c|}{\cellcolor{condensedcolor}$\condensedr$} & \multicolumn{2}{c|}{\cellcolor{idcolor}$\idr$} & \multicolumn{1}{c|}{\cellcolor{valuecolor}$\valuer$}   \\
                \hline
                \rowcolor{RowGray}
                                                                               &
                \textbf{?vng}                                                  &
                \textbf{?b}                                                    &
                \textbf{?height}                                               &                                                                                                                                                                        \\
                \hline
                                                                               & (6, \{v:1, v:2\})                                            & 1                                              & 2                                                    & \\
                                                                               & (7, \{v:2\})                                                 & 1                                              & 2                                                    & \\
                                                                               & (7, \{v:1\})                                                 & 1                                              & 3                                                    & \\
                                                                               & (6, \{v:2\})                                                 & 4                                              & 5                                                    & \\
                \hline
            \end{tabular}
        \end{table}
    \end{example}
\end{definition}

\begin{definition}[Condensed aggregation function]
    Let's define the function $\aggregationCFunc$ that creates a list of result of an aggregation function. It is defined as:
    \begin{equation}
        \begin{split}
             & \aggregationCFunc =                                                                                                  \\
             & \aggFuncC([\evalLE{[\expression'_1, \dots, \expression'_n, \card], \mapping} \mid \mapping \in \Omega], \scalarvals)
        \end{split}
    \end{equation}
\end{definition}

\begin{definition}[Condensed aggregate join function]
    \label{def:aggregate-join-c-func}
    After the group by operation, we can apply the aggregation function to the results list. The function is defined as:
    \begin{equation}
        \begin{split}
             & \aggregateJoinCFunc[ \expressionsagg_n ] = [ agg_i \to val_i \mid                    \\
             & \quad k \in \{ k' \mid \exists 1 \leq j \leq n, k' \in \domain{\aggregationC[j]} \}, \\
             & \quad \forall 1 \leq i \leq n, (k \to val_i) \in \aggregationC[i] ]
        \end{split}
    \end{equation}
\end{definition}

\begin{definition}[Condensed compatibility function]
    It checks compatibility by requiring that, for each component of the mappings, the values or identifiers are equal for shared variables, condensed elements have the same graph and overlapping version sets, and unbound graphs are equal for common keys. In summary, two mappings are compatible if all their shared domains agree according to these criteria.
    It is defined as:
    \begin{equation}
        \begin{split}
            \cptcfunc{\mapping[1]}{\mapping[2]} & = \forall{x} \in \domain{\mvaluer{1}{}} \cap \domain{\mvaluer{2}{}}, \mvaluer{1}{}(x) = \mvaluer{2}{}(x) \land                                                  \\
                                                & \forall{x} \in \domain{\midr{1}{}} \cap \domain{\midr{2}{}}, \midr{1}{}(x) = \midr{2}{}(x),                                                                     \\
                                                & \forall{x} \in \domain{\mcondensedr{1}{}} \cap \domain{\mcondensedr{2}{}},                                                                                      \\
                                                & \mcondensedr{1}{}(x) = (\nnamedgraph{1}, V_1), \mcondensedr{2}{}(x) = (\nnamedgraph{2}, V_2), \nnamedgraph{1} = \nnamedgraph{2},                                \\
                                                & V_1 \cap V_2 \neq \emptyset, \forall{x} \in \domain{\munboundgraphr{1}{}} \cap \domain{\munboundgraphr{2}{}}, \munboundgraphr{1}{}(x) = \munboundgraphr{2}{}(x)
        \end{split}
    \end{equation}

    A basic graph pattern inside a graph can be interpreted as a join of two quad patterns. In order to evaluate the join of two quad patterns, we need to define the combination of two mappings $\mapping[1], \mapping[2]$. This equation describes how to merge two mappings by defining the resulting values for each common domain based on the values of the original mappings. The combination function is defined as:
    \begin{equation}
        \begin{split}
             & \mergec{\mapping[1], \mapping[2]} = \mapping \mid \cptcfunc{\mapping[1]}{\mapping[2]} \land \forall{x} \in \domain{\mvaluer{1}{}} \cap \domain{\mvaluer{2}{}}, \\
             & \mvaluer{}{}(x) =
            \begin{cases}
                \mvaluer{1}{}(x), & \text{if } x \in \domain{\mvaluer{1}{}} \\
                \mvaluer{2}{}(x), & \text{otherwise}
            \end{cases}                                                                                             \\
             & \forall{x} \in \domain{\midr{1}{}} \cap \domain{\midr{2}{}},                                                                                                   \\
             & \midr{}{}(x) =                            \begin{cases}
                                                             \midr{1}{}(x), & \text{if } x \in \domain{\midr{1}{}} \\
                                                             \midr{2}{}(x), & \text{otherwise}                     \\
                                                         \end{cases}                                                   \\
             & \forall{x} \in \domain{\mcondensedr{1}{}} \cap \domain{\mcondensedr{2}{}},                                                                                     \\
             & \mcondensedr{}{}(x) =
            \begin{cases}
                \mcondensedr{1}{}(x), \text{if } x \in \domain{\mcondensedr{1}{}} \setminus \domain{\mcondensedr{2}{}} \\
                \mcondensedr{2}{}(x), \text{if } x \in \domain{\mcondensedr{2}{}} \setminus \domain{\mcondensedr{1}{}} \\
                (\nnamedgraph{1}, \versions_1 \cap \versions_2), \text{otherwise}                                      \\
                \text{where } \mcondensedr{1}{}(x) = (\nnamedgraph{1}, \versions_1) \land \mcondensedr{2}{}(x) = (\nnamedgraph{2}, \versions_2)
            \end{cases}                                    \\
             & \forall{x} \in \domain{\munboundgraphr{1}{}} \cap \domain{\munboundgraphr{2}{}}, \munboundgraphr{}{}(x) = \munboundgraphr{1}{}(x)
        \end{split}
    \end{equation}
\end{definition}

The compatibility function $\cpt$ and the merge function $\merge_\condense$ are used to evaluate the join expression. The evaluation functions of those graph patterns are defined in the appendix \ref{appendix:condensed-graph-patterns}.

\subsubsection{Results equivalence}

\begin{definition}[Represents]
    Let $\valsfunc{}{\var}: \varset \to 2^\termfset$ be a function that returns the set of terms for a variable in the mapping $\mapping$. We consider the injective function $\idtoval$ to be given as the evaluation of a $\midr{}{}$ to a $\mvaluer{}{}$. The function $\valsfunc{}{\var}$ is defined as: $\valsfunc{}{\var} =$
    \begin{equation}
        \begin{cases}
            \{ \mvaluer{}{}(\var) \},                                                                                                     & \text{if } \var \in \domain{\mvaluer{}{}}        \\
            \{\idtovalfunc{\midr{}{}(\var)} \},                                                                                           & \text{if } \var \in \domain{\midr{}{}}           \\
            \{ \vis(\version,\namedgraphterm) \mid (\namedgraphterm, \versions) = \mcondensedr{}{}(\var) \land \version \in \versions \}, & \text{if } \var \in \domain{\mcondensedr{}{}}    \\
            \{ \vis(\version,\namedgraphterm) \mid \version \in \versionset \land \namedgraphterm \in \namedgraphs \},                    & \text{if } \var \in \domain{\munboundgraphr{}{}}
        \end{cases}
    \end{equation}
\end{definition}

\begin{example}
    Let $\mapping = \{\mvaluer{}{} \mapsto (), \midr{}{} \mapsto (?b \mapsto 4, ?height \mapsto 5), \mcondensedr{}{} \mapsto (?vng \mapsto (6, \{\text{v:2}\})),\munboundgraphr{}{} \mapsto () \}$.

    Then $\valsfunc{}{?b} = \{ \text{ex:bldg\#3} \}$, $\valsfunc{}{?height} = \{ 15 \}$, $\valsfunc{}{?vng} = \{ \vis(\text{v:2}, 6) = vi:3 \}$.
\end{example}

\begin{definition}[Flattening of a condensed solution mapping]
    \label{def:flattening-condensed-solution-mapping}
    We write $\mapping[\flatten]$ the mapping that is of the form $\mapping[\flatten] \in \flatmodel\versionedrdfquad$. Let $\expand$ be a function that returns the sequence of all flat solution mappings for a condensed solution mapping. The function $\expand$ is defined as:
    \begin{equation}
        \begin{split}
            \expandfunc{\mapping} = [ & \mapping[\flatten] \mid \domain{\mapping[\flatten]} = \domain{\mvaluer{}{}} \union \domain{\midr{}{}} \union \domain{\mcondensedr{}{}} \union \domain{\munboundgraphr{}{}} \land \\
                                      & \forall \var \in \domain{\mapping[\flatten]}, \mapping[\flatten](\var) \in \valsfunc{}{\var}]
        \end{split}
    \end{equation}
\end{definition}

\begin{definition}[Permutation of a sequence]
    \label{def:permutation}
    Let $\permut(\solutionsequence)$ the function that returns all the permutations of a sequence where $S_n$ is the symmetric group on $n$ elements, i.e., the set of all permutations from the set $\{1, \dots, n\}$ to itself. The function $\permut$ is defined as: $\permutfunc{\solutionsequence} = \permutfunc{[\mapping[1], \dots, \mapping[n]]} = \{ [\mapping[\sigma(1)], \dots, \mapping[\sigma(n)]] \mid \sigma \in S_n \}$.
\end{definition}

\begin{lemma}[Canonical representation of an evaluation]
    If $\solutionsequence_{\flatten} \equivalent \evalc{\condensedmodel, \expression}$ then $\solutionsequence_{\flatten} \equivalent \evalc{\condensedmodel, \uprepresentationFunc(\environement, \environement')(\expression)}$.
\end{lemma}
The proof is provided in Appendix~\ref{proof:eval-canonical}.

\begin{theorem}[Evaluation equivalence]
    Let a flat algebra $\algebraf$, a condensed algebra $\algebrac$ such that $\algebrac = \translate(\algebraf)$ and a dataset $\rdfdataset$ and its associated representations $\flatmodel$ and $\condensedmodel$ then $\evalf{\flatmodel, \algebraf} \equivalent \evalc{\condensedmodel, \translate(\algebraf)}$.
\end{theorem}
The proof is provided in Appendix~\ref{proof:eval-equivalence}.

%% file: content/chapters/results.tex
\section{From theoretical foundations to benchmarks}
\label{section:benchmarks}

\subsection{Implementation of ConVer-G}
The ConVer-G system is composed of several components that work together to provide a comprehensive solution for versioned knowledge representation and querying.
The Quads loaDer (QuaDer) component is responsible for loading versions of graphs into the system.
The Quads Query (QuaQue) component handles the querying of versioned graphs using either the condensed or the flat algebra.

In its current implementation, core operators are available, including the condensed quad pattern, join, filter, union, optional and aggregation.
The translation function is implemented to convert SPARQL queries involving these patterns from the flat algebra into the condensed algebra for execution.

Rather than storing version sets as explicit lists, we encode them as bitstrings, where each bit position corresponds to a specific version.
A set bit indicates the presence of that version in the set.
This representation enables set operations—such as intersection (bitwise AND) and union (bitwise OR)—which are usefull for the condensed join operator.
For example, joining two quads valid in versions \{v:1, v:5\} and \{v:1, v:10\} involves a simple bitwise AND of their bitstrings, instantly yielding the intersection \{v:1\}.
This approach eliminates the need for iteration over version lists.

To compare different methods for querying multiple versions of knowledge, we benchmarked our condensed representation.
This allowed us to evaluate and contrast the approaches for querying multiple versions simultaneously.

The benchmark measured query times, memory usage, and space consumption for both the condensed and flat representations.
These tests provided insights into the strengths and weaknesses of each model regarding resource usage and performance.

Experiments were conducted on the LIRIS platform PAGoDA \url{https://projet.liris.cnrs.fr/pagoda/latest/}, which offers a stable, high-performance, and controlled environment to ensure reliable results.
We used Argo Workflows, deployed on the PAGoDA Kubernetes cluster, to automate and manage experiment pipelines efficiently.
A dataset with fixed numbers of versions, products, and steps was used to ensure consistency across runs, and a fixed memory limit was set for comparability.

The benchmarking environment provided 156 virtual CPU cores and 896GB of RAM, distributed across three HPE Proliant servers (DL380 and DL385 models), each with two AMD EPYC 7443 processors (24 cores per processor, 2.85GHz, 200W).

The environment was configured as follows:
\begin{itemize}
    \item \textbf{Cluster provider}: RKE2.
    \item \textbf{Kubernetes version}: 1.30.7+rke2r1.
    \item \textbf{Architecture}: Amd64.
\end{itemize}

The benchmark used these parameters:
\begin{itemize}
    \item \textbf{Number of versions in the dataset (Ve)}: [5,20,50,100].
    \item \textbf{Number of steps (St)}: [0,5,10].
    This determined the number of products (i.e., quads) added to the dataset in each version.
\end{itemize}

These parameters were chosen to cover a broad range of scenarios and ensure the results reflect real-world use cases.
A cartesian product of the parameters generated the configurations.
For each configuration, an experiment was run and results recorded.
The benchmark results are available in the following sections, and for reproducibility, the source code used to conduct the benchmark is available at \url{https://github.com/VCityTeam/UD-knowledge-evolution-experiment}.

Each experiment deployed the components in a Kubernetes-controlled environment, allowing precise management of memory and CPU resources.
Resource allocations for each component were:
\begin{itemize}
    \item \textbf{Processor}: Virtual CPU core request: 1, Virtual CPU core limit: 2.
    \item \textbf{Memory allocation}: Request - 4GB, Limit - 8GB.
\end{itemize}

\subsection{Model - Space usage}
The condensed representation may use less space than the flat model.
We conducted experiments to compare the space usage of both models.
Three plots illustrate performance and resource usage for different types of knowledge graphs: non-evolutive, evolutive, and highly evolutive.
These visualizations provide further insight into the comparative analysis.
To ensure validity, only experiments completed without errors were included.

Results show that the Postgres-condensed approach consistently uses the least space across all configurations.
For instance, with 50 versions and 10 steps, Postgres-condensed uses about 841MB, compared to 2333MB for Postgres-flat, 3091MB for Blazegraph, and 13148MB for Jena.
This trend holds for smaller datasets: with 5 versions and 10 steps, Postgres-condensed uses 34MB, while Postgres-flat uses 74MB, Blazegraph 86MB, and Jena 195MB.

As the number of versions and steps increases, space usage grows for all systems, but much more slowly for Postgres-condensed.
Blazegraph and Jena show the steepest increases, especially for highly evolutive datasets.

\subsection{Dataset query time}
For the same query, the condensed representation is expected to be faster than the flat one.
We conducted experiments to compare query times for both models.

A set of 17 queries was designed to evaluate system performance, covering both non-aggregative and aggregative use cases.
Each query was executed 200 times, with the first 20 runs discarded to account for system warm-up, ensuring only stabilized measurements were analyzed.
Each experiment ran every query multiple times on each component to ensure statistical significance and account for variability.

Benchmark results for both non-aggregative and aggregative queries reveal clear trends.
For non-aggregative queries, QuaQue-flat consistently achieves the lowest median query times in all configurations, often outperforming other systems by a significant margin.
Blazegraph and Jena TDB generally have higher query times, with performance degrading as the number of versions and steps increases.
QuaQue-condensed, while not as fast as QuaQue-flat for non-aggregative queries, remains competitive and scales better than Blazegraph and Jena TDB.

For aggregative queries, performance is more nuanced.
QuaQue-flat often provides the best median times, but in some configurations, QuaQue-condensed approaches or surpasses its performance.
This suggests that the choice between condensed and flat representations depends on the query type and dataset characteristics.
As datasets become more complex, the performance gap between QuaQue-flat and other systems widens, especially for non-aggregative queries.

In summary, the flat representation is generally optimal for non-aggregative queries, while the condensed representation can be advantageous for certain aggregative queries and scales better in terms of space usage.
Blazegraph and Jena TDB are less efficient in both query time and scalability compared to the specialized QuaQue implementations.
QuaQue-flat queries are faster because they operate on a simple, pre-processed table and use \textbf{Hash Joins}, ideal for bulk processing.
QuaQue-condensed queries are slower due to complex data transformations performed on the fly, which can mislead the query planner and result in inefficient \textbf{Nested Loop} plans, leading to suboptimal performance under real workloads.

%% file: content/chapters/discussion.tex
\section{Discussion}
\label{section:discussions}
Our formalization provides a solid foundation for understanding and handling RDF versioning in a condensed representation. While our primary motivation and examples have focused on urban data management, the approach is generalizable to a wide range of domains where versioned knowledge graphs are relevant, such as biomedical informatics, digital humanities, software engineering, and scientific data management. Any application that requires querying, and provenance tracking of evolving RDF datasets with different POVs can benefit from the condensed representation and associated algebra. Furthermore, several theoretical limitations need to be addressed in future works.

\subsection{Integration with Existing RDF Infrastructure}
Our formalization is designed to be compatible with existing RDF standards and infrastructure. The condensed algebra can be implemented as an extension to current RDF stores and SPARQL engines, enabling backward compatibility and incremental adoption. Integration with standards such as PROV \cite{W3CPROV} would facilitate interoperability and provenance tracking. Furthermore, the translation layer from standard SPARQL to the condensed algebra allows users to leverage familiar query languages while benefiting from the performance and storage optimizations of the condensed model. By aligning our ontology and implementation with established RDF tools and protocols, our approach can be seamlessly incorporated into existing workflows and infrastructures, broadening its impact across multiple domains.

\subsection{RDF standards and expressiveness}
\subsubsection{Integration with Existing RDF Standards}
Our formalization has not been fully integrated with existing RDF standards and tools. This would involve aligning our versioning ontology with standards like PROV \cite{W3CPROV} and implementing our model in popular RDF query engines.

\subsubsection{Handling of Variable Clashes}
Our current formalization does not address variable clashes, which can lead to ambiguity and errors in query evaluation. A more robust approach would involve mechanisms to detect and resolve such clashes.

\subsubsection{Blank node handling}
Our formalization does not address the handling of blank nodes, which are a common feature of RDF data. A more comprehensive model would need to include mechanisms for dealing with blank nodes in versioned data.

\subsubsection{Limited Expressiveness of the Versioning Model}
Our model is based on a snapshot-based versioning approach, which may not be suitable for all use cases. For example, it cannot handle continuous changes or temporal intervals effectively.

\subsection{Solution Sequences and modifiers needs}
Projection modifier has been considered in our formalization. However, other modifiers such as Offset, Limit, Order, Distinct, and Reduced are not yet supported. These modifiers are essential for many query scenarios and will be considered in future work.

\subsection{Condensed representation limitations}
\subsubsection{Complexity of the Condensed Representation}
While the condensed representation offers potential storage and query optimization benefits, it also introduces complexity in terms of model manipulation and query translation.

\subsubsection{Abstraction of the level of compression}
\paragraph{Finite set of compression levels}
Our formalization uses a finite set of compression levels. This means that we have defined a limited number of compression levels that our system can handle. While this approach is sufficient for many use cases, it has certain limitations.
A formalization that generalizes the typing of variables for any representation could be more advantageous. By allowing more flexible typing, we could support a wider range of compression levels, including intermediate levels or levels defined dynamically based on the specific needs of the application.

\paragraph{From a bounded total order to a bounded lattice}
Our formalization uses a bounded total order to represent the variable representations. While this approach is sufficient for many use cases, it has certain limitations.
A formalization that uses a bounded lattice could be more advantageous. By allowing more flexible typing, we could support a wider range of compression levels, including intermediate levels or levels defined dynamically based on the specific needs of the application.

%% file: content/acknowledments.tex
\begin{acks}
This work \emph{Condensed representation of RDF and its application on graph versioning} is supported and funded by the IADoc@UDL (Université de Lyon, Universite Claude Bernard Lyon 1) and LIRIS UMR 5205.
We also acknowledge the BD team and the Virtual City Project \cite{VCITYWebsite} members for their invaluable advice and assistance.
We thank the anonymous reviewers for their constructive comments that helped improve the quality of this paper.
\end{acks}

%% file: content/appendix.tex
\appendix

\clearpage

\onecolumn
\section{State of the art - Versioning}
\subsection{Dataset versioning}

\begin{table}[h]
      \caption{Comparison of versioning tools by paradigms and query capabilities and area of application adapted from \cite{pelgrinefficient}}
      \label{table:versioning-tools-queries}
\begin{tabular}{|c||c||c|c|c|}
            \hline
            \textbf{Tool}                                                                                                     & \textbf{Paradigms} & \textbf{Queries} & \textbf{Operators} \\
            \hline
            \href{https://github.com/VCityTeam/ConVer-G}{\textbf{ConVer-G}}\cite{gil2024convergconcurrentversioningknowledge} & TB                 & all              & all                \\
            Drydra\cite{anderson2016transaction}                                                                              & TB                 & all              & all                \\
            \href{https://github.com/rdfostrich/ostrich}{OSTRICH}\cite{taelman2018ostrich}                                    & IC/CB/TB           & VM/DM/VQ         & TP/BGP/Agg         \\
            \href{https://github.com/AKSW/QuitStore}{Quit Store}\cite{arndt-2020-dissertation}                                & FB                 & all              & all                \\
            RDF-TX\cite{gao2016rdf}                                                                                           & TB                 & all              & TP/BGP/Agg         \\
            R43ples\cite{graube2014r43ples}                                                                                   & CB                 & all              & all                \\
            R\&WBase\cite{vander2013r}                                                                                        & CB                 & all              & TP/BGP             \\
\hline
      \end{tabular}
      \vspace{1ex}

      {\raggedright \textbf{Operators:} [TP] Triple pattern, [BGP] Basic graph pattern, [Graph] Graph name filter, [Agg] Aggregation.\par}
\end{table}

\section{Contributions}

\subsection{Dataset models}
\begin{table}[h]
      \caption{Association of the identifiers with their values}
      \label{ids-to-values}
      \begin{tabular}{ |l|l| }
            \hline
            identifier & Value      \\
            \hline
            1          & ex:bldg\#1 \\ \hline
            2          & 10.5       \\ \hline
            3          & 11         \\ \hline
            4          & ex:bldg\#3 \\ \hline
            5          & 15         \\ \hline
            6          & ng:Gr-Lyon \\ \hline
            7          & ng:IGN     \\ \hline
            8          & ex:bldg\#2 \\ \hline
            9          & height     \\
            \hline
      \end{tabular}
\end{table}

\subsection{Versioning ontology}

\begin{figure*}[h!]
      \includegraphics[angle=90,width=0.75\textwidth]{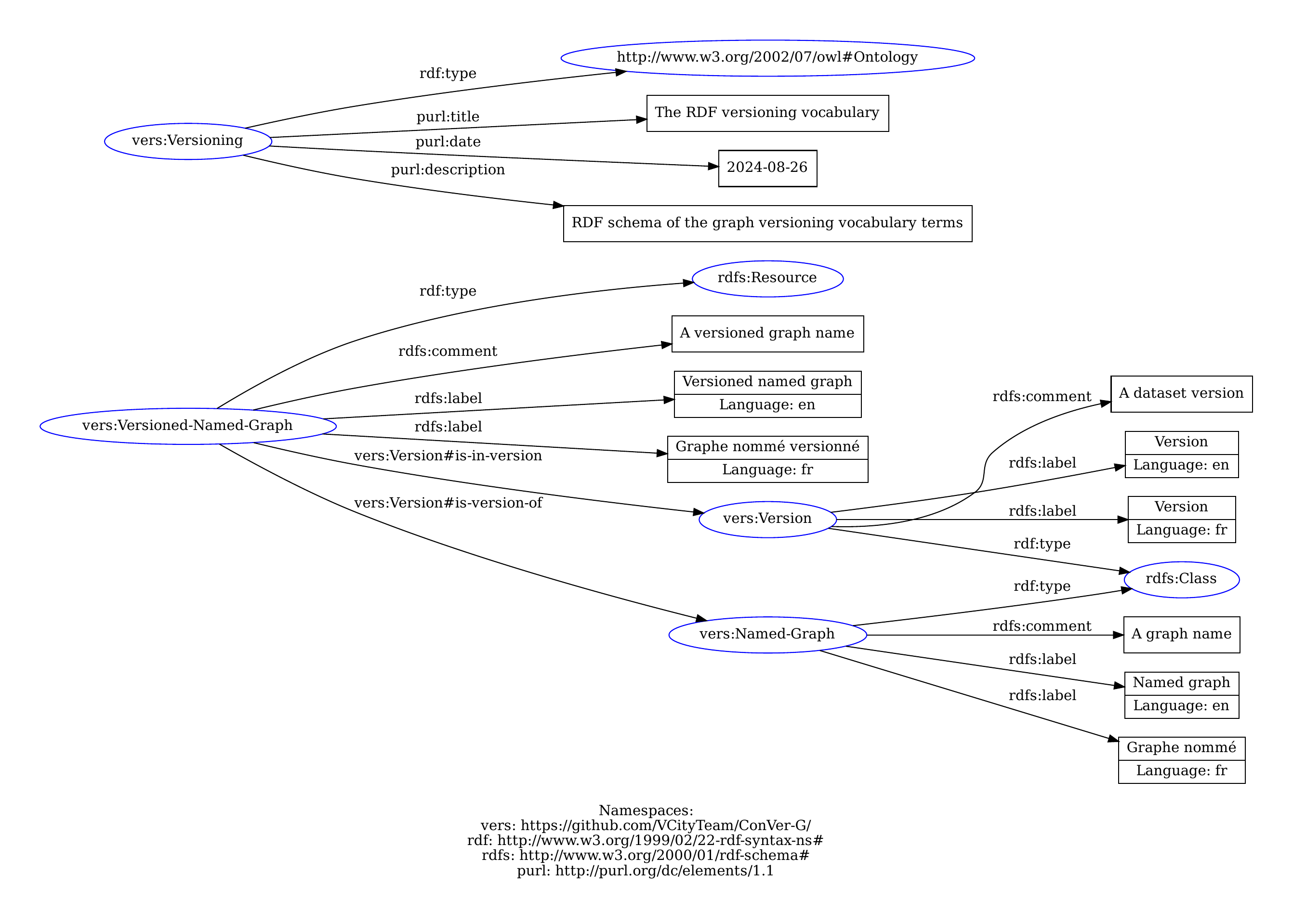}
      \caption{Versioning ontology}
      \Description{Versioning ontology of the proposition}
      \label{vers:ontology}
\end{figure*}

\subsection{Algebra semantics}
\subsubsection{Flat algebra semantics}
\paragraph{Quad pattern}
\label{appendix:graph-patterns}
%

\paragraph{Join}
\begin{flalign}
       & \evalf{\flatmodel, \join} = \evalf{\flatmodel, \joinf{\expression_1, \expression_2}} = \Omega                                                                                                    &  & \\ \nonumber
       & \domain{\Omega} = [ \mapping \mid \exists{\mapping[1]}, \exists{\mapping[2]} \mapping[1] \in \Omega_1, \mapping[2] \in \Omega_2, \mapping = \mergef{\mapping[1], \mapping[2]}]                   &  & \\ \nonumber
       & \Omega(\mapping) = \sum_{\mapping[1] \in \domain{\Omega_1}, \mapping[2] \in \domain{\Omega_2}, \mapping = \mergef{\mapping[1], \mapping[2]}}{\Omega_1(\mapping[1]) \times \Omega_2(\mapping[2])} &  &
\end{flalign}

\paragraph{Union}
\begin{flalign}
       & \evalf{\flatmodel, \union} = \evalf{\flatmodel, \unionf{\expression_1, \expression_2}} = \concatenate{\evalf{\flatmodel, \expression_1}}{\evalf{\flatmodel, \expression_2}} &  &
\end{flalign}

\paragraph{Filter}
\begin{flalign}
       & \evalf{\flatmodel, \filter} = \evalf{\flatmodel, \filterf{\expressionset, \expression}} = [ \mapping \in \evalf{\flatmodel, \expression} \land \applyEBV{\evalE{\expressionset,\mapping}}{true} ] &  &
\end{flalign}

\paragraph{Difference}
\begin{flalign}
       & \evalf{\flatmodel, \diff} = \evalf{\flatmodel, \difff{\expression_1, \expression_2, \expression_3}} = \Omega                                                                                         &  &                                                                        \\ \nonumber
       & \domain{\Omega} = [ \mapping \mid \mapping \in \Omega_1, \forall \mapping[2] \in \Omega_2, \Omega_1 = \evalf{\flatmodel, \expression_1}, \Omega_2 = \evalf{\flatmodel, \expression_2}, \begin{cases}
                                                                                                                                                                                                      \neg(\cptffunc{\mapping}{\mapping[2]}) \\
                                                                                                                                                                                                      \cptffunc{\mapping}{\mapping[2]} \land \applyEBV{\evalE{\{\expression_3\}, \mergef{\mapping, \mapping[2]}}}{false}
                                                                                                                                                                                                \end{cases} ] &  & \\ \nonumber
       & \Omega(\mapping) = \difff{\Omega_1, \Omega_2, \expression_3}(\mapping) = \Omega_1(\mapping)                                                                                                          &  &
\end{flalign}

\paragraph{Left outer join (optional)}
\begin{flalign}
       & \evalf{\flatmodel, \leftouterjoin} = \evalf{\flatmodel, \leftouterjoinf{\expression_1, \expression_2, \expression_3}} = \evalf{\flatmodel, \unionf{\filterf{\{\expression_3\}, \joinf{\expression_1, \expression_2}}, \difff{\expression_1, \expression_2, \expression_3}}} &  &
\end{flalign}

\paragraph{Group by}
\begin{flalign}
       & \evalf{\flatmodel, \group} = \evalf{\flatmodel, \groupFFunc[\expressionset_n, \Omega]} = \evalf{\flatmodel, \groupFFunc[\expressionset_n, \evalf{\flatmodel, \expression}]} &  &
\end{flalign}

\paragraph{Aggregation}
\begin{flalign}
       & \evalf{\flatmodel, \aggregation} = \evalf{\flatmodel, \aggregationFFunc} = \evalf{\flatmodel, \aggregationFFunc[\expressionset_n, \aggFuncF, \scalarvals, \evalf{\flatmodel, \groupF}]} &  &
\end{flalign}

\paragraph{Aggregate join}
\begin{flalign}
      \evalf{\flatmodel, \aggregateJoin} & = \evalf{\flatmodel, \aggregateJoinFFunc} = \evalf{\flatmodel, \aggregateJoinFFunc[\aggregation_1, \dots, \aggregation_k]} &  & \\ \nonumber
                                         & = \evalf{\flatmodel, \aggregateJoinFFunc[\evalf{\flatmodel, \aggregation_1}, \dots, \evalf{\flatmodel, \aggregation_k}]}
\end{flalign}

\subsubsection{Condensed algebra semantics}
\label{appendix:condensed-graph-patterns}
\paragraph{Quad pattern}
\begin{flalign}
                                                                                                                                                                   & \evalc{\condensedmodel, \quadpattern} =
      \begin{cases}
            \evalc{\condensedmodel, \quadpatternc{?s, ?p, ?o, ?ag}} = \begin{cases}
                                                                            [ \mapping \mid \exists{\namedgraphterm} \exists{\versions}, \mcondensedr{}{}(?ag) = (\namedgraphterm, \versions) \land \condensedmodel(\midr{}{}(?s), \midr{}{}(?p), \midr{}{}(?o), \namedgraphterm) = \versions \land \versions \neq \emptyset ] \\
                                                                            \text{if } ?ag \neq ?s \lor ?ag \neq ?p \lor ?ag \neq ?o                                                                                                                                                                                           \\
                                                                            [ \mapping \mid \exists{\namedgraphterm} \exists{\version}, \midr{}{}(?ag) = \vis(\version, \namedgraphterm) \land \version \in \condensedmodel(\midr{}{}(?s), \midr{}{}(?p), \midr{}{}(?o), \namedgraphterm) ]                                    \\
                                                                            \text{ otherwise}
                                                                      \end{cases} \\ \text{if } ?ag \text{ is a variable} \\
            \evalc{\condensedmodel, \quadpatternc{?s, ?p, ?o, ag}} = [ \mapping \mid \exists{\namedgraphterm} \exists{\version}, \vis(\version, \namedgraphterm) = ag \land
            (\midr{}{}(?s), \midr{}{}(?p), \midr{}{}(?o), \namedgraphterm, \version) \in \flatmodel] \quad \text{if } ag \text{ is an IRI}
      \end{cases} &                                         &
\end{flalign}

\paragraph{Join}
\begin{flalign}
       & \evalc{\condensedmodel, \join} = \evalc{\condensedmodel, \joinc{\expression_1, \expression_2}} = \Omega                                                                                          &  & \\ \nonumber
       & \domain{\Omega} = [ \mapping \mid \exists{\mapping[1]}, \exists{\mapping[2]}, \mapping[1] \in \Omega_1, \mapping[2] \in \Omega_2, \mapping = \mergec{\mapping[1], \mapping[2]}]                  &  & \\ \nonumber
       & \Omega(\mapping) = \sum_{\mapping[1] \in \domain{\Omega_1}, \mapping[2] \in \domain{\Omega_2}, \mapping = \mergec{\mapping[1], \mapping[2]}}{\Omega_1(\mapping[1]) \times \Omega_2(\mapping[2])} &  &
\end{flalign}

\paragraph{Union}
\begin{flalign}
       & \evalc{\condensedmodel, \union} = \evalc{\condensedmodel, \unionc{\expression_1, \expression_2}} = \concatenate{\evalc{\condensedmodel, \expression_1}}{\evalc{\condensedmodel, \expression_2}} &  &
\end{flalign}

\paragraph{Filter}
\begin{flalign}
       & \evalc{\condensedmodel, \filter} = \evalc{\condensedmodel, \filterc{\expressionset, \expression}} = [ \mapping \mid \mapping \in \evalc{\condensedmodel, \expression} \land \applyEBV{\evalE{\expressionset,\mapping}}{true}] &  &
\end{flalign}

\paragraph{Difference}
\begin{flalign}
       & \evalc{\condensedmodel, \diff} = \evalc{\condensedmodel, \diffc{\expression_1, \expression_2, \expression_3}} = \Omega                                                                                               &  &                                                                                         \\ \nonumber
       & \domain{\Omega} = [ \mapping[1] \mid \mapping[1] \in \Omega_1, \forall \mapping[2] \in \Omega_2, \Omega_1 = \evalc{\condensedmodel, \expression_1}, \Omega_2 = \evalc{\condensedmodel, \expression_2}, \begin{cases}
                                                                                                                                                                                                                      \neg(\cptcfunc{\mapping[1]}{\mapping[2]}) \\
                                                                                                                                                                                                                      \cptcfunc{\mapping[1]}{\mapping[2]} \land \applyEBV{\evalE{\{\expression_3\}, \mergec{\mapping[1], \mapping[2]}}}{false}
                                                                                                                                                                                                                \end{cases} ] &  & \\ \nonumber
       & \Omega(\mapping) = \diffc{\Omega_1, \Omega_2, \expression_3}(\mapping) = \Omega_1(\mapping)                                                                                                                          &  &
\end{flalign}

\paragraph{Left outer join (optional)}
\begin{flalign}
       & \evalc{\condensedmodel, \leftouterjoin} = \evalc{\condensedmodel, \leftouterjoinc{\expression_1, \expression_2, \expression_3}} = \evalc{\condensedmodel, \unionc{\filterc{\{\expression_3\}, \joinc{\expression_1, \expression_2}}, \diffc{\expression_1, \expression_2, \expression_3}}} &  &
\end{flalign}

\paragraph{Group by}
\begin{flalign}
       & \evalc{\condensedmodel, \group} = \evalc{\condensedmodel, \groupCFunc[\expressionset_n, \Omega]} = \evalc{\condensedmodel, \groupCFunc[\expressionset_n, \evalc{\condensedmodel, \expression}]} &  &
\end{flalign}

\paragraph{Aggregation}
\begin{flalign}
       & \evalc{\condensedmodel, \aggregation} = \evalc{\condensedmodel, \aggregationCFunc} = \evalc{\condensedmodel, \aggregationCFunc[\expressionset_n, \aggFuncC, \scalarvals, \evalc{\condensedmodel, \groupC}]} &  &
\end{flalign}

\paragraph{Aggregate join}
\begin{flalign}
      \evalc{\condensedmodel, \aggregateJoin} & = \evalc{\condensedmodel, \aggregateJoinCFunc} = \evalc{\condensedmodel, \aggregateJoinCFunc[\aggregation_1, \dots, \aggregation_k]}    &  & \\ \nonumber
                                              & = \evalc{\condensedmodel, \aggregateJoinCFunc[\evalc{\condensedmodel, \aggregation_1}, \dots, \evalc{\condensedmodel, \aggregation_k}]}
\end{flalign}

\subsubsection{Results equivalence}
\begin{proof}[Canonical representation of an evaluation]
      \label{proof:eval-canonical}
      Let's use the condensed algebra $\algebrac = \translate(\algebraf)$ be the condensed algebra expression derived from a flat algebra expression $\algebraf$ after its translation. The lemma states: If $\solutionsequence_{\flatten} \equivalent \evalc{\condensedmodel, \algebrac}$ then $\solutionsequence_{\flatten} \equivalent \evalc{\condensedmodel, \uprepresentationFunc(\environement, \environement')(\algebrac)}$.

      The proof proceeds by induction on the number of elementary variable representation transformations required to get from $\environement$ to $\environement'$. An elementary transformation changes the representation of a single variable to the next representation in the defined order $\unboundgraphr \infrp \condensedr \infrp \idr \infrp \valuer$.

      \textbf{Base Case:} $\environement = \environement'$.
      In this case, the number of transformations is 0. The environment transformation function $\uprepresentationFunc(\environement, \environement')$ is the identity function when $\environement = \environement'$, because $\{\var \mid \environement(\var) \neq \environement'(\var)\} = \emptyset$. Thus, $\uprepresentationFunc(\environement, \environement')(\expression) = \expression$.
      The lemma statement becomes: If $\solutionsequence_{\flatten} \equivalent \evalc{\condensedmodel, \expression}$ then $\solutionsequence_{\flatten} \equivalent \evalc{\condensedmodel, \expression}$, which is trivially true.

      \textbf{Inductive Hypothesis (IH):} Assume that for any pair of environments $(\environement, \environement')$ such that $\environement \infrp \environement'$ and the transformation $\uprepresentationFunc(\environement, \environement')$ involves $k$ elementary steps, the lemma holds. That is, if $\solutionsequence_{\flatten} \equivalent \evalc{\condensedmodel, \expression}$, then $\solutionsequence_{\flatten} \equivalent \evalc{\condensedmodel, \uprepresentationFunc(\environement, \environement')(\expression)}$.

      \textbf{Inductive Step:} Consider a transformation $\uprepresentationFunc(\environement, \environement')$ that involves $k+1$ elementary steps, where $\environement \infrp \environement'$.
      We can decompose this transformation by identifying an intermediate environment $\environement_1$ such that $\environement \infrp \environement_1 \leqrp \environement'$, where $\uprepresentationFunc(\environement, \environement_1)$ represents a single elementary transformation (e.g., changing the representation of the variable $\var_0$ from $\environement(\var_0)$ to $\environement_1(\var_0)$), and $\uprepresentationFunc(\environement_1, \environement')$ involves $k$ elementary steps.
      Thus, $\uprepresentationFunc(\environement, \environement')(\expression) = \uprepresentationFunc(\environement_1, \environement')(\uprepresentationFunc(\environement, \environement_1)(\expression))$.

      Let $\expression_{1} = \uprepresentationFunc(\environement, \environement_1)(\expression)$.
      The evaluation $\evalc{\condensedmodel, \expression_{1}}$ produces a solution sequence that is equivalent to applying the type downgrade transformation $\translateresult{\var_0}{\environement \to \environement_1}$ to the solution sequence resulting from $\evalc{\condensedmodel, \expression}$.
      Let $\solutionsequence_{\condense} = \evalc{\condensedmodel, \expression}$ and $\solutionsequence_{\condense1} = \evalc{\condensedmodel, \expression_{1}}$.
      By definition of the operator $\uprepresentationFunc$ on expressions, $\solutionsequence_{\condense1}$ is equivalent to $\translateresult{\var_0}{\environement \to \environement_1}(\solutionsequence_{\condense})$ and $\translateresult{\var_0}{\environement \to \environement_1}(\solutionsequence_{\condense}) = \bigcirc_{\mapping \in \solutionsequence_{\condense}} (\downgrade{\var_0}{\environement \to \environement_1}{\mapping})$.

      We get $\solutionsequence_{\flatten} \equivalent \solutionsequence_{\condense}$. By Definition \ref{def:flattening-condensed-solution-mapping} (Flattening) and Definition \ref{def:permutation} (Permutation), this means $\solutionsequence_{\flatten}$ is permutation-equivalent to $\bigcup_{\mapping \in \solutionsequence_{\condense}} \expandfunc{\mapping}$.

      We need to show that $\solutionsequence_{\condense} \equivalent \solutionsequence_{\condense1}$ in terms of their flat representation equivalence to $\solutionsequence_{\flatten}$. Here we use the fact that flattening a $\mapping$ match gives the same set of flattened matches as if we first applied type reduction to $\mapping(\var_0)$ to get a set of (potentially multiple) mappings $\{\mapping'\}$ (one for each value in $\translatevalue{\environement(\var_0) \leadsto \environement_1(\var_0)}{\mapping(\var_0)}$), and then flattening each $\mapping'$ and taking their union.

      That is: $\expandfunc{\mapping} = \bigcup_{\mapping' \in \downgrade{\var_0}{\environement \to \environement_1}{\mapping}} \expandfunc{\mapping'}$.
      Therefore, $\bigcup_{\mapping \in \solutionsequence_{\condense}} \expandfunc{\mapping} = \bigcup_{\mapping \in \solutionsequence_{\condense}} \left( \bigcup_{\mapping' \in \downgrade{\var_0}{\environement(\var_0) \to \environement_1(\var_0)}{\mapping}} \expandfunc{\mapping'} \right)$. This means $\solutionsequence_{\flatten} \equivalent \solutionsequence_{\condense1}$. So, $\solutionsequence_{\flatten} \equivalent \evalc{\condensedmodel, \expression_{1}}$.

      We have $\solutionsequence_{\flatten} \equivalent \evalc{\condensedmodel, \expression_{1}}$, and $\expression_{1} = \uprepresentationFunc(\environement, \environement_1)(\expression)$. The remaining part of the transformation is $\uprepresentationFunc(\environement_1, \environement')$, which involves $k$ elementary steps. By the IH (applied to $\expression_{1}$ with environments $\environement_1$ and $\environement'$): if $\solutionsequence_{\flatten} \equivalent \evalc{\condensedmodel, \expression_{1}}$, then $\solutionsequence_{\flatten} \equivalent \evalc{\condensedmodel, \uprepresentationFunc(\environement_1, \environement')(\expression_{1})}$.

      Since we have established that $\solutionsequence_{\flatten} \equivalent \evalc{\condensedmodel, \expression_{1}}$, we can conclude:
      $\solutionsequence_{\flatten} \equivalent \evalc{\condensedmodel, \uprepresentationFunc(\environement_1, \environement')(\expression_{1})}$.
      Substituting $\expression_{1} = \uprepresentationFunc(\environement, \environement_1)(\expression)$, we get:
      $\solutionsequence_{\flatten} \equivalent \evalc{\condensedmodel, \uprepresentationFunc(\environement_1, \environement')(\uprepresentationFunc(\environement, \environement_1)(\expression))}$.
      By the compositional nature of $\uprepresentationFunc$, this is equivalent to:
      $\solutionsequence_{\flatten} \equivalent \evalc{\condensedmodel, \uprepresentationFunc(\environement, \environement')(\expression)}$.
      This completes the inductive step. Therefore, by the principle of induction, the lemma holds for any transformation $\uprepresentationFunc(\environement, \environement')$ where $\environement \infrp \environement'$.
\end{proof}

\begin{proof}[Evaluation equivalence]
      \label{proof:eval-equivalence}
      We want to prove $\evalf{\flatmodel, \algebraf} \equivalent \evalc{\condensedmodel, \translate(\algebraf)}$. Let's proceed with the proof by induction on the structure of SPARQL algebra operators.

      \textbf{Base Case:} we want to show that $\evalf{\flatmodel, \quadpatternf{\subject, \predicate, \object, \ngraph}} \equivalent \evalc{\condensedmodel, \translate(\quadpatternf{\subject, \predicate, \object, \ngraph})}$. We rewrite this formula by applying the $\translate$ function: $\evalf{\flatmodel, \quadpatternf{\subject, \predicate, \object, \ngraph}} \equivalent \evalc{\condensedmodel, \quadpatternc{\subject, \predicate, \object, \ngraph}}$.

      Let $\mapping[\flatten] \in \evalf{\flatmodel, \quadpatternf{\subject, \predicate, \object, \ngraph}}$, by definition, there exists $\versionedrdfquad \in \flatmodel$ such that: $\mapping[\flatten](\subject) = \subjectterm, \mapping[\flatten](\predicate) = \predicateterm, \mapping[\flatten](\object) = \objectterm, \mapping[\flatten](\ngraph) = \vis(\version, \namedgraphterm)$. We have $\version \in \condensedmodel\rdfquad$ then $\rdfquad \in \domain{\condensedmodel}$. This means that $\mapping \in \evalc{\condensedmodel, \quadpatternc{\subject, \predicate, \object, \ngraph}}$ such that $\midr{}{}(\subject) = \subjectterm, \midr{}{}(\predicate) = \predicateterm, \midr{}{}(\object) = \objectterm$ and $\mcondensedr{}{}(\ngraph) = (\namedgraphterm, \versions)$ with $\versions = \condensedmodel\rdfquad$.

      Let $\var \in \domain{\mapping[\flatten]}$ then:
      \begin{itemize}
            \item if $\var \in \{\subject, \predicate, \object\}, \mapping[\flatten](\var) = \midr{}{}(\var) \land \midr{}{}(\var) \in \valsfunc{}{\var}$.
            \item if $\var = \ngraph, \mapping[\flatten] = \vis(\version, \namedgraphterm), \version \in \versions \land \mcondensedr{}{}(\var) = (\namedgraphterm, \versions)$ then $\mapping[\flatten](\var) \in \valsfunc{}{\var}$ in addition, $\domain{\mapping[\flatten]} = \domain{\mvaluer{}{}} \union \domain{\midr{}{}} \union \domain{\mcondensedr{}{}} \union \domain{\munboundgraphr{}{}}$. This means that $\mapping[\flatten] \in \expandfunc{\mapping}$.
      \end{itemize}

      \textbf{IH:} we now assume that the theorem holds for subexpressions and prove it for each SPARQL algebra operator.

      \textbf{Inductive Step:} we want to show that $\evalf{\flatmodel, \expression} \equivalent \evalc{\condensedmodel, \translate(\expression)}$ for each operator $\expression$ in the SPARQL algebra. We will show that the theorem holds for each operator in the SPARQL algebra.

      \textbf{Join operator:} we need to show that $\evalf{\flatmodel, \joinf{\expression_1, \expression_2}} \equivalent \evalc{\condensedmodel, \translate(\joinf{\expression_1, \expression_2})}$. The rewriting of this formula is: $\evalf{\flatmodel, \joinf{\expression_1, \expression_2}} \equivalent \evalc{\condensedmodel, \joinc{\translate(\expression_1), \translate(\expression_2)}}$.

      Let $\mapping[\flatten] \in \evalf{\flatmodel, \joinf{\expression_1, \expression_2}}, \exists \mapping[\flatten1] \in \evalf{\flatmodel, \expression_1}, \exists \mapping[\flatten2] \in \evalf{\flatmodel, \expression_2}$, with $\mapping[\flatten] = \mergef{\mapping[\flatten1], \mapping[\flatten2]}$. By the IH: $\evalf{\flatmodel, \expression_1} \equivalent \evalc{\condensedmodel, \translate(\expression_{1})}$. So, $\exists \mapping[1] \in \evalc{\condensedmodel, \translate(\expression_1)}, \mapping[\flatten1] \in \expandfunc{\mapping[1]}$.

      It is given that $\evalf{\flatmodel, \expression_2} \equivalent \evalc{\condensedmodel, \translate(\expression_2)}$. So, $\exists \mapping[2] \in \evalc{\condensedmodel, \translate(\expression_2)}, \mapping[\flatten2] \in \expandfunc{\mapping[2]}$. Since $\mapping[\flatten] = \mergef{\mapping[\flatten1], \mapping[\flatten2]}$, the variables common to $\mapping[\flatten1]$ and $\mapping[\flatten2]$ map to the same value in $\mapping[\flatten]$.

      If $\mapping[\flatten1] \in \expandfunc{\mapping[1]}$ and $\mapping[\flatten2] \in \expandfunc{\mapping[2]}$ and $\cptffunc{\mapping[\flatten1]}{\mapping[\flatten2]}$, then it must be that $\cptcfunc{\mapping[1]}{\mapping[2]}$ (compatibility in the condensed model holds if it holds in the flat model for the corresponding expanded mappings). Then, $\mapping = \mergec{\mapping[1], \mapping[2]}$ (the merge in the condensed model) is defined.

      Moreover, $\mapping[\flatten] \in \expandfunc{\mapping}$, since the merge operations in both models are consistent with respect to the flattening function $\expand$. Thus, $\mapping \in \evalc{\condensedmodel, \joinc{\translate(\expression_1), \translate(\expression_2)}}$, and then $\evalf{\flatmodel, \joinf{\expression_1, \expression_2}} \equivalent \evalc{\condensedmodel, \joinc{\translate(\expression_1), \translate(\expression_2)}}$.

      \textbf{Union operator:} we want to show that $\evalf{\flatmodel, \unionf{\expression_1, \expression_2}} \equivalent \evalc{\condensedmodel, \translate(\unionf{\expression_1, \expression_2})}$ We rewrite this formula: $\evalf{\flatmodel, \unionf{\expression_1, \expression_2}} \equivalent \evalc{\condensedmodel, \unionc{\translate(\expression_1), \translate(\expression_2)}}$.

      Let $\mapping[\flatten] \in \evalf{\flatmodel, \unionf{\expression_1, \expression_2}}$. This implies that either $\mapping[\flatten] \in \evalf{\flatmodel, \expression_1}$ or $\mapping[\flatten] \in \evalf{\flatmodel, \expression_2}$.

      \begin{itemize}
            \item Case $\mapping[\flatten] \in \evalf{\flatmodel, \expression_1}$:
                  By the IH, $\evalf{\flatmodel, \expression_1} \equivalent \evalc{\condensedmodel, \translate(\expression_1)}$.
                  So, $\exists \mapping[1] \in \evalc{\condensedmodel, \translate(\expression_1)}$ such that $\mapping[\flatten] \in \expandfunc{\mapping[1]}$.
                  Since $\mapping[1] \in \evalc{\condensedmodel, \translate(\expression_1)}$, it is also true that:
                  $\mapping[1] \in \evalc{\condensedmodel, \unionc{\translate(\expression_1), \translate(\expression_2)}}$ by the definition of the union operator in the condensed algebra.
                  Thus, $\mapping[\flatten] \in \expandfunc{\mapping[1]} \Rightarrow \mapping[\flatten] \in \evalc{\condensedmodel, \unionc{\translate(\expression_1), \translate(\expression_2)}}$.
            \item Case $\mapping[\flatten] \in \evalf{\flatmodel, \expression_2}$:
                  By the IH, $\evalf{\flatmodel, \expression_2} \equivalent \evalc{\condensedmodel, \translate(\expression_2)}$.
                  So, $\exists \mapping[2] \in \evalc{\condensedmodel, \translate(\expression_2)}$ such that $\mapping[\flatten] \in \expandfunc{\mapping[2]}$.
                  Since $\mapping[2] \in \evalc{\condensedmodel, \translate(\expression_2)}$, it is also true that:
                  $\mapping[2] \in \evalc{\condensedmodel, \unionc{\translate(\expression_1), \translate(\expression_2)}}$ by the definition of the union operator in the condensed algebra.
                  Thus, $\mapping[\flatten] \in \expandfunc{\mapping[2]} \Rightarrow \mapping[\flatten] \in \evalc{\condensedmodel, \unionc{\translate(\expression_1), \translate(\expression_2)}}$.
      \end{itemize}

      We suppose $\mapping \in \evalc{\condensedmodel, \unionc{\translate(\expression_1), \translate(\expression_2)}}$. Then, either $\mapping \in \evalc{\condensedmodel, \translate(\expression_1)}$ or $\mapping \in \evalc{\condensedmodel, \translate(\expression_2)}$.

      If $\mapping \in \evalc{\condensedmodel, \translate(\expression_1)}$, then by IH, for any $\mapping[\flatten] \in \expandfunc{\mapping}$, we have $\mapping[\flatten] \in \evalf{\flatmodel, \expression_1}$, and $\mapping[\flatten] \in \evalf{\flatmodel, \unionf{\expression_1, \expression_2}}$. If $\mapping \in \evalc{\condensedmodel, \translate(\expression_2)}$, then by IH, for any $\mapping[\flatten] \in \expandfunc{\mapping}$, we have $\mapping[\flatten] \in \evalf{\flatmodel, \expression_2}$, and $\mapping[\flatten] \in \evalf{\flatmodel, \unionf{\expression_1, \expression_2}}$.

      Then, $\evalf{\flatmodel, \unionf{\expression_1, \expression_2}} \equivalent \evalc{\condensedmodel, \unionc{\translate(\expression_1), \translate(\expression_2)}}$.

      \textbf{Filter operator:} $\evalf{\flatmodel, \filterf{\expressionset, \expression}} \equivalent \evalc{\condensedmodel, \translate(\filterf{\expressionset, \expression})}$ must be demonstrated. We rewrite this formula with the result of the $\translate$ function: $\evalf{\flatmodel, \filterf{\expressionset, \expression}} \equivalent \evalc{\condensedmodel, \filterc{\expressionset, \translate(\expression)}}$.

      Let $\mapping[\flatten] \in \evalf{\flatmodel, \filterf{\expressionset, \expression}}$. By definition, this means $\mapping[\flatten] \in \evalf{\flatmodel, \expression}$ and $\applyEBV{\evalE{\expressionset, \mapping[\flatten]}}{true}$.
      By the IH for the sub-expression $\expression$, $\evalf{\flatmodel, \expression} \equivalent \evalc{\condensedmodel, \translate(\expression)}$.
      So, there exists $\mapping \in \evalc{\condensedmodel, \translate(\expression)}$ such that $\mapping[\flatten] \in \expandfunc{\mapping}$.

      Because the filter condition $\expressionset$ depends only on the values of variables, and the flattening function $\expand$ preserves these values (i.e., for any $\mapping'$ represented by $\mapping$, $\evalE{\expressionset, \mapping[\flatten]}$ is equal to $\evalE{\expressionset, \mapping'}$ for the variables in $\expressionset$), it follows that if $\applyEBV{\evalE{\expressionset, \mapping[\flatten]}}{true}$, then $\applyEBV{\evalE{\expressionset, \mapping}}{true}$ also holds.

      Thus, $\mapping$ will also satisfy the filter in the condensed algebra: $\mapping \in \evalc{\condensedmodel, \filterc{\expressionset, \translate(\expression)}}$.
      This means that for every $\mapping[\flatten]$ satisfying the filter in the flat model, there's a corresponding $\mapping$ satisfying the filter in the condensed model, and $\mapping[\flatten] \in \expandfunc{\mapping}$.

      Conversely, let $\mapping \in \evalc{\condensedmodel, \filterc{\expressionset, \translate(\expression)}}$. This means $\mapping \in \evalc{\condensedmodel, \translate(\expression)}$ and $\applyEBV{\evalE{\expressionset, \mapping}}{true}$.
      By IH, for any $\mapping[\flatten] \in \expandfunc{\mapping}$, we have $\mapping[\flatten] \in \evalf{\flatmodel, \expression}$.
      Since $\applyEBV{\evalE{\expressionset, \mapping}}{true}$, and the values relevant to $\expressionset$ are preserved by $\expand$, it holds that $\applyEBV{\evalE{\expressionset, \mapping[\flatten]}}{true}$.
      Then, $\mapping[\flatten] \in \evalf{\flatmodel, \filterf{\expressionset, \expression}}$.
      This establishes that $\evalf{\flatmodel, \filterf{\expressionset, \expression}} \equivalent \evalc{\condensedmodel, \filterc{\expressionset, \translate(\expression)}}$

      \textbf{Diff operator:} $\evalf{\flatmodel, \difff{\expression_1, \expression_2, \expression_3}} \equivalent \evalc{\condensedmodel, \translate(\difff{\expression_1, \expression_2, \expression_3})}$ must be demonstrated.
      We rewrite this formula with the result of the $\translate$ function: $\evalf{\flatmodel, \difff{\expression_1, \expression_2, \expression_3}} \equivalent \evalc{\condensedmodel, \diffc{\translate(\expression_1), \translate(\expression_2), \translate(\expression_3)}}$.

      Let $\mapping[\flatten] \in \evalf{\flatmodel, \difff{\expression_1, \expression_2, \expression_3}}$.
      By definition, $\mapping[\flatten] \in \evalf{\flatmodel, \expression_1}$ and $\forall \mapping[\flatten]' \in \evalf{\flatmodel, \expression_2}$, either $\cptffunc{\mapping[\flatten]}{\mapping[\flatten]'}$ does not hold, or if it does, then it must give $\applyEBV{\evalE{\{\expression_3\}, \mergef{\mapping[\flatten], \mapping[\flatten]'}}}{false}$.

      By the IH for $\expression_1$, $\evalf{\flatmodel, \expression_1} \equivalent \evalc{\condensedmodel, \translate(\expression_1)}$.
      So, there exists $\mapping[1] \in \evalc{\condensedmodel, \translate(\expression_1)}$ such that $\mapping[\flatten] \in \expandfunc{\mapping[1]}$.

      Also, by IH for $\expression_2$, $\evalf{\flatmodel, \expression_2} \equivalent \evalc{\condensedmodel, \translate(\expression_2)}$.
      So for every $\mapping[\flatten]' \in \evalf{\flatmodel, \expression_2}$, $\exists \mapping[2] \in \evalc{\condensedmodel, \translate(\expression_2)}$ such that $\mapping[\flatten]' \in \expandfunc{\mapping[2]}$.

      It remains to show that for $\mapping[1]$, $\forall \mapping[2] \in \evalc{\condensedmodel, \translate(\expression_2)}$, either $\cptcfunc{\mapping[1]}{\mapping[2]}$ does not hold, or, if it does, then $\applyEBV{\evalE{\{\translate(\expression_3)\}, \mergec{\mapping[1], \mapping[2]}}}{false}$ holds.
      Consider any $\mapping[2] \in \evalc{\condensedmodel, \translate(\expression_2)}$. For any $\mapping[\flatten]' \in \expandfunc{\mapping[2]}$:
      if $\neg(\cptffunc{\mapping[\flatten]}{\mapping[\flatten]'})$ for all $\mapping[\flatten]' \in \expandfunc{\mapping[2]}$, then it implies $\neg(\cptcfunc{\mapping[1]}{\mapping[2]})$.
      If there exists a $\mapping[\flatten]'$ such that $\cptffunc{\mapping[\flatten]}{\mapping[\flatten]'}$, then it must also hold that $\applyEBV{\evalE{\{\expression_3\}, \mergef{\mapping[\flatten], \mapping[\flatten]'}}}{false}$.
      Because the compatibility and evaluation of $\expression_3$ are preserved between the flat and condensed models (as established in the previous proofs for Join and Filter), this entails that $\applyEBV{\evalE{\{\translate(\expression_3)\}, \mergec{\mapping[1], \mapping[2]}}}{false}$.
      It follows that $\mapping[1]$ belongs to $\evalc{\condensedmodel, \diffc{\translate(\expression_1), \translate(\expression_2), \translate(\expression_3)}}$.

      Conversely, let $\mapping \in \evalc{\condensedmodel, \diffc{\translate(\expression_1), \translate(\expression_2), \translate(\expression_3)}}$.

      This means $\mapping \in \evalc{\condensedmodel, \translate(\expression_1)}$ and $\forall \mapping' \in \evalc{\condensedmodel, \translate(\expression_2)}$, either $\neg(\cptcfunc{\mapping}{\mapping'})$ or $(\cptcfunc{\mapping}{\mapping'} \land \applyEBV{\evalE{\{\translate(\expression_3)\}, \mergec{\mapping, \mapping'}}}{false})$.

      Let $\mapping[\flatten] \in \expandfunc{\mapping}$. By IH, $\mapping[\flatten] \in \evalf{\flatmodel, \expression_1}$.
      Consider any $\mapping[\flatten]' \in \evalf{\flatmodel, \expression_2}$. By IH, there is a $\mapping'' \in \evalc{\condensedmodel, \translate(\expression_2)}$ such that $\mapping[\flatten]' \in \expandfunc{\mapping''}$.

      If $\neg(\cptcfunc{\mapping}{\mapping''})$, this implies $\forall \mapping[\flatten]^* \in \expandfunc{\mapping}$ and $\mapping[\flatten]^{**} \in \expandfunc{\mapping''}$, that $\neg(\cptffunc{\mapping[\flatten]^*}{\mapping[\flatten]^{**}})$. Thus, $\neg(\cptffunc{\mapping[\flatten]}{\mapping[\flatten]'})$.

      If $\cptcfunc{\mapping}{\mapping''}$ and $\applyEBV{\evalE{\{\translate(\expression_3)\}, \mergec{\mapping, \mapping''}}}{false}$, then for each corresponding pair of flat mappings $\mapping[\flatten] \in \expandfunc{\mapping}$ and $\mapping[\flatten]' \in \expandfunc{\mapping''}$, it follows that $\cptffunc{\mapping[\flatten]}{\mapping[\flatten]'}$ and also $\applyEBV{\evalE{\{\expression_3\}, \mergef{\mapping[\flatten], \mapping[\flatten]'}}}{false}$.

      Therefore, $\mapping[\flatten] \in \evalf{\flatmodel, \difff{\expression_1, \expression_2, \expression_3}}$. Then, we proved that $\evalf{\flatmodel, \difff{\expression_1, \expression_2, \expression_3}} \equivalent \evalc{\condensedmodel, \diffc{\translate(\expression_1), \translate(\expression_2), \translate(\expression_3)}}$.

      \textbf{Optional operator:} $\evalf{\flatmodel, \leftouterjoinf{\expression_1, \expression_2, \expression_3}} \equivalent \evalc{\condensedmodel, \translate(\leftouterjoinf{\expression_1, \expression_2, \expression_3})}$ must be demonstrated. We rewrite this formula with the result of the $\translate$ function: $\evalf{\flatmodel, \leftouterjoinf{\expression_1, \expression_2, \expression_3}} \equivalent \evalc{\condensedmodel, \leftouterjoinc{\translate(\expression_1), \translate(\expression_2), \translate(\expression_3)}}$.

      Let $\mapping[\flatten] \in \evalf{\flatmodel, \leftouterjoinf{\expression_1, \expression_2, \expression_3}}$. This implies, by definition, that $\mapping[\flatten] \in \evalf{\flatmodel, \unionf{\expression, \expression'}}$, where $\expression = \filterf{\{\expression_3\}, \joinf{\expression_1, \expression_2}}$ and $\expression' = \difff{\expression_1, \expression_2, \expression_3}$.
      As a result, either $\mapping[\flatten] \in \evalf{\flatmodel, \expression}$ or $\mapping[\flatten] \in \evalf{\flatmodel, \expression'}$.

      \begin{itemize}
            \item Case $\mapping[\flatten] \in \evalf{\flatmodel, \expression}$:
                  by the IH applied to the Filter and Join operators, we have $\evalf{\flatmodel, \expression} \equivalent \evalc{\condensedmodel, \translate(\expression)}$.
                  Therefore, there exists $\mapping \in \evalc{\condensedmodel, \translate(\expression)}$ such that $\mapping[\flatten] \in \expandfunc{\mapping}$.
                  Since $\translate(\expression) = \filterc{\{\translate(\expression_3)\}, \joinc{\translate(\expression_1), \translate(\expression_2)}}$, it follows that $\mapping \in \evalc{\condensedmodel, \translate(\expression)}$.
                  Thus, $\mapping[\flatten]$ is included in $\evalc{\condensedmodel, \leftouterjoinc{\translate(\expression_1), \translate(\expression_2), \translate(\expression_3)}}$.
            \item Case $\mapping[\flatten] \in \evalf{\flatmodel, \expression'}$:
                  by IH applied to the Diff operator, $\evalf{\flatmodel, \expression'} \equivalent \evalc{\condensedmodel, \translate(\expression')}$.
                  Therefore, there exists $\mapping \in \evalc{\condensedmodel, \translate(\expression')}$ such that $\mapping[\flatten] \in \expandfunc{\mapping}$.
                  Since $\translate(\expression') = \difff{\translate(\expression_1), \translate(\expression_2), \translate(\expression_3)}$, so $\mapping \in \evalc{\condensedmodel, \unionc{\translate(\expression), \translate(\expression')}}$.

                  Thus, $\mapping[\flatten]$ is covered by $\evalc{\condensedmodel, \leftouterjoinc{\translate(\expression_1), \translate(\expression_2), \translate(\expression_3)}}$.
      \end{itemize}

      Conversely, let $\mapping[\condense] \in \evalc{\condensedmodel, \leftouterjoinc{\translate(\expression_1), \translate(\expression_2), \translate(\expression_3)}}$.
      This implies $\mapping[\condense] \in \evalc{\condensedmodel, \unionc{\translate(A_\flatten), \translate(B_\flatten)}}$.

      Thus, either $\mapping[\condense] \in \evalc{\condensedmodel, \translate(A_\flatten)}$ or $\mapping[\condense] \in \evalc{\condensedmodel, \translate(B_\flatten)}$.

      \begin{itemize}
            \item If $\mapping[\condense] \in \evalc{\condensedmodel, \translate(A_\flatten)}$:
                  then for all $\mapping[\flatten] \in \expandfunc{\mapping[C]}$, by IH, $\mapping[\flatten] \in \evalf{\flatmodel, A_\flatten}$.
                  Thus, $\mapping[\flatten] \in \evalf{\flatmodel, \unionf{A_\flatten, B_\flatten}}$, which means $\mapping[\flatten] \in \evalf{\flatmodel, \leftouterjoinf{\expression_1, \expression_2, \expression_3}}$.
            \item If $\mapping[\condense] \in \evalc{\condensedmodel, \translate(B_\flatten)}$:
                  then for all $\mapping[\flatten] \in \expandfunc{\mapping[C]}$, by IH, $\mapping[\flatten] \in \evalf{\flatmodel, B_\flatten}$.
                  Thus, $\mapping[\flatten] \in \evalf{\flatmodel, \unionf{A_\flatten, B_\flatten}}$, which means $\mapping[\flatten] \in \evalf{\flatmodel, \leftouterjoinf{\expression_1, \expression_2, \expression_3}}$.
      \end{itemize}

      Therefore, $\evalf{\flatmodel, \leftouterjoinf{\expression_1, \expression_2, \expression_3}} \equivalent \\
            \evalc{\condensedmodel, \leftouterjoinc{\translate(\expression_1), \translate(\expression_2), \translate(\expression_3)}}$.

      \textbf{Group operator:} $\evalf{\flatmodel, \groupFFunc[\expressionset_n, \expression]} \equivalent \evalc{\condensedmodel, \translate(\groupFFunc[\expressionset_n, \expression])}$ must be demonstrated. We rewrite this formula with the result of the $\translate$ function: $\evalf{\flatmodel, \groupFFunc[\expressionset_n, \expression]} \equivalent \evalc{\condensedmodel, \groupC([\translate(\expression_1), \dots, \translate(\expression_n)], \translate(\expression))}$.

      Let's consider a solution sequence $\Omega_\flatten = \evalf{\flatmodel, \expression}$ and its condensed equivalent $\Omega_\condense = \evalc{\condensedmodel, \translate(\expression)}$. By the inductive hypothesis, we have $\Omega_\flatten \equivalent \Omega_\condense$. This means that for every flat mapping $\mapping[\flatten] \in \Omega_\flatten$, there is a corresponding condensed mapping $\mapping \in \Omega_\condense$ such that $\mapping[\flatten] \in \expandfunc{\mapping}$, and vice versa.

      The Group operator partitions the solution sequence based on the values of the grouping expressions $\expressionset_n$. In the flat model, the grouping is performed on $\Omega_\flatten$, while in the condensed model, it is performed on $\Omega_\condense$. We need to show that the resulting grouped partitions are equivalent.

      Let $G_\flatten = \groupFFunc[\expressionset_n, \Omega_\flatten]$ and $G_\condense = \groupCFunc[[\translate(\expression_1), \dots, \translate(\expression_n)], \Omega_\condense]$.

      For any group key $k$ in the flat model, the corresponding group is $\{ \mapping[\flatten] \mid \mapping[\flatten] \in \Omega_\flatten \land \evalLE{\expressionset_n, \mapping[\flatten]} = k \}$.
      In the condensed model, the group key is the same, and the group is $\{ \mapping \mid \mapping \in \Omega_\condense \land \evalLE{[\translate(\expression_1), \dots, \translate(\expression_n)], \mapping} = k \}$.

      Since the evaluation of expressions is preserved between the flat and condensed mappings (as shown in Lemma \ref{lemma:evalE}), a flat mapping $\mapping[\flatten]$ and its corresponding condensed mapping $\mapping$ will produce the same group key $k$. Therefore, the set of flat mappings in a group in $G_{\flatten}$ corresponds exactly to the set of flat mappings that can be generated from the condensed mappings in the corresponding group in $G_\condense$.

      Thus, the grouped structures are equivalent, and the theorem holds for the Group operator.

      \textbf{Aggregation operator:} We must demonstrate that: $\evalf{\flatmodel, \aggregationFFunc[\expressionset_n, \aggFuncF, \scalarvals, \groupF]} \equivalent \evalc{\condensedmodel, \translate(\aggregationFFunc[\expressionset_n, \aggFuncF, \scalarvals, \groupF])}$.
      This can be rewritten using the translation function $\translate$ as: $\evalf{\flatmodel, \aggregationFFunc} \equivalent \evalc{\condensedmodel, \aggregationCFunc}$.

      From the IH for the \texttt{Group} operator, we know that the grouped structures are equivalent: $\evalf{\flatmodel, \groupF} \equivalent \evalc{\condensedmodel, \groupC}$.

      Let $G_{\flatten}$ and $G_{\condense}$ be the grouped solution sequences for the flat and condensed models, respectively. By the inductive hypothesis, for each group $g_F \in G_{\flatten}$ defined by a key $k$, there is a corresponding group $g_{\condense} \in G_{\condense}$ for the same key $k$, such that $\bigcup_{\mapping \in g_{\condense}} \expandfunc{\mapping}$ is a permutation of $g_{\flatten}$.

      The flat aggregation function $\aggFuncF$ is applied to each group $g_{\flatten}$. The condensed aggregation function $\aggFuncC$ is applied to each group $g_{\condense}$. We need to show that for each group, the result of the aggregation is the same.

      For the \texttt{COUNT} aggregation, the flat model computes the number of mappings in $g_{\flatten}$. In the condensed model, for each mapping $\mapping \in g_{\condense}$, we compute $\evalE{\card, \mapping}$, which represents the number of flat mappings encoded by $\mapping$. Summing $\evalE{\card, \mapping}$ over all $\mapping \in g_{\condense}$ yields the same total as counting the mappings in $g_{\flatten}$, ensuring equivalence between the two models.

      For other aggregations like \texttt{SUM}, \texttt{AVG}, \texttt{MIN}, \texttt{MAX}, the values are extracted from the variables. Since the $\expand$ function preserves the values of the variables, the set of values passed to the aggregation function is the same for both the flat and condensed models. For example, for each mapping $\mapping[\flatten] \in g_{\flatten}$, there's a corresponding $\mapping \in g_{\condense}$ such that $\mapping[\flatten] \in \expandfunc{\mapping}$ and the values of the relevant variables are the same.

      This equivalence is formally supported by the principles of rewriting queries with aggregation functions using views, as established in the paper by Sarah Cohen \cite{cohen2006rewriting}, which shows that aggregations over base data can be equivalently computed over pre-aggregated views.

      Therefore, for any given group key, the aggregation results will be identical. Since this holds for all groups, the entire result of the aggregation operation is equivalent.

      \textbf{Aggregate join operator:} We must demonstrate that $\evalf{\flatmodel, \aggregateJoinFFunc[\expressionsagg_n]} \equivalent \evalc{\condensedmodel, \translate(\aggregateJoinFFunc[\expressionsagg_n])}$.
      This can be rewritten as: $\evalf{\flatmodel, \aggregateJoinFFunc[Agg_{F1}, \dots, Agg_{Fk}]} \equivalent \evalc{\condensedmodel, \aggregateJoinCFunc[\translate(Agg_{F1}), \dots, \translate(Agg_{Fk})]}$.

      The \texttt{AggregateJoin} operator, as per Definitions \ref{def:aggregate-join-func} and \ref{def:aggregate-join-c-func}, combines the results from multiple aggregation operations based on their common group keys.
      From the proof for the \texttt{Aggregation} operator above, we have established that the result of each flat aggregation $\aggregationF[i]$ is equivalent to the result of its translated condensed counterpart $\translate(\aggregationF[i]) = \aggregationC[i]$.
      This means that for each aggregation, the set of $key \to val$ pairs produced by the flat evaluation is identical to the set produced by the condensed evaluation.
      Since both $\aggregateJoinFFunc$ and $\aggregateJoinCFunc$ operate on an identical set of input pairs and perform the same function of joining these pairs by their keys, their final output must also be equivalent.
      Therefore, the theorem holds for the \texttt{Aggregate Join} operator.
\end{proof}

\subsection{Model - Space usage}
\subsubsection{Space usage}
\label{app:space-usage}
The tables below compare the storage requirements for Blazegraph, Jena TDB, QuaQue-flat and QuaQue-condensed implementations.

\begin{figure}[h]
      \centering
      \includegraphics[width=0.90\textwidth]{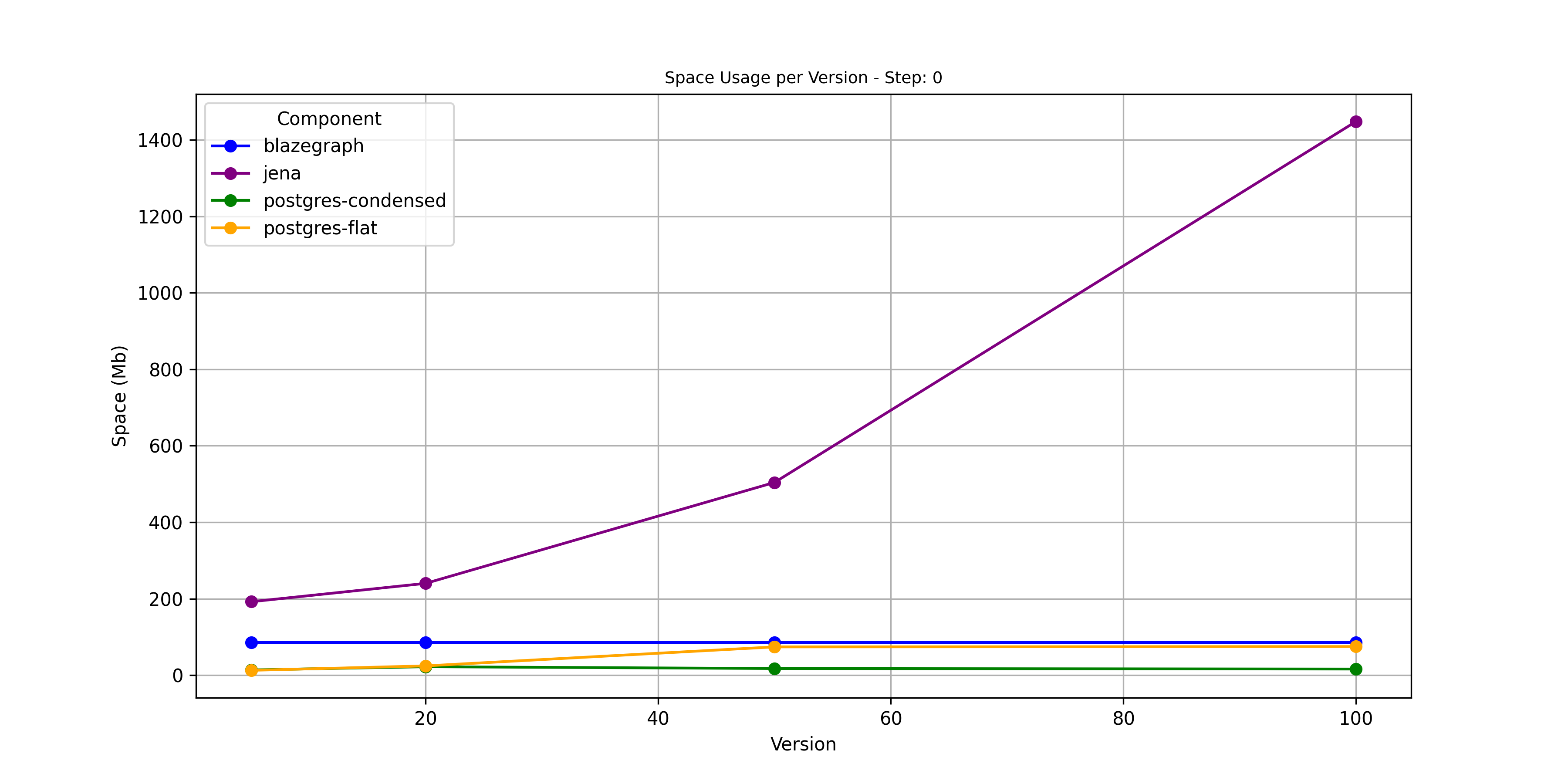}
      \caption{Comparison of space usage between condensed and flat representations for a non-evolutive dataset.}
      \Description{This figure illustrates the differences in space usage between the two representation models across various configurations.}
      \label{fig:space-usage-0}
\end{figure}

\begin{figure}[h]
      \centering
      \includegraphics[width=0.90\textwidth]{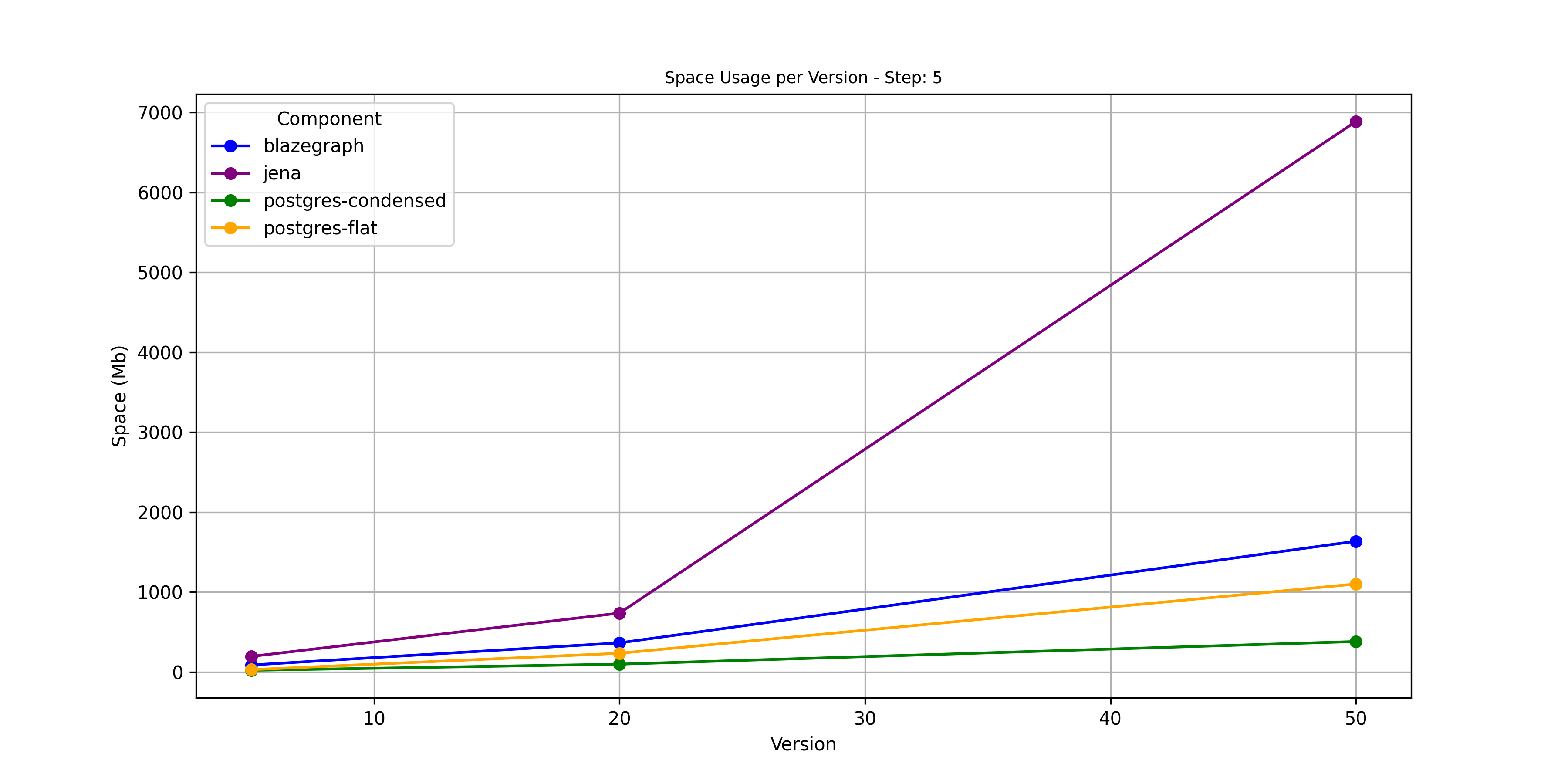}
      \caption{Comparison of space usage between condensed and flat representations for a evolutive dataset.}
      \Description{This figure illustrates the differences in space usage between the two representation models across various configurations.}
      \label{fig:space-usage-5}
\end{figure}

\begin{figure}[h]
      \centering
      \includegraphics[width=0.90\textwidth]{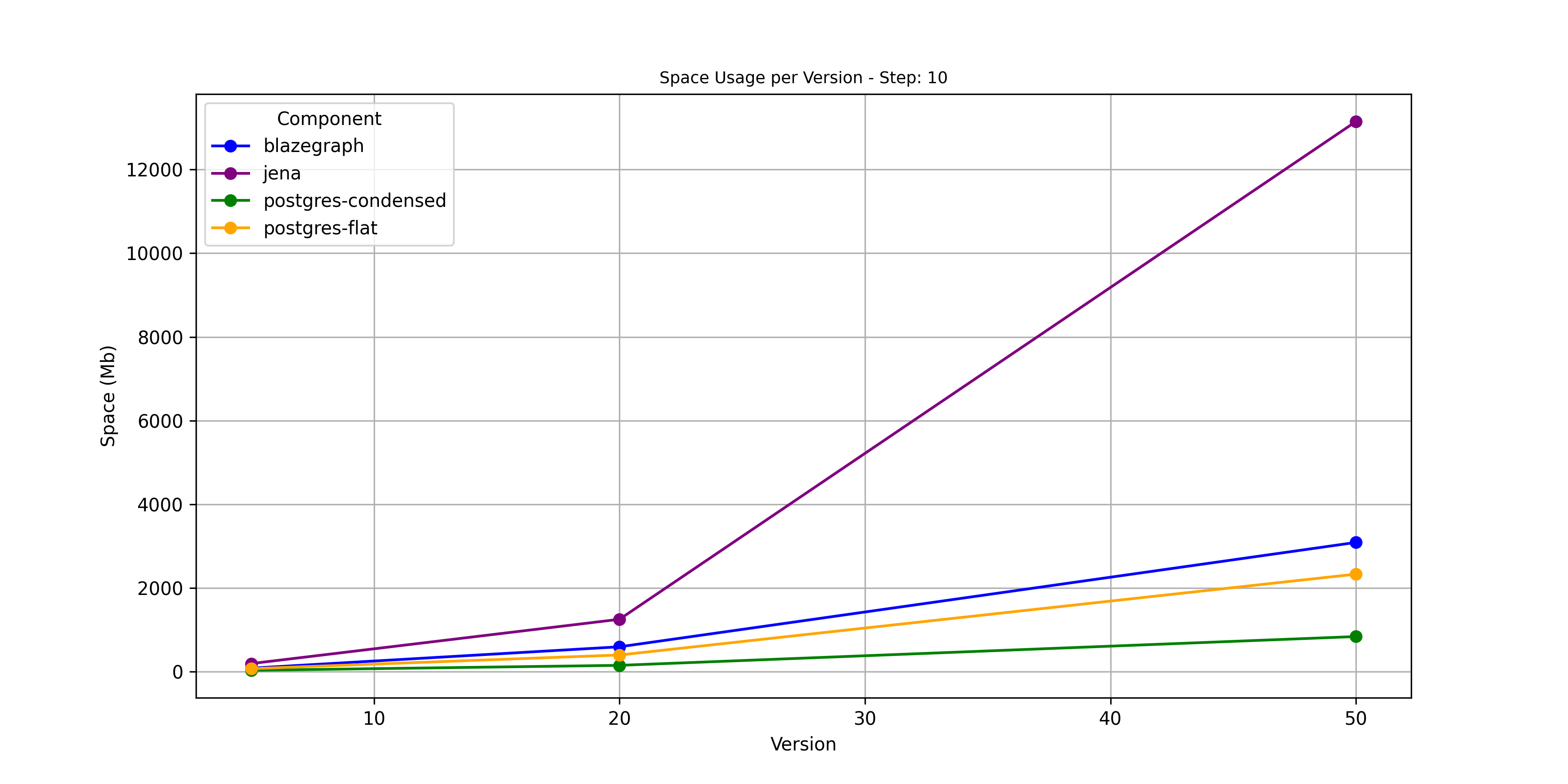}
      \caption{Comparison of space usage between condensed and flat representations for a highly evolutive dataset.}
      \Description{This figure illustrates the differences in space usage between the two representation models across various configurations.}
      \label{fig:space-usage-10}
\end{figure}

\subsection{Dataset query}
\subsubsection{Query time}
\label{appendix:query-times}

This section presents the results of dataset query performance across different systems and configurations.
The tables below compare query execution times for both non-aggregative and aggregative queries, measured on Blazegraph, Jena TDB, QuaQue-flat, and QuaQue-condensed implementations.
Metrics include the 75th percentile, 95th percentile, and median query times for varying dataset sizes and version counts, providing insight into the scalability and efficiency of each approach.

\begin{table}[h]
      \centering
      \small
      \caption{Comparison of average times for non-aggregative queries across all components.}
      \begin{tabular}{|c|c||c|c|c||c|c|c||c|c|c||c|c|c|}
            \hline
            \multicolumn{2}{|c||}{Configuration} & \multicolumn{3}{c||}{\cellcolor{blazegraphcolor}Blazegraph} & \multicolumn{3}{c||}{\cellcolor{jenacolor}Jena TDB} & \multicolumn{3}{c||}{\cellcolor{quaquecondensedcolor}QuaQue-condensed} & \multicolumn{3}{c|}{\cellcolor{quaqueflatcolor}QuaQue-flat}                                                                                                  \\
            \hline
            Step                                 & Version                                                     & 75th P.                                             & 95th P.                                                                & Median                                                      & 75th P. & 95th P. & Median  & 75th P. & 95th P. & Median  & 75th P. & 95th P. & Median         \\
            \hline
            0                                    & 5                                                           & 86.30                                               & 121.40                                                                 & 69.40                                                       & 83.33   & 104.81  & 71.40   & 75.30   & 98.47   & 59.85   & 64.20   & 101.57  & \textbf{46.50} \\
            0                                    & 20                                                          & 79.75                                               & 106.26                                                                 & 66.65                                                       & 101.70  & 130.23  & 88.35   & 116.55  & 141.84  & 100.35  & 70.40   & 90.86   & \textbf{50.65} \\
            0                                    & 50                                                          & 114.28                                              & 143.27                                                                 & 91.40                                                       & 188.93  & 222.07  & 170.65  & 197.40  & 224.91  & 180.30  & 56.88   & 76.23   & \textbf{42.30} \\
            0                                    & 100                                                         & 174.43                                              & 215.38                                                                 & 153.00                                                      & 233.60  & 277.34  & 218.35  & 330.38  & 357.05  & 307.80  & 74.50   & 115.66  & \textbf{56.00} \\
            5                                    & 5                                                           & 94.00                                               & 139.58                                                                 & 69.60                                                       & 86.55   & 121.19  & 72.20   & 131.90  & 163.56  & 109.55  & 57.27   & 82.99   & \textbf{48.85} \\
            5                                    & 20                                                          & 241.60                                              & 283.02                                                                 & 220.25                                                      & 326.52  & 369.75  & 289.05  & 457.15  & 496.12  & 429.40  & 79.03   & 124.72  & \textbf{59.05} \\
            5                                    & 50                                                          & 933.20                                              & 1000.39                                                                & 885.45                                                      & 1341.42 & 1460.86 & 1283.75 & 1724.75 & 1844.65 & 1648.70 & 59.48   & 82.65   & \textbf{44.10} \\
            10                                   & 5                                                           & 110.20                                              & 138.01                                                                 & 96.20                                                       & 97.28   & 129.05  & 85.20   & 162.43  & 208.91  & 138.75  & 57.45   & 82.13   & \textbf{43.15} \\
            10                                   & 20                                                          & 426.07                                              & 530.04                                                                 & 377.35                                                      & 535.70  & 637.97  & 486.65  & 640.15  & 721.86  & 599.40  & 86.28   & 135.91  & \textbf{65.20} \\
            10                                   & 50                                                          & 1889.95                                             & 1975.68                                                                & 1826.15                                                     & 2358.30 & 2497.15 & 2236.75 & 3056.75 & 3200.38 & 2893.05 & 65.20   & 91.95   & \textbf{47.60} \\
            \hline
      \end{tabular}
\end{table}

\begin{table}[h]
      \centering
      \small
      \caption{Comparison of average times for aggregative queries across all components.}
      \begin{tabular}{|c|c||c|c|c||c|c|c||c|c|c||c|c|c|}
            \hline
            \multicolumn{2}{|c||}{Configuration} & \multicolumn{3}{c||}{\cellcolor{blazegraphcolor}Blazegraph} & \multicolumn{3}{c||}{\cellcolor{jenacolor}Jena TDB} & \multicolumn{3}{c||}{\cellcolor{quaquecondensedcolor}QuaQue-condensed} & \multicolumn{3}{c|}{\cellcolor{quaqueflatcolor}QuaQue-flat}                                                                                                                \\
            \hline
            Step                                 & Version                                                     & 75th P.                                             & 95th P.                                                                & Median                                                      & 75th P. & 95th P. & Median         & 75th P. & 95th P. & Median         & 75th P. & 95th P. & Median         \\
            \hline
            0                                    & 5                                                           & 71.75                                               & 113.89                                                                 & 56.93                                                       & 72.00   & 103.51  & 53.86          & 60.79   & 103.02  & \textbf{45.36} & 67.21   & 104.49  & 48.86          \\
            0                                    & 20                                                          & 45.86                                               & 64.11                                                                  & \textbf{41.50}                                              & 57.18   & 95.02   & 42.29          & 74.68   & 122.38  & 50.57          & 77.36   & 117.25  & 53.86          \\
            0                                    & 50                                                          & 71.54                                               & 136.28                                                                 & 53.57                                                       & 69.00   & 98.30   & 53.21          & 48.04   & 81.31   & \textbf{37.00} & 51.71   & 79.16   & 37.21          \\
            0                                    & 100                                                         & 48.43                                               & 70.59                                                                  & \textbf{41.71}                                              & 83.71   & 137.97  & 58.43          & 74.96   & 126.25  & 53.50          & 70.75   & 103.44  & 46.93          \\
            5                                    & 5                                                           & 77.36                                               & 152.74                                                                 & 52.71                                                       & 59.54   & 96.30   & \textbf{42.07} & 70.61   & 100.35  & 49.14          & 60.68   & 92.64   & 46.29          \\
            5                                    & 20                                                          & 101.39                                              & 128.54                                                                 & 88.71                                                       & 97.64   & 154.01  & 73.07          & 101.79  & 139.36  & 71.64          & 81.86   & 137.03  & \textbf{56.57} \\
            5                                    & 50                                                          & 404.00                                              & 459.72                                                                 & 368.50                                                      & 222.89  & 268.16  & 203.07         & 232.71  & 280.91  & 218.57         & 58.86   & 95.07   & \textbf{44.93} \\
            10                                   & 5                                                           & 81.29                                               & 131.74                                                                 & 65.36                                                       & 59.39   & 87.23   & 42.93          & 79.68   & 152.66  & 51.29          & 46.82   & 75.23   & \textbf{34.43} \\
            10                                   & 20                                                          & 235.43                                              & 303.52                                                                 & 170.43                                                      & 147.86  & 207.18  & 109.07         & 123.39  & 163.47  & 103.50         & 80.61   & 138.06  & \textbf{53.36} \\
            10                                   & 50                                                          & 730.89                                              & 803.31                                                                 & 704.14                                                      & 353.11  & 379.76  & 332.86         & 560.25  & 617.20  & 536.29         & 74.25   & 106.71  & \textbf{63.79} \\
            \hline
      \end{tabular}
\end{table}

%% file: bibliography.bib
@String{Computing = "Computing" }

@String{Computer = "{IEEE} Computer" }

@String{Springer = "Springer-Verlag" }

@artifactsoftware{R,
  title        = {R: A Language and Environment for Statistical Computing},
  author       = {{R Core Team}},
  organization = {R Foundation for Statistical Computing},
  address      = {Vienna, Austria},
  year         = {2019},
  url          = {https://www.R-project.org/}
}

@article{samuel_representation_2020,
  title    = {Representation of concurrent points of view of urban changes for city models},
  issn     = {1435-5930, 1435-5949},
  url      = {http://link.springer.com/10.1007/s10109-020-00319-1},
  doi      = {10.1007/s10109-020-00319-1},
  language = {en},
  urldate  = {2020-03-06},
  journal  = {Journal of Geographical Systems},
  author   = {Samuel, John and Servigne, Sylvie and Gesquière, Gilles},
  month    = feb,
  year     = {2020}
}

@incollection{chaturvedi_managing_2017,
  address   = {Cham},
  title     = {Managing {Versions} and {History} {Within} {Semantic} {3D} {City} {Models} for the {Next} {Generation} of {CityGML}},
  isbn      = {9783319256917},
  url       = {https://doi.org/10.1007/978-3-319-25691-7_11},
  language  = {en},
  urldate   = {2025-02-07},
  booktitle = {Advances in {3D} {Geoinformation}},
  publisher = {Springer International Publishing},
  author    = {Chaturvedi, Kanishk and Smyth, Carl Stephen and Gesquière, Gilles and Kutzner, Tatjana and Kolbe, Thomas H.},
  editor    = {Abdul-Rahman, Alias},
  year      = {2017},
  doi       = {10.1007/978-3-319-25691-7_11},
  pages     = {191--206}
}

@inproceedings{Swierstra2014,
  author    = {Swierstra, Wouter and L\"{o}h, Andres},
  title     = {The Semantics of Version Control},
  year      = {2014},
  isbn      = {9781450332101},
  publisher = {Association for Computing Machinery},
  address   = {New York, NY, USA},
  url       = {https://doi.org/10.1145/2661136.2661137},
  doi       = {10.1145/2661136.2661137},
  booktitle = {Proceedings of the 2014 ACM International Symposium on New Ideas, New Paradigms, and Reflections on Programming \& Software},
  pages     = {43–54},
  numpages  = {12},
  location  = {Portland, Oregon, USA},
  series    = {Onward! 2014}
}

@article{kutzner_citygml_2020,
  title      = {{CityGML} 3.0: {New} {Functions} {Open} {Up} {New} {Applications}},
  volume     = {88},
  issn       = {2512-2789, 2512-2819},
  shorttitle = {{CityGML} 3.0},
  url        = {http://link.springer.com/10.1007/s41064-020-00095-z},
  doi        = {10.1007/s41064-020-00095-z},
  language   = {en},
  number     = {1},
  urldate    = {2025-02-07},
  journal    = {PFG – Journal of Photogrammetry, Remote Sensing and Geoinformation Science},
  author     = {Kutzner, Tatjana and Chaturvedi, Kanishk and Kolbe, Thomas H.},
  month      = feb,
  year       = {2020},
  pages      = {43--61}
}

@inproceedings{Bulteau2023,
  author    = {Laurent Bulteau and Pierre-Yves David and Florian Horn},
  title     = {The Problem of Discovery in Version Control Systems},
  booktitle = {Proceedings of the XII Latin-American Algorithms, Graphs and Optimization Symposium (LAGOS 2023)},
  year      = {2023},
  pages     = {209--216},
  publisher = {Elsevier},
  address   = {Huatulco, Oaxaca, Mexico},
  doi       = {10.1016/j.procs.2023.08.231},
  url       = {https://hal.science/hal-03830513/document}
}

@inproceedings{Yilmaz2025,
  author    = {Gunce Su Yilmaz and Jens Dittrich},
  title     = {Generic Version Control: Configurable Versioning for Application-Specific Requirements},
  booktitle = {Proceedings of the 15th Annual Conference on Innovative Data Systems Research (CIDR 2025)},
  year      = {2025},
  pages     = {24},
  publisher = {CIDR},
  address   = {Amsterdam, The Netherlands},
  url       = {https://vldb.org/cidrdb/papers/2025/p24-yilmaz.pdf}
}

@article{sveen_geomdiff_2020,
  title    = {{GeomDiff} — an algorithm for differential geospatial vector data comparison},
  volume   = {5},
  issn     = {2363-7501},
  url      = {https://opengeospatialdata.springeropen.com/articles/10.1186/s40965-020-00076-4},
  doi      = {10.1186/s40965-020-00076-4},
  language = {en},
  number   = {1},
  urldate  = {2025-02-07},
  journal  = {Open Geospatial Data, Software and Standards},
  author   = {Sveen, Atle Frenvik},
  month    = dec,
  year     = {2020},
  pages    = {3}
}

@article{samuel2018urbanco2fab,
  title     = {Urbanco2fab: comprehension of concurrent viewpoints of urban fabric based on git},
  author    = {Samuel, John and Servigne, Sylvie and Gesquiere, Gilles},
  journal   = {ISPRS Annals of the Photogrammetry, Remote Sensing and Spatial Information Sciences},
  volume    = {4},
  pages     = {65--72},
  year      = {2018},
  publisher = {Copernicus Publications G{\"o}ttingen, Germany}
}

@article{cyganiak2005relational,
  title   = {A relational algebra for SPARQL},
  author  = {Cyganiak, Richard},
  journal = {Digital Media Systems Laboratory HP Laboratories Bristol. HPL-2005-170},
  volume  = {35},
  number  = {9},
  year    = {2005}
}

@inproceedings{pelgrin2023glenda,
  title        = {GLENDA: querying RDF archives with full SPARQL},
  author       = {Pelgrin, Olivier and Taelman, Ruben and Gal{\'a}rraga, Luis and Hose, Katja},
  booktitle    = {European Semantic Web Conference},
  pages        = {75--80},
  year         = {2023},
  organization = {Springer}
}

@article{pelgrinefficient,
  title  = {Efficient Management of Large RDF Archives},
  author = {Pelgrin, Olivier Paul}
}

@article{10.14778/1920841.1920877,
  author     = {Neumann, Thomas and Weikum, Gerhard},
  title      = {x-RDF-3X: fast querying, high update rates, and consistency for RDF databases},
  year       = {2010},
  issue_date = {September 2010},
  publisher  = {VLDB Endowment},
  volume     = {3},
  number     = {1–2},
  issn       = {2150-8097},
  url        = {https://doi.org/10.14778/1920841.1920877},
  doi        = {10.14778/1920841.1920877},
  journal    = {Proc. VLDB Endow.},
  month      = sep,
  pages      = {256–263},
  numpages   = {8}
}

@inproceedings{7786197,
  author    = {Cerdeira-Pena, Ana and Fariña, Antonio and Fernández, Javier D. and Martínez-Prieto, Miguel A.},
  booktitle = {2016 Data Compression Conference (DCC)},
  title     = {Self-Indexing RDF Archives},
  year      = {2016},
  volume    = {},
  number    = {},
  pages     = {526-535},
  doi       = {10.1109/DCC.2016.40}
}

@inproceedings{volkel2005semversion,
  title={Semversion: A versioning system for rdf and ontologies},
  author={Volkel, Max and Winkler, Wolf and Sure, York and Kruk, S Ryszard and Synak, Marcin},
  booktitle={Proc. of ESWC},
  year={2005}
}

@article{vander2013r,
  title={R\&Wbase: git for triples.},
  author={Vander Sande, Miel and Colpaert, Pieter and Verborgh, Ruben and Coppens, Sam and Mannens, Erik and Van de Walle, Rik},
  journal={LDOW},
  volume={996},
  year={2013}
}

@inproceedings{graube2014r43ples,
  title={R43ples: Revisions for triples},
  author={Graube, Markus and Hensel, Stephan and Urbas, Leon},
  booktitle={Proceedings of the 1st Workshop on Linked Data Quality co-located with 10th International Conference on Semantic Systems (SEMANTiCS 2014)},
  year={2014}
}

@inproceedings{gao2016rdf,
  title={RDF-TX: A Fast, User-Friendly System for Querying the History of RDF Knowledge Bases.},
  author={Gao, Shi and Gu, Jiaqi and Zaniolo, Carlo},
  booktitle={EDBT},
  pages={269--280},
  year={2016}
}

@inproceedings{vinasco2021towards,
  title     = {Towards a semantic web representation from a 3D geospatial urban data model},
  author    = {Vinasco-Alvarez, Diego and Samuel, John and Servigne, Sylvie and Gesqui{\`e}re, Gilles},
  booktitle = {SAGEO 2021, 16{\`e}me Conf{\'e}rence Internationale de la G{\'e}omatique, de l'Analyse Spatiale et des Sciences de l'Information G{\'e}ographique.},
  pages     = {227--238},
  url       = {https://hal.science/hal-03240567/file/SAGEO_2021.pdf},
  year      = {2021}
}

@article{eriksson_comparison_2021,
  title    = {Comparison of versioning methods to improve the information flow in the planning and building processes},
  volume   = {25},
  issn     = {1361-1682, 1467-9671},
  url      = {https://onlinelibrary.wiley.com/doi/10.1111/tgis.12672},
  doi      = {10.1111/tgis.12672},
  language = {en},
  number   = {1},
  urldate  = {2025-02-07},
  journal  = {Transactions in GIS},
  author   = {Eriksson, Helen and Sun, Jing and Tarandi, Väino and Harrie, Lars},
  month    = feb,
  year     = {2021},
  pages    = {134--163}
}

@article{hogan_everything_2014,
  series   = {Semantic {Web} {Challenge} 2013},
  title    = {Everything you always wanted to know about blank nodes},
  volume   = {27-28},
  issn     = {1570-8268},
  url      = {https://www.sciencedirect.com/science/article/pii/S1570826814000481},
  doi      = {10.1016/j.websem.2014.06.004},
  urldate  = {2025-02-07},
  journal  = {Journal of Web Semantics},
  author   = {Hogan, Aidan and Arenas, Marcelo and Mallea, Alejandro and Polleres, Axel},
  month    = aug,
  year     = {2014},
  pages    = {42--69}
}

@misc{gil2024convergconcurrentversioningknowledge,
  title         = {ConVer-G: Concurrent versioning of knowledge graphs},
  author        = {Jey Puget Gil and Emmanuel Coquery and John Samuel and Gilles Gesquiere},
  year          = {2024},
  eprint        = {2409.04499},
  archiveprefix = {arXiv},
  primaryclass  = {cs.DB},
  url           = {https://arxiv.org/abs/2409.04499}
}

@inproceedings{bayoudhi2020survey,
  title        = {A survey on versioning approaches and tools},
  author       = {Bayoudhi, Leila and Sassi, Najla and Jaziri, Wassim},
  booktitle    = {International Conference on Intelligent Systems Design and Applications},
  pages        = {1155--1164},
  year         = {2020},
  organization = {Springer}
}

@article{taelman2019triple,
  title     = {Triple storage for random-access versioned querying of RDF archives},
  author    = {Taelman, Ruben and Vander Sande, Miel and Van Herwegen, Joachim and Mannens, Erik and Verborgh, Ruben},
  journal   = {Journal of Web Semantics},
  volume    = {54},
  pages     = {4--28},
  year      = {2019},
  publisher = {Elsevier}
}

@article{cohen2006rewriting,
  title     = {Rewriting queries with arbitrary aggregation functions using views},
  author    = {Cohen, Sara and Nutt, Werner and Sagiv, Yehoshua},
  journal   = {ACM Transactions on Database Systems (TODS)},
  volume    = {31},
  number    = {2},
  pages     = {672--715},
  year      = {2006},
  publisher = {ACM New York, NY, USA}
}

@article{klump2021versioning,
  title     = {Versioning data is about more than revisions: A conceptual framework and proposed principles},
  author    = {Klump, Jens and Wyborn, Lesley and Wu, Mingfang and Martin, Julia and Downs, Robert R and Asmi, Ari},
  journal   = {Data Science Journal},
  volume    = {20},
  number    = {1},
  pages     = {12},
  year      = {2021},
  publisher = {Ubiquity Press}
}

@phdthesis{arndt-2020-dissertation,
  author    = {Arndt, Natanael},
  doi       = {10.33968/9783966270205-00},
  isbn      = {978-3-96627-019-9},
  publisher = {{Open-Access-Hochschulverlag Leipzig}},
  school    = {Universität Leipzig},
  title     = {{Distributed Collaboration on Versioned Decentralized RDF Knowledge Bases}},
  url       = {https://oa-hochschulverlag.htwk-leipzig.de/fileadmin/portal/m_oa_hochschulverlag/Katalog/Arndt/2020-08-04-natanael-diss-final-pdfx.pdf},
  year      = 2020
}

@inproceedings{taelman2018ostrich,
  title={OSTRICH: versioned random-access triple store},
  author={Taelman, Ruben and Vander Sande, Miel and Verborgh, Ruben},
  booktitle={Companion Proceedings of the The Web Conference 2018},
  pages={127--130},
  year={2018}
}

@article{anderson2016transaction,
  title={Transaction-Time Queries in Dydra.},
  author={Anderson, James and Bendiken, Arto},
  journal={MEPDaW/LDQ@ ESWC},
  volume={1585},
  pages={11--19},
  year={2016}
}

@incollection{jensen2009temporal,
  title     = {Temporal Database},
  author    = {Jensen, Christian S{\o}ndergaard and Snodgrass, Richard T},
  booktitle = {Encyclopedia of Database Systems},
  pages     = {2957--2960},
  year      = {2009},
  publisher = {Springer}
}

@article{10.1145/2380776.2380786,
  author     = {Kulkarni, Krishna and Michels, Jan-Eike},
  title      = {Temporal features in SQL:2011},
  year       = {2012},
  issue_date = {September 2012},
  publisher  = {Association for Computing Machinery},
  address    = {New York, NY, USA},
  volume     = {41},
  number     = {3},
  issn       = {0163-5808},
  url        = {https://doi.org/10.1145/2380776.2380786},
  doi        = {10.1145/2380776.2380786},
  journal    = {SIGMOD Rec.},
  month      = {oct},
  pages      = {34–43},
  numpages   = {10}
}

@article{ZOLKIFLI2018408,
  title    = {Version Control System: A Review},
  journal  = {Procedia Computer Science},
  volume   = {135},
  pages    = {408-415},
  year     = {2018},
  note     = {The 3rd International Conference on Computer Science and Computational Intelligence (ICCSCI 2018) : Empowering Smart Technology in Digital Era for a Better Life},
  issn     = {1877-0509},
  doi      = {https://doi.org/10.1016/j.procs.2018.08.191},
  url      = {https://www.sciencedirect.com/science/article/pii/S1877050918314819},
  author   = {Nazatul Nurlisa Zolkifli and Amir Ngah and Aziz Deraman}
}

@misc{VCITYWebsite,
  author = {VCity Team},
  title  = {VCity project website},
  year   = {2024},
  month  = {September},
  day    = {27},
  url    = {https://projet.liris.cnrs.fr/vcity/}
}

@misc{W3CSPARQL,
  author = {W3C},
  title  = {SPARQL 1.1 Query Language},
  year   = {2024},
  month  = {September},
  day    = {27},
  url    = {https://www.w3.org/TR/sparql11-query/}
}

@misc{W3CConcepts,
  author = {W3C},
  title  = {RDF 1.1 Concepts and Abstract Syntax},
  year   = {2024},
  month  = {September},
  day    = {27},
  url    = {https://www.w3.org/TR/rdf11-concepts/}
}

@misc{W3CQuery,
  author = {W3C},
  title  = {SPARQL Query Language for RDF},
  year   = {2024},
  month  = {September},
  day    = {27},
  url    = {https://www.w3.org/TR/rdf-sparql-query/}
}

@misc{W3CPROV,
  author = {W3C},
  title  = {PROV-Overview},
  year   = {2025},
  month  = {January},
  day    = {16},
  url    = {https://www.w3.org/TR/prov-overview/}
}

@misc{QRI,
  author = {Qri team},
  title  = {Qri CLI},
  year   = {2024},
  month  = {September},
  day    = {27},
  url    = {https://github.com/qri-io/qri}
}

@misc{GeoGig,
  author = {GeoGig team},
  title  = {GeoGig - Geospatial Distributed Version Control System},
  year   = {2024},
  month  = {September},
  day    = {27},
  url    = {https://github.com/locationtech/geogig}
}

@inproceedings{cuevas2020versioned,
  title     = {Versioned Queries over RDF Archives: All You Need is SPARQL?},
  author    = {Cuevas, Ignacio and Hogan, Aidan},
  booktitle = {MEPDaW@ ISWC},
  pages     = {43--52},
  year      = {2020}
}
